\documentclass[12pt]{iopart}

\usepackage{graphicx}

\begin{document}
\title[A three-component description of multiplicity distributions]
{A three-component description of multiplicity distributions 
in $pp$ collisions at the LHC}

\author{I Zborovsk\'{y}}
\address{ 
Nuclear Physics Institute, Academy of Sciences of the Czech Republic,\\ 
250 68 \v{R}e\v{z}, Czech Republic\\ 
}

\ead{zborovsky@ujf.cas.cz}

\begin{abstract}
A possible signal of new phenomena emerging in the global characteristics
of multiparticle production in hadron interactions at TeV energies is studied.
The multiplicity distributions of charged particles measured in proton-proton collisions 
at the Large Hadron Collider at CERN 
are analyzed in the phenomenological framework of the weighted superposition of
three negative binomial distribution functions.
The examination of the experimental data indicates the existence of a narrow peak at low multiplicities 
which can be described by a separate component in the total distribution.
The multiplicity characteristics of the third component reveal approximate energy and pseudorapidity 
invariance, of which a physical explanation represents a challenging problem
in high energy multiple particle production.  
\end{abstract}
\maketitle

\section{Introduction}

The multiple production of particles in high energy hadron interactions 
has received a lot of attention over the years.
A renewal in interest in this topic coincided with the operation
of the Large Hadron Collider (LHC) at CERN.
Among the first results obtained at the LHC were the multiplicity measurements
in proton-proton collisions. The ATLAS, CMS, ALICE  and
LHCb Collaborations provided high statistic data on multiplicity distributions (MDs)
of charged particles produced in the new super high energy domain.

The study of particle production as a function of multiplicity
has revealed the very popular Koba-Nielsen-Olesen (KNO) scaling \cite{KNO}.
In $pp/p\bar{p}$ collisions, scaling
in the full phase-space holds up to the highest energy of the CERN
Intersecting Storage Rings 
but it is clearly violated \cite{UA5a}
in the energy region of the CERN Super Proton Synchrotron collider and beyond.
The strong violation of KNO scaling
was observed also at the LHC energy $\sqrt{s}=7$~TeV
in the limited pseudorapidity intervals $(|\eta|<2.4)$ though for
small pseudorapidity
windows $(|\eta|<0.5)$ the scaling is approximately valid \cite{CMS}.
A related phenomenon is the so-called negative
binomial regularity  which is the occurrence of the negative
binomial distribution (NBD) in different interactions
over a wide range of collision energies.
The UA5 Collaboration showed that MDs 
of charged particles  
in non-single-diffractive (NSD) $pp/p\bar{p}$ collisions
can be described by NBD up to the energy $\sqrt{s}=546$~GeV
both in the full phase-space \cite{UA5b}
and in symmetric pseudorapidity windows \cite{UA5c}.
First analysis of MDs measured by the ALICE Collaboration \cite{ALICE1} indicated
that NBD describes the data in the small pseudorapidity window $|\eta|<0.5$
up to the energy $\sqrt{s}=2360$~GeV \cite{Mizoguchi}.
The accuracy of such a description is however subject to further disputes and investigations \cite{Prorok}.
Much effort has been made to explain the negative binomial form of MDs observed
in many situations, still its physical origin has not been fully
understood \cite{Grosse}.

The shape of MD of charged particles produced in sufficiently large windows of pseudorapidity
in high energy hadron-hadron collisions is quite different.
The full phase-space data on multiplicities \cite{UA5d}
obtained from the NSD events in $p\bar{p}$ collisions at $\sqrt{s}= 900$~GeV
indicated that besides KNO, the
negative binomial regularity is violated as well.
The measurements of MD at the energy $\sqrt{s}= 1800$~GeV
by the E735 Collaboration \cite{E735} at the Fermilab Tevatron
showed even stronger deviation of data from NBD.
Though with large mutual discrepancies at high multiplicities,
both data demonstrate a narrow maximum and a shoulder-like structure around $n\sim~2<n>$.
Despite its correct qualitative behavior, single NBD is not sufficient to
describe the experimental data.
More functions were suggested \cite{Fowler,Walker} to investigate 
the properties of high energy multiple particle production. 
A systematic study of the complex form of MDs was performed in the framework of the
two-component model \cite{GiovUgo1} and its three-component modification \cite{GiovUgo2}.
The multicomponent quark gluon string model (QGSM) \cite{Kaidalov} was used \cite{Matinyan}
to study the role of multiparton collisions.  
There also exist other approaches to particle
production at high energies (for a review see \cite{Grosse}).

The ATLAS \cite{ATLAS}, CMS \cite{CMS} and ALICE \cite{ALICE2}  
Collaborations measured MDs of charged particles in proton-proton collisions 
at the LHC energies $\sqrt{s}=0.9$, 2.36 and 7~TeV 
using different kinematic ranges and different classes of events.
The measurements in the central pseudorapidity windows allow to study evolution of the narrow maximum
and the broad shoulder of the distributions at high $n$
both with collision energy and pseudorapidity. The shoulder-like structure of MD
was investigated in the framework of independent
pair parton interaction (IPPI) \cite{Dremin1}.
The IPPI model was applied \cite{Dremin2} to experimental data measured by the CMS Collaboration \cite{CMS}
at $ \sqrt{s}=7$~TeV in various pseudorapidity intervals.
The CMS data were analyzed \cite{Ghosh} by weighted superposition of two NBDs 
representing soft and semihard components of particle production.

In this paper we study the three-component description of data on MD of charged particles produced  
in proton-proton collisions at the LHC. 
Each component is represented by a single NBD and parametrized by two parameters $\bar{n}$ and $k$. 
The total distribution is the weighted sum of three NBDs with
eight free parameters altogether. 
The high statistic data on MD measured at the LHC allow us to extract
these parameters in most cases reasonably well. The analysis of the ATLAS measurements \cite{ATLAS}
shows us that the two-component description of MD is unsatisfactory. Within the
multi-component NBD parametrization, the structure of the data indicates the necessity
of a third component in the region of low multiplicities.
The shape of the ATLAS data at low $n$ is critical to the IPPI model and to the two-component 
superposition of NBDs. 
The goal of our analysis is to obtain a detailed description of the high statistic
data on MD in $pp$ collisions at the LHC in the framework of weighted superposition 
of three NBDs and study its properties. 
We demonstrate that, besides the shoulder-like structure at high $n$, the data 
manifest a distinct peak at a maximum which can be described by the third  component of the total distribution. 
We show that the obtained parametrization of the LHC data on MD reveals some invariant features.

\section{Weighted superposition of NBDs}

The multiple production of particles in the soft processes at high energies
includes highly non-perturbative effects of QCD. A comparison of experimental data with various 
distributions relies mostly on phenomenological approaches in this field.
The prominent role in the phenomenological description of data on MD in 
$ pp/\bar{p}p$ collisions is the two-parameter NBD
\begin{equation}
P(n,\bar{n},k)=\frac{\Gamma(n+k)}{\Gamma(k)\Gamma(n+1)}
\left[\frac{\bar{n}}{k+\bar{n}}\right]^n
\left[\frac{k}{k+\bar{n}}\right]^k.
\label{eq:r1}
\end{equation}
The parameter $\bar{n}$ is the average multiplicity and $k$ characterizes the width of
the distribution.
The Poisson distribution is obtained for $k=\infty$.
A remarkable property is the convolution of NBDs 
which results again in NBD with modified parameters. 
A superposition of NBDs is a well-known idea exploited by different approaches
to multiple particle production in hadron collisions.
It is based on the assumption that MD of the created particles
can be expressed as 
\begin{equation}
P(n)=
\sum_{i=1}^N \alpha_i P(n,\bar{n}_i,k_i), \ \ \ \
\sum_{i=1}^N \alpha_i = 1,
\label{eq:r2}
\end{equation}
where $P(n,\bar{n}_i,k_i)$ is given by (\ref{eq:r1}).
In the general case, the corresponding number of free parameters is $3N-1$.
In specific models of this type, the number of parameters can be smaller.
In the IPPI model there are three parameters, $\alpha$, $m$ and $k$,  given by the conditions
$\alpha_i=\alpha^i$, $\bar{n}_i=im$ and $k_i=ik$.
The parameter $\alpha$ is defined by the number $N$ of the independent
active parton pairs via the normalization of the probabilities $ \alpha_i$ to unity.
The value of $N=6$ was used in the analysis \cite{Dremin2} of the CMS data \cite{CMS} at the energy
$\sqrt{s}=7$~TeV.
In the QGSM model the probabilities $\alpha_i$ are proportional to the cross
sections for simultaneous production of $2i$ chains created in $i$-fold parton collisions \cite{Matinyan}.
The MD of the $i$ cut-pomerons contribution, 
$P(n,\bar{n}_i,k_i)$, is assumed to have a Poisson-like form.
The characteristic feature of both models is an increase of the mean multiplicities 
$ \bar{n}_i=i\bar{n}_1$ ($\bar{n}_i=i[a+b \ln (s/s_0i^2)]$) with decreasing values of $\alpha_i$.
In this paper we consider the function (\ref{eq:r2}) for $N=3$ in a formal way
and try to establish values of the eight independent parameters ($\bar{n}_i$, $k_i$,  $\alpha_2$ and $\alpha_3$)  
to obtain a detailed description of data on MD measured with high accuracy in  
the experiments at the LHC.

\section{Analysis of data}

The MDs of charged particles produced in proton-proton collisions at the LHC
were measured both in the central and forward pseudorapidity region.
The phenomenological studies of the measurements have been mainly 
carried out \cite{Dremin2,Ghosh,Ghosh1,Praszalowicz}
with the CMS data.
In this paper we present results of a combined analysis of data on MD obtained by the
ATLAS, CMS and  ALICE Collaborations in the central interaction region and by the 
LHCb Collaboration in the forward region.

The ATLAS Collaboration measured \cite{ATLAS} the charged particle MD in different
phase-space regions using various multiplicity cuts in the pseudorapidity window $|\eta|<2.5$.
The distributions are based on the experimental analysis of a large number of events
which accumulated in particular at the energies $\sqrt{s}=7$ and 0.9 TeV.
The most inclusive  phase-space region covered by the measurements corresponds to
the conditions $p_T > 100$~MeV/c and  $n_{ch}\ge 2$.
The record number of the analyzed events in this region exceeds 10M and 0.3M
at $\sqrt{s}=7$ and 0.9 TeV, respectively.
A similarly large number of events was considered in a data sample with the
higher transverse momentum cut $p_T > 500$~MeV/c and $n_{ch}\ge 1$.
The measurements in both regions give severe restriction to the models 
of multiparticle production in proton-proton collisions at high energies.
This particularly concerns the weighted superposition of two NBDs which 
is usually attributed to the classification of events into soft and semihard with respect to the momentum 
transfer in parton-parton scatterings.

\begin{figure}
\begin{center}
\vskip 0cm
\hspace*{0mm}
\includegraphics[width=78mm,height=78mm]{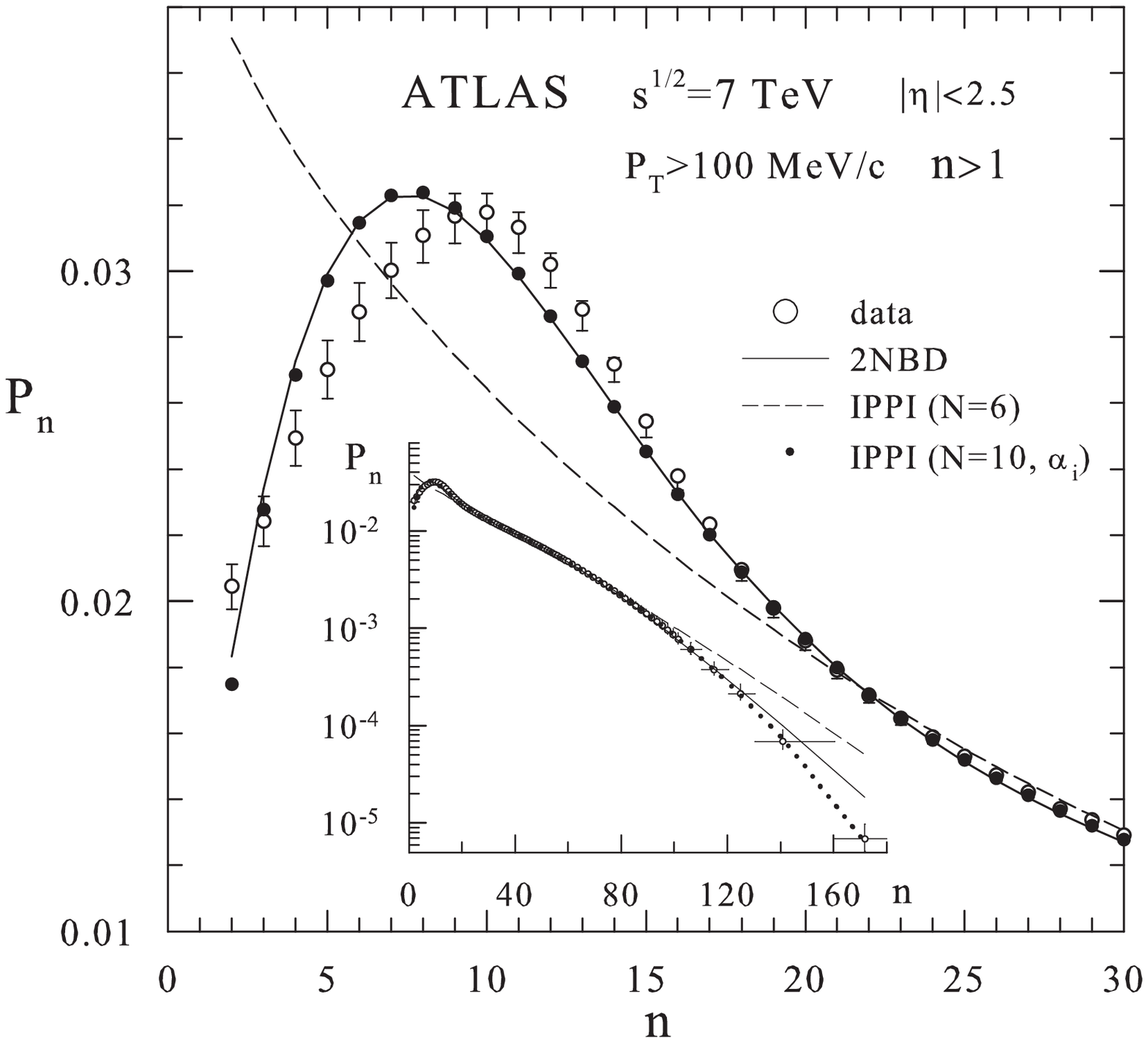}
\hspace{-0.5cm}
\includegraphics[width=78mm,height=78mm]{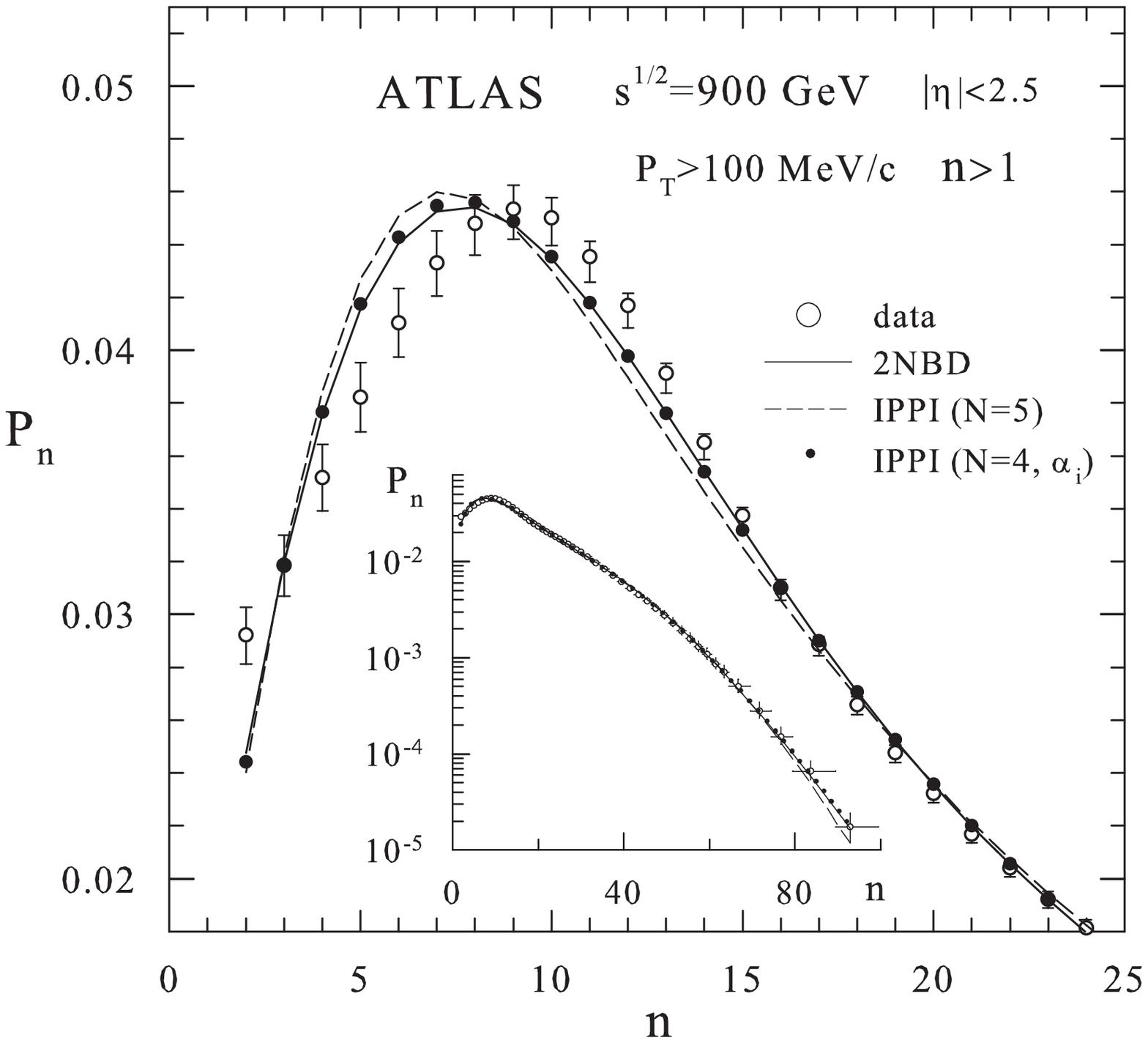}
\vskip -10mm
\hspace{1.cm}    (a) \hspace*{70mm} (b)
\caption{\label{Fig1}
Data on MD of charged particles in $pp$ collisions (open circles) measured 
by the ATLAS Collaboration \protect \cite{ATLAS}
in the interval $|\eta|<2.5$ for $p_T > 100$~MeV/c and  $n_{ch}\ge 2$ at 
(a) $\sqrt s=7$~TeV and (b) $\sqrt s=900$~GeV.
The solid and dashed lines represent fits using the superposition of two NBDs and the IPPI model 
with the indicated number of components (N), respectively. The black dots correspond to the IPPI model 
with fitted probabilities $\alpha_i$ and the restriction $ \alpha_{i+1}\le\alpha_i$. 
}  
\end{center}
\end{figure}

Figure \ref{Fig1} shows data on MD measured by the ATLAS Collaboration \cite{ATLAS}
in the pseudorapidity window $|\eta|<2.5$ and the phase-space 
region $p_T > 100$~MeV/c, $n_{ch}\ge 2$ at the energies $\sqrt{s}=$~7 and 0.9~TeV. 
The distributions are displayed around a maximum 
with the insets illustrating their overall shape. 
The experimental data are depicted as open circles and the error bars correspond to the quadratic sum 
of statistical and systematic uncertainties.
The solid lines connect the points representing best
fits to the data using function (\ref{eq:r2}) with N=2 and five free parameters.
Terms of the fitting procedure are explained in more detail in the appendix.
The visual contradiction of the experimental data at $\sqrt{s}=$~7 TeV (0.9 TeV) 
with the solid curves for $n<17$ ($n<15$)
and the corresponding large values of
$\chi^2/dof=128.7/80$ ($\chi^2/dof=70.3/46$) exclude description of the ATLAS data with two NBDs. 
As can be seen from figure \ref{Fig1}, the maximum of the measured MD is shifted
toward higher multiplicities at both energies in comparison with the two-component NBD model.

The dashed lines in figure \ref{Fig1} represent the results of fitting data 
in the framework of the IPPI model 
with the probabilities for $i$ pairs of colliding partons taken in the form $\alpha_i=\alpha^i$.
In that case, the parameter $\alpha$ is given by the maximum number of the active parton pairs $N$. 
The IPPI model with three adjustable parameters $m$, $k$ and $N$ does not reproduce the ATLAS data well. 
The minimal value of $\chi^2/dof=2133/82$ obtained for $N=6$ at $\sqrt{s}=$~7 TeV and  
$\chi^2/dof=117/48$ for $N=5$ at $\sqrt{s}=$~900 GeV is unsatisfactory. 
Here we do not include additional conditions 
in the minimization procedure which were applied in \cite{Dremin2}.
Such conditions impose constraints on the parameters $m$ and $k$ of the model and result 
in even larger values of $\chi^2$ (\ref{eq:a1}). 
A better approximation of the data can be obtained if one allows freedom in the choice of the 
probabilities $\alpha_i$ while preserving their monotonic decrease $\alpha_{i+1}\le\alpha_i$. 
This condition follows from the physical picture of the IPPI model characterized 
by the relations $\bar{n}_i=im$ and $k_i=ik$. 
With the increase in the number of independent adjustable parameters ($N+1$ in this case), 
one can get smaller value of   
$\chi^2/dof=268/78$ for six active parton pairs at $\sqrt{s}=$~7 TeV.
The minimal value of $ \chi^2/dof=91.7/74$ at this energy is obtained  for ten components
with the parameters quoted in the appendix. 
The corresponding distribution is depicted by the black dots in figure \ref{Fig1}(a).
The same symbols in figure \ref{Fig1}(b) represent the best approximation of the data at $\sqrt{s}=$~900 GeV
with $ \chi^2/dof=74.7/46$ and with four components taken into account.  
One can see from figure \ref{Fig1} that the distributions within the IPPI model with $N+1$ adjustable parameters 
show similar discrepancies with the ATLAS data for $n < 15$ at both energies,  $\sqrt{s}=$~7 TeV and 900~GeV, 
as it is in the case of superposition of two independent NBDs. 
Somewhat smaller $\chi^2/dof=91.7/74$ for the IPPI model at $\sqrt{s}=$~7 TeV 
is attributed to a better description of the data in the tail of the distribution. 
Regardless of this, neither of these models can describe the ATLAS data around the maximum.
The experimental distributions at maximum are shifted toward higher multiplicities relative to
the best-fitted parametrizations based on both models.

The CMS measurements \cite{CMS} in a similar pseudorapidity window are less critical 
to the mentioned hypotheses.
\begin{figure}[b]
\begin{center}
\vskip 0cm
\hspace*{0mm}
\includegraphics[width=78mm,height=78mm]{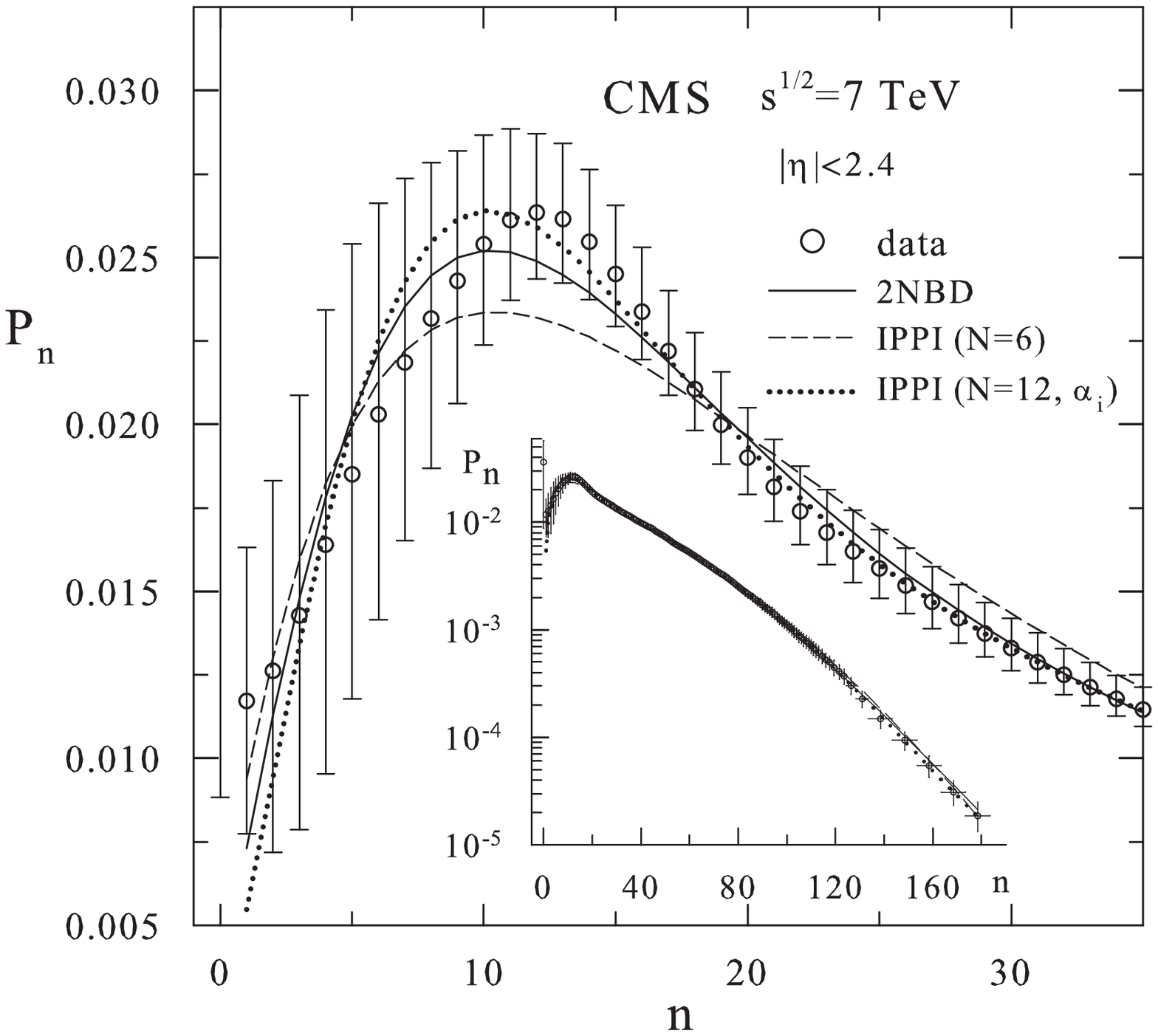}
\hspace{-0.5cm}
\includegraphics[width=78mm,height=78mm]{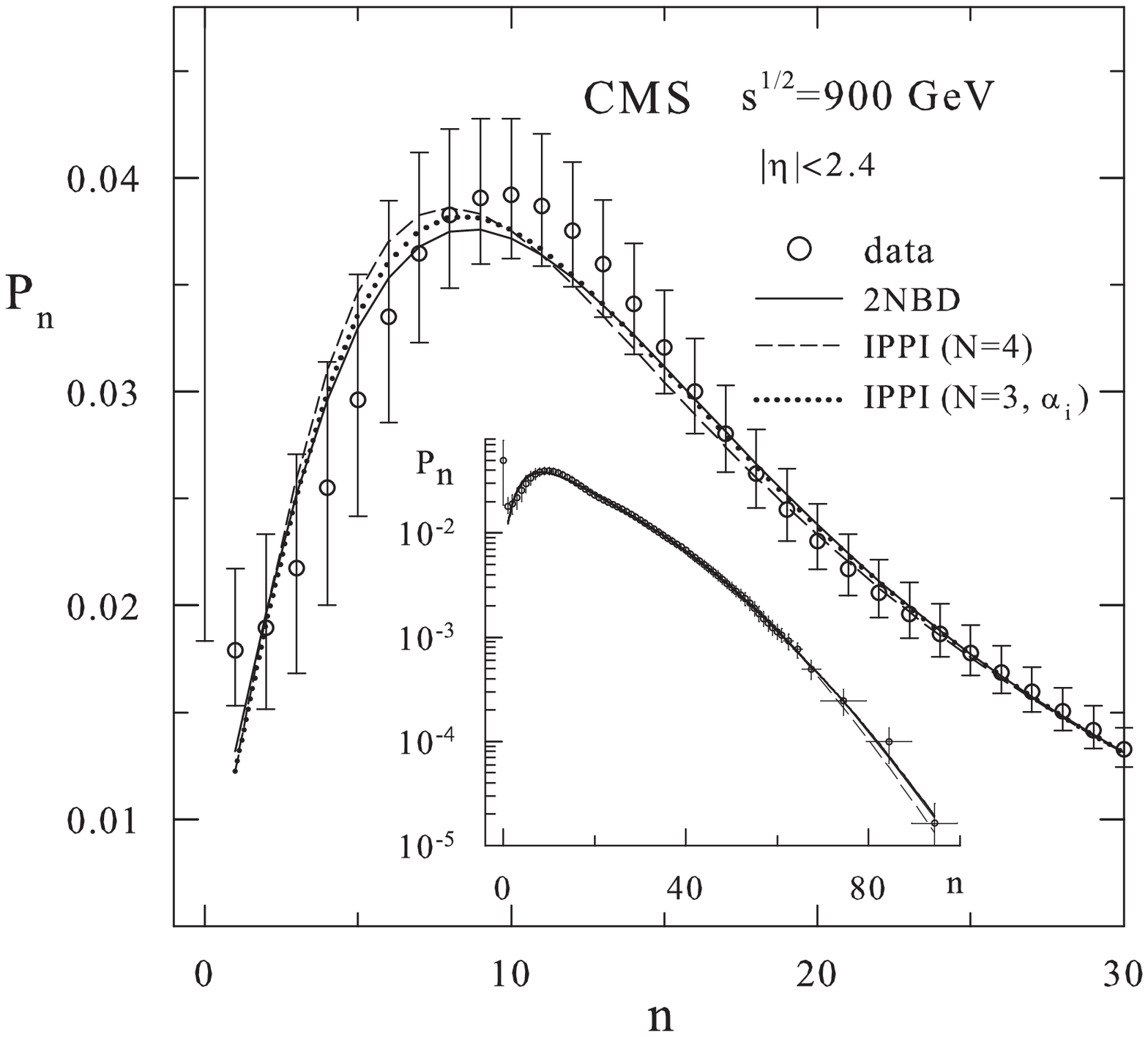}
\vskip -10mm
\hspace{1.cm}    (a) \hspace*{70mm} (b)
\caption{\label{Fig2}
Data on MD of charged particles in $pp$ collisions (open circles) measured 
by the CMS Collaboration \protect \cite{CMS}
in the interval $|\eta|<2.4$ at 
(a) $\sqrt s=7$~TeV and (b) $\sqrt s=900$~GeV.
The solid and dashed lines represent fits using the superposition of two NBDs and the IPPI model 
with the indicated number of components (N), respectively.
The dotted lines correspond to the IPPI model 
with fitted probabilities $\alpha_i$ and the restriction $ \alpha_{i+1}\le\alpha_i$. 
}  
\end{center}
\end{figure} 
This is illustrated in figure \ref{Fig2} where data on MD of charged particles measured  
by the CMS Collaboration \cite{CMS} in the interval $|\eta|<2.4$
at the energies $\sqrt{s}=7$~TeV and 900~GeV  are shown. 
One can see that the CMS data  are more or less well-described by the two-NBD superposition  
(solid lines). 
The corresponding
values of $\chi^2/dof=13.4/121$ and $\chi^2/dof=$10.7/62 are at both energies satisfactory. 
Somewhat larger values of $\chi^2/dof=49.6/124$ ($\chi^2/dof=14.7/64$)
are obtained within the IPPI model (dashed lines) with the probabilities 
$\alpha_i=\alpha^i$ and six (four) components at the energies $\sqrt{s}=7$~TeV (900~GeV).
At closer inspection, however, one can see that the dashed line 
in figure \ref{Fig2}(a) deviates from the shape
of the experimental distribution for $8<n<35$. 
This can be improved if one allows freedom in the choice of the 
probabilities $\alpha_i$ while preserving the condition $\alpha_{i+1}\le\alpha_i$. 
Unlike similar fittings of the ATLAS data, an acceptable fit and much better $\chi^2/dof=6.6/113$ can be obtained 
within the IPPI model at $\sqrt{s}=7$~TeV with twelve components 
and thirteen parameters quoted in the appendix.
The modification of the parameters $\alpha_i$ has a smaller impact on the 
fitted distribution at lower energy. At $\sqrt{s}=$900~GeV one can get $\chi^2/dof=11.5/63$ within the IPPI model
with three components taken into account. 
The dotted lines shown in figure \ref{Fig2} connect the points which represent best fits to the CMS data 
using the IPPI model with $N+1$ adjustable parameters.

The reason for the acceptable values of the least-squares tests are significant systematic
uncertainties of the CMS data as compared with the experimental data presented by the ATLAS Collaboration. 
Despite the large systematic errors, the values of $P_n$ measured with high statistics 
allow observation of a certain tendency in the CMS data.  
One can notice from figure \ref{Fig2} that open circles representing the mean values of the measured probabilities 
are shifted toward 
higher multiplicities at the maximum of the distribution as compared to the lines depicted in this figure.
Moreover, the shift values at the maximum are roughly the same as are the corresponding displacements ($\Delta n \sim 2$) 
clearly visible in figure \ref{Fig1}.
The observation supports the consistency of both data in the considered sense.
On the other hand, the systematic errors of the CMS data influence
inference from the relatively small values of $\chi^2/dof$ and can lead to conclusions
which do not comply with the ATLAS data. 
As for errors, similar considerations concerning the hypothesis 
of single NBD and the measurements of the UA5 and ALICE Collaborations
in small pseudorapidity windows were expressed in \cite{Prorok}.

Motivated by the discrepancies shown above, we study the high statistic data on 
MDs obtained at the LHC in the framework of weighted superposition of three NBDs.
The aim is to achieve a detailed description of the data including the shape
of the distributions in the region of maximal values of $P_n$.
The results are compared with two-NBD parametrization of the same data.
Hereafter we refer to the notion 
\lq three NBD superposition\rq\ 
also for the situation if the third NBD is reduced to its Poisson limit with 
$k_3=\infty$.    
The limiting case of two NBDs plus one Poisson distribution is used in our analysis where 
stated below.

\subsection{Three-component description of ATLAS data}

\begin{figure}[t]
\begin{center}
\vskip 0cm
\hspace*{0mm}
\includegraphics[width=78mm,height=78mm]{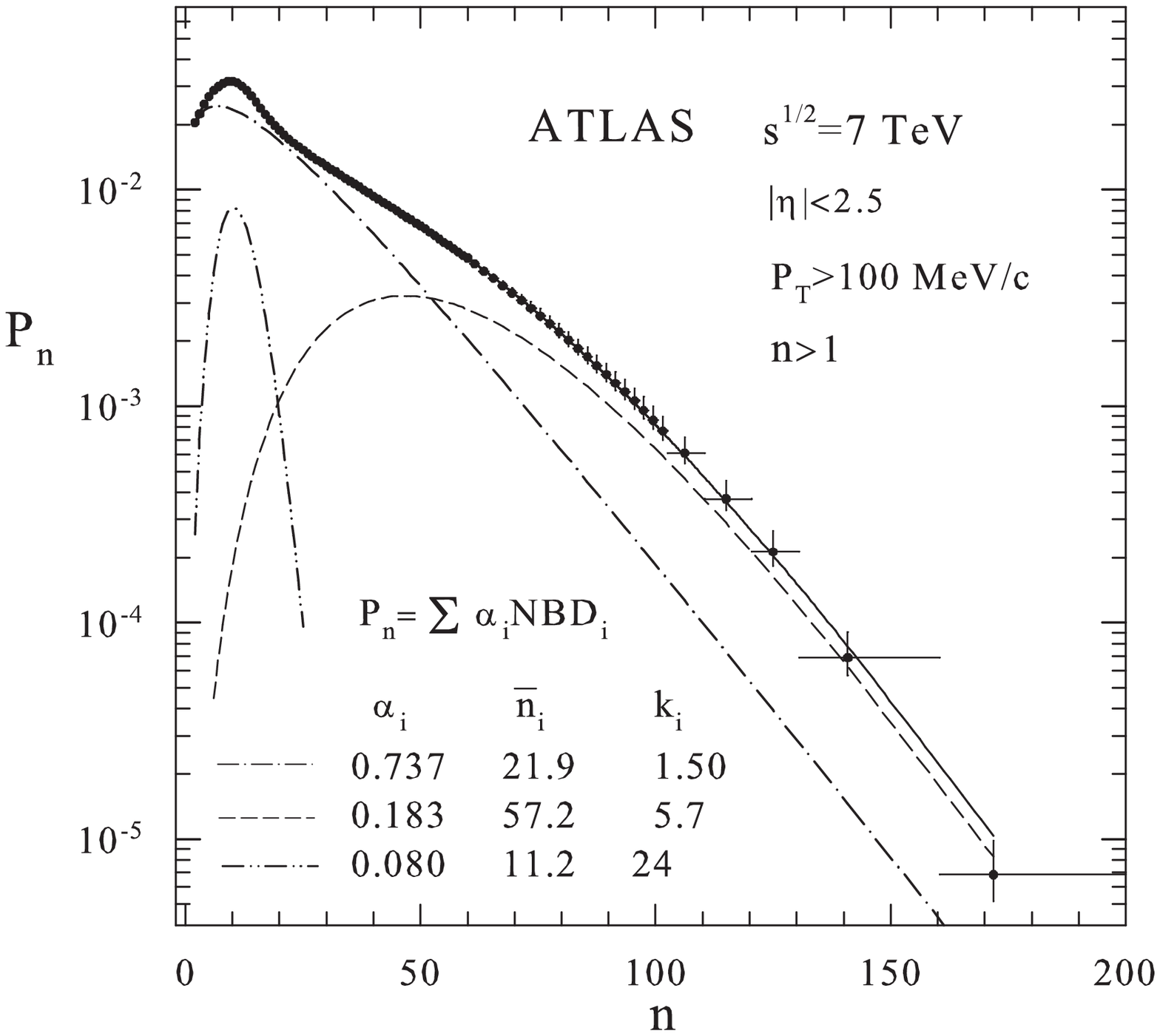}
\hspace{-0.5cm}
\includegraphics[width=78mm,height=78mm]{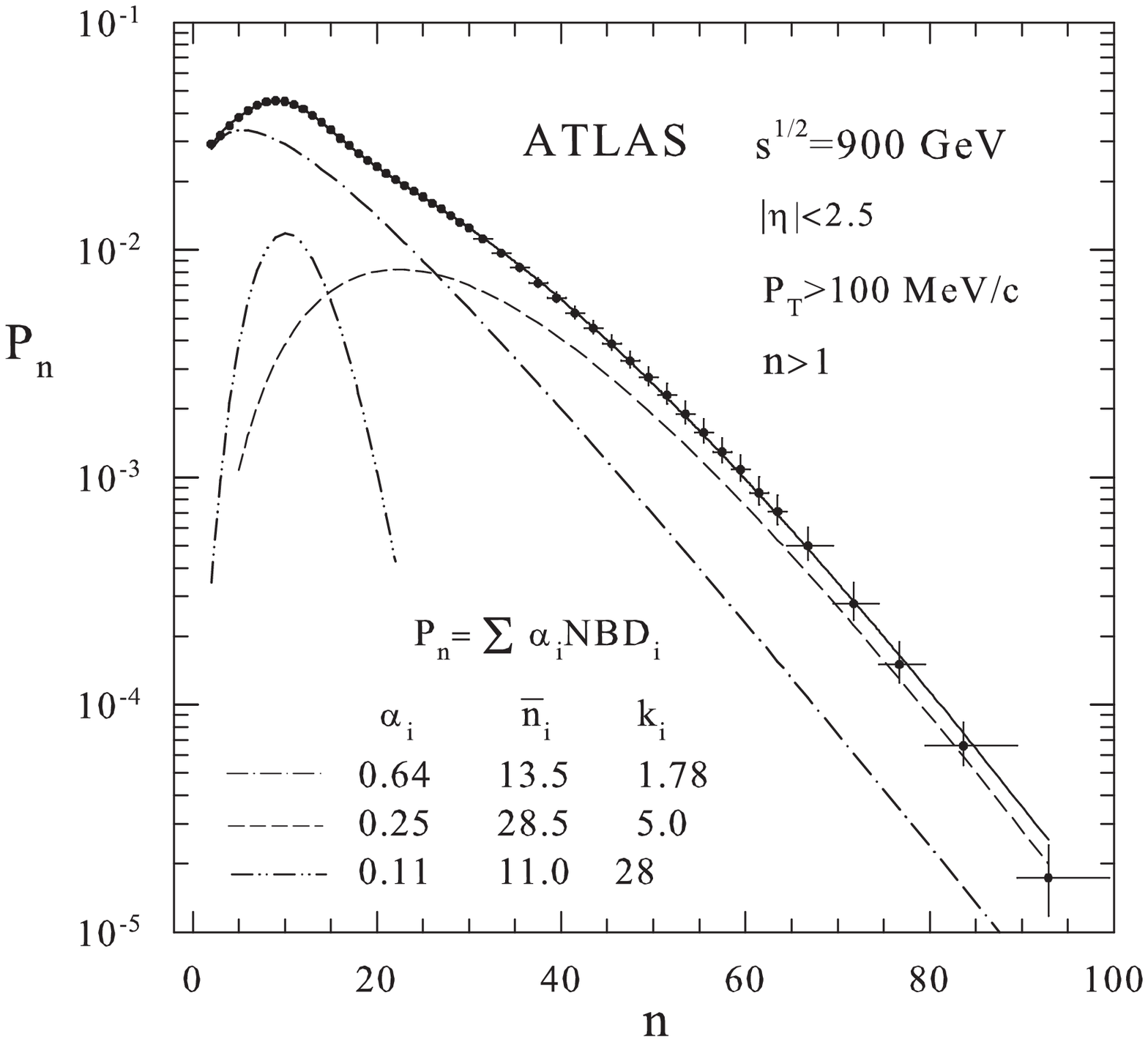}
\vskip -10mm
\hspace{1.cm}    (a) \hspace*{70mm} (b)
\caption{\label{Fig3}
MD of charged particles
measured by the ATLAS Collaboration \protect\cite{ATLAS}
in the pseudorapidity interval $|\eta|<2.5$ for $p_T > 100$~MeV/c, $n>1$
at  (a) $\sqrt s=7$~TeV and (b) $\sqrt s=0.9$~TeV.
The solid lines represent fits to the data by three-component
superposition of NBDs. The dash-dot, dash and dash-dot-dot lines show
single components corresponding to the indicated parameters.  }
\end{center}
\end{figure}

We have fitted data \cite{ATLAS} on MDs measured by the ATLAS Collaboration 
in the interval $ |\eta|<2.5$ with weighted superposition of three NBDs.
The decomposition of the total distribution in the
most inclusive phase-space region corresponding to
the transverse momentum cut $p_T > 100$~MeV/c and  $n_{ch}\ge 2$ 
is shown in figures \ref{Fig3}(a) and (b) at the energies
$\sqrt s=7$~TeV and 0.9~TeV, respectively.
The experimental data are indicated by symbols and the fitted three-component 
function (\ref{eq:r2}) is depicted by the solid line.
The dash-dot, dash and dash-dot-dot lines represent  
single NBD components of the total distribution.  
The corresponding parameters and values of $\chi^2$ are quoted in table~\ref{tab1}.

\begin{table}[b]
\lineup 
\caption{\label{tab1}
The parameters of superposition of three and two NBDs  
obtained from fits to  data \protect\cite{ATLAS} on MD measured by the ATLAS
Collaboration in the pseudorapidity window $|\eta|<2.5$ with the cut $p_T>100$~MeV/c, $n>1$ at 
$ \sqrt{s}=7$ and 0.9 TeV. }
\begin{indented}
\item[]
\begin{tabular}{@{}p{0.7cm}@{}ll@{}l@{}p{0.5cm}@{}ll@{}l@{}}
\br
  & \multicolumn{3}{c}{$\sqrt{s}$ = 7 TeV} & & \multicolumn{3}{c}{$\sqrt{s}$ = 0.9 TeV} \\ 
\cline{2-4}\cline{6-8} \\[-0.30cm]
{\it i}  & $\alpha_i$ & $\bar{n}_i$ & $k_i$ &   & $\alpha_i$ & $\bar{n}_i$ & $k_i$  \\
\cline{1-4}\cline{6-8} \\[-0.25cm]
1 &  0.737$^{+0.062}_{-0.089}$  &  21.9$^{+2.0  }_{-2.9  }$ &  \01.50$^{+0.12 }_{-0.06 }$  &           
  &   0.64$^{+0.20 }_{-0.36 }$  &  13.5$^{+2.1  }_{-5.6  }$ &  \01.78$^{+0.23 }_{-0.45 }$  \\[0.15cm]  
2 &  0.183$^{+0.080}_{-0.050}$  &  57.2$^{+2.1  }_{-3.7  }$ & \05.7\0$^{+0.9  }_{-0.8  }$  &           
  &   0.25$^{+0.27 }_{-0.15 }$  &  28.5$^{+5.5  }_{-4.4  }$ & \05.0\0$^{+3.2  }_{-1.3  }$  \\[0.15cm]  
3 &  0.080$^{+0.009}_{-0.012}$  &  11.2$^{+0.2  }_{-0.2  }$ &   24.\0$^{\,+14. }_{\0-6.}$  &           
  &   0.11$^{+0.09 }_{-0.05 }$  &  11.0$^{+0.3  }_{-0.4  }$ &   28.\0$^{\,+\infty}_{\,-17.}$\\[0.08cm] 
\cline{2-4}\cline{6-8}\\[-0.35cm]
  & \multicolumn{3}{c}{$\chi^2/dof$ = 6.6/(85-8)}  & &  \multicolumn{3}{c}{$\chi^2/dof$ = 5.5/(51-8)} \\  
\cline{1-4}\cline{6-8} \\[-0.35cm] 
1 &  0.412$\pm$0.016   &   11.04$\pm$0.16\0&    \02.71$\pm$0.09  &    
  &  0.774$\pm$0.031   &   12.63$\pm$0.44\0&    \02.70$\pm$0.12  \\   
2 &  0.588$\pm$0.016   &   39.50$\pm$0.58  &    \02.65$\pm$0.11  &    
  &  0.226$\pm$0.031   &   32.9\0$\pm$1.2  &    \07.53$\pm$0.77  \\   
\cline{2-4}\cline{6-8} \\[-0.32cm]
  & \multicolumn{3}{c}{$\chi^2/dof$ = 128.7/(85-5)} & & \multicolumn{3}{c}{$\chi^2/dof$ = 70.3/(51-5)} \\ 
\br
\end{tabular} 
\end{indented}
\end{table}

One can see from figure~\ref{Fig3} and table~\ref{tab1} 
that the decomposition into three components is similar at both energies.
The dominant component with the largest probability gives the main contribution $ \alpha_1\bar{n}_1 $ 
to the total average multiplicity.
The other two components contribute to the high and low multiplicity region, respectively.
The average multiplicities of the first and the second component, $\bar{n}_1$ and  $\bar{n}_2$,
increase with energy. This results in a broadening of the total distribution.
The average multiplicity $\bar{n}_3 \simeq 11$ of the third component 
is nearly energy independent.
Within the errors quoted in table~\ref{tab1}, 
the probabilities $ \alpha_i $ show weak energy dependence as well.
The values of the parameters $k_i$ increase with decreasing probabilities $\alpha_i$.
The first NBD with the largest probability $\alpha_1$ is characterized by the smallest parameter $k_1$.
The component under the peak of the distribution at low multiplicities
is narrow with large values of $k_3$.

\begin{figure}[t]
\begin{center}
\vskip 0cm
\hspace*{0mm}
\includegraphics[width=78mm,height=78mm]{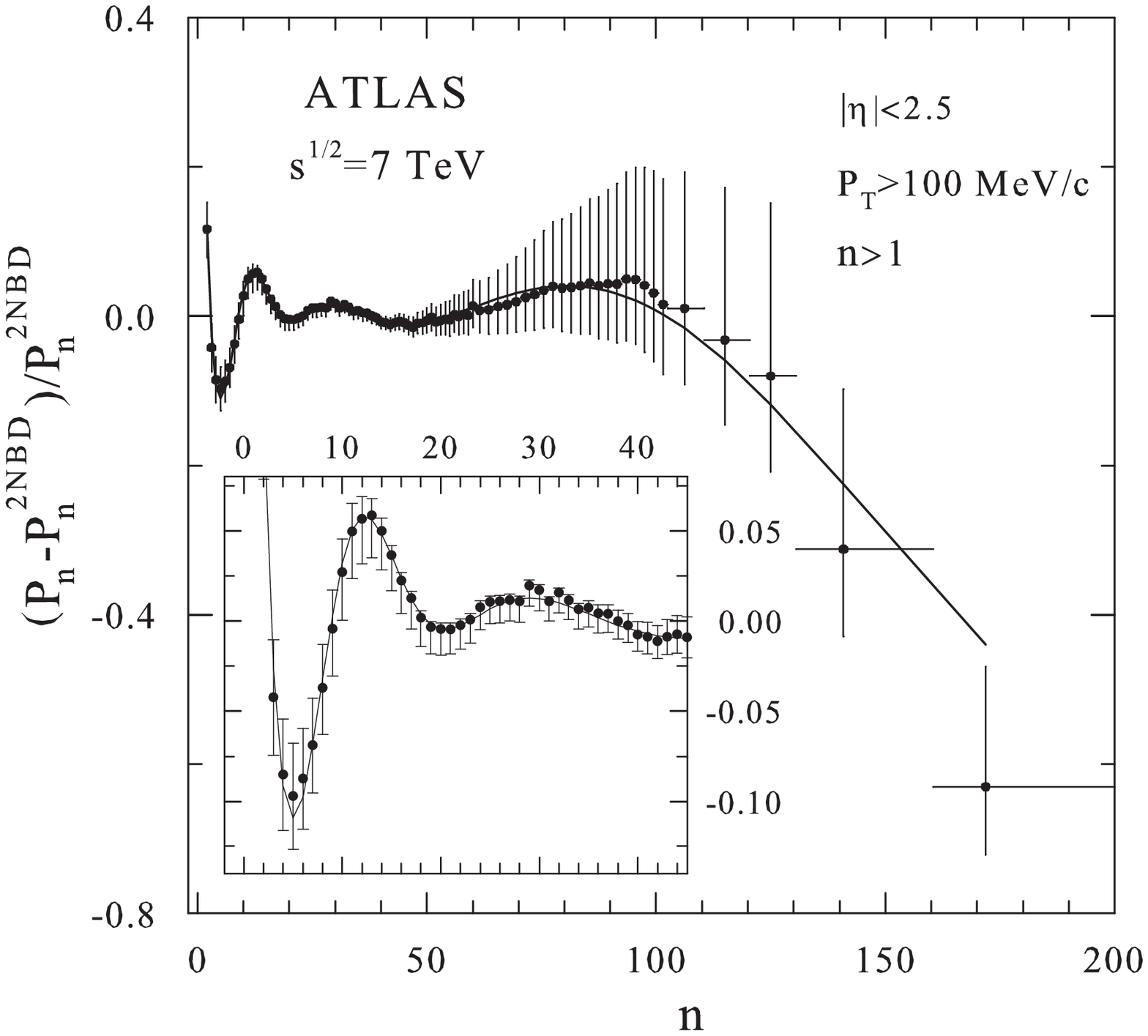}
\hspace{-0.5cm}
\includegraphics[width=78mm,height=78mm]{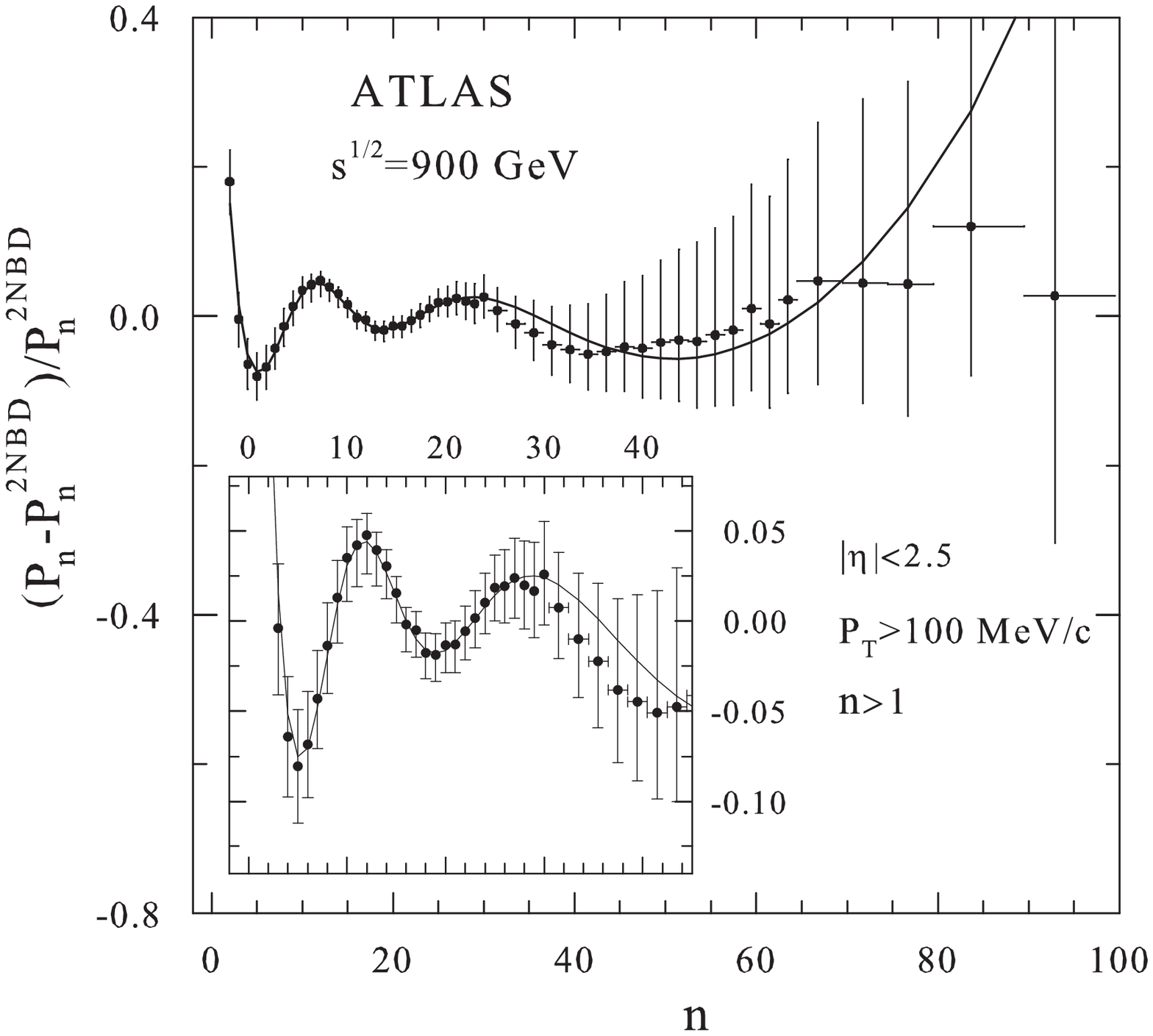}
\vskip -10mm
\hspace{1.cm}    (a) \hspace*{70mm} (b)
\caption{\label{Fig4}
Normalized residues of MD relative to the weighted superposition of two NBDs $(\mathrm{P_n^{2NBD}})$
with the parameters listed in table~\ref{tab1}.
The symbols correspond to data on MD measured by the ATLAS Collaboration \protect\cite{ATLAS}
in the pseudorapidity interval $|\eta|<2.5$ with the cut $p_T > 100$~MeV/c, $n>1$
at (a) $\sqrt s=7$~TeV and (b) $\sqrt s=0.9$~TeV.
The lines correspond to fits to the data by three-component 
superposition of NBDs. The insets show the detailed behavior of the residues at 
low $n$.  }
\end{center}
\vskip -5mm
\end{figure} 

We have compared the ATLAS data with the best-fitted weighted superposition of two NBDs.
The values of the parameters corresponding to the two-component hypothesis  
are stated in table~\ref{tab1}.
The relative residues with respect to the two-NBD fits are depicted in figure \ref{Fig4}.
The points correspond to data  measured by the ATLAS Collaboration
in the pseudorapidity interval $|\eta|<2.5$ for $p_T > 100$~MeV/c, $n>1$
at the energy $\sqrt s=7$~TeV and $\sqrt s=0.9$~TeV.
The insets show the detailed structure of the residues at low multiplicities.
The solid lines are given by the three-component description of the data shown in figure \ref{Fig3}.
One can see from figure \ref{Fig4} that the two-NBD approximation of the data is unsatisfactory at
both energies. 
The corresponding values of $ \chi^2/dof $ quoted in table~\ref{tab1} are too large.
The high statistic ATLAS data manifest a distinct peak around $n\sim 11$.
The description of the peak clearly seen in the residues  in figure \ref{Fig4}
was obtained by the third negative binomial component with $\bar{n}_3 \simeq 11$. The component is depicted 
by the dash-dot-dot line in figure \ref{Fig3}.

\begin{figure}[t]
\begin{center}
\vskip 0cm
\hspace*{0mm}
\includegraphics[width=78mm,height=78mm]{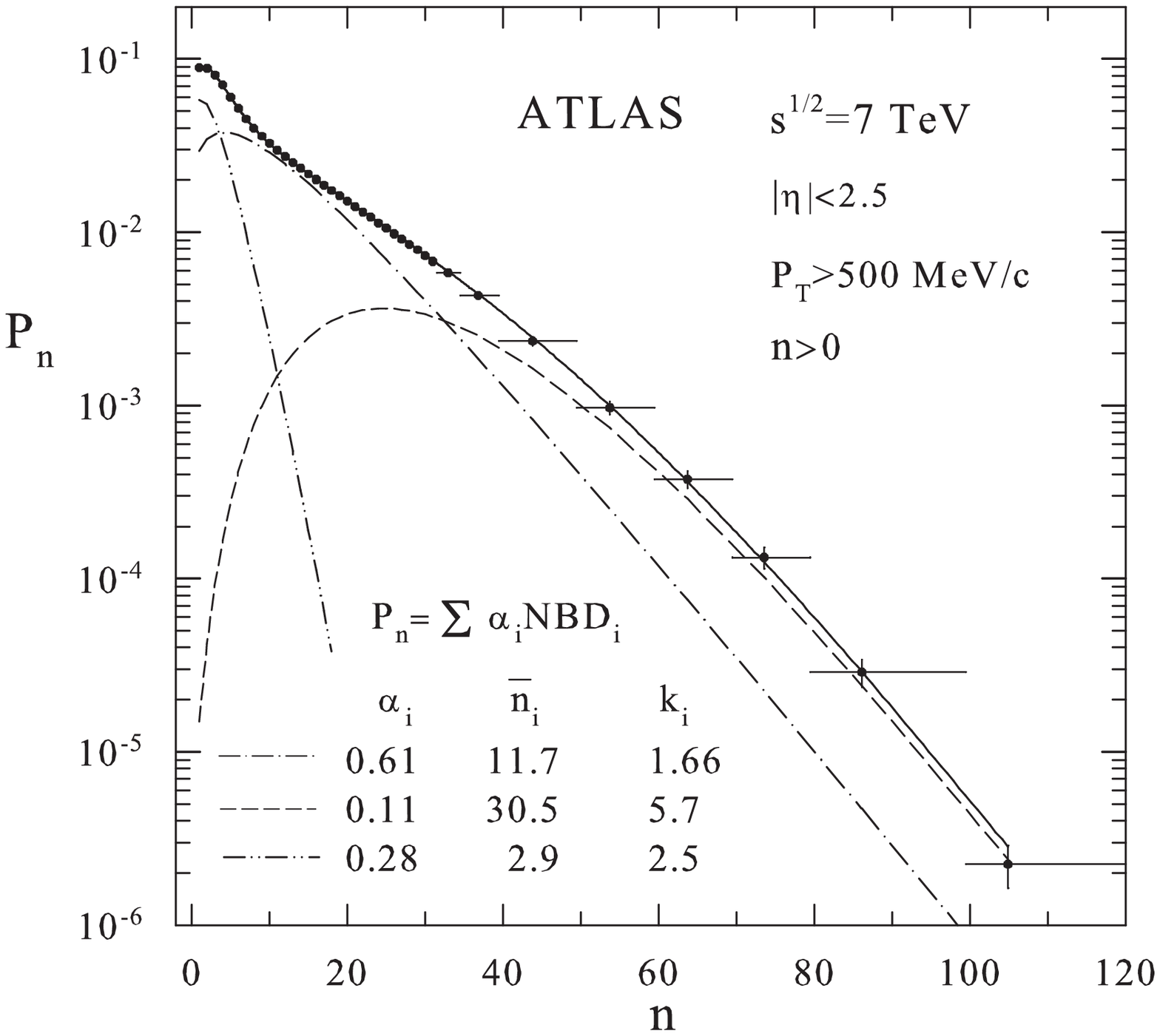}
\hspace{-0.5cm}
\includegraphics[width=78mm,height=78mm]{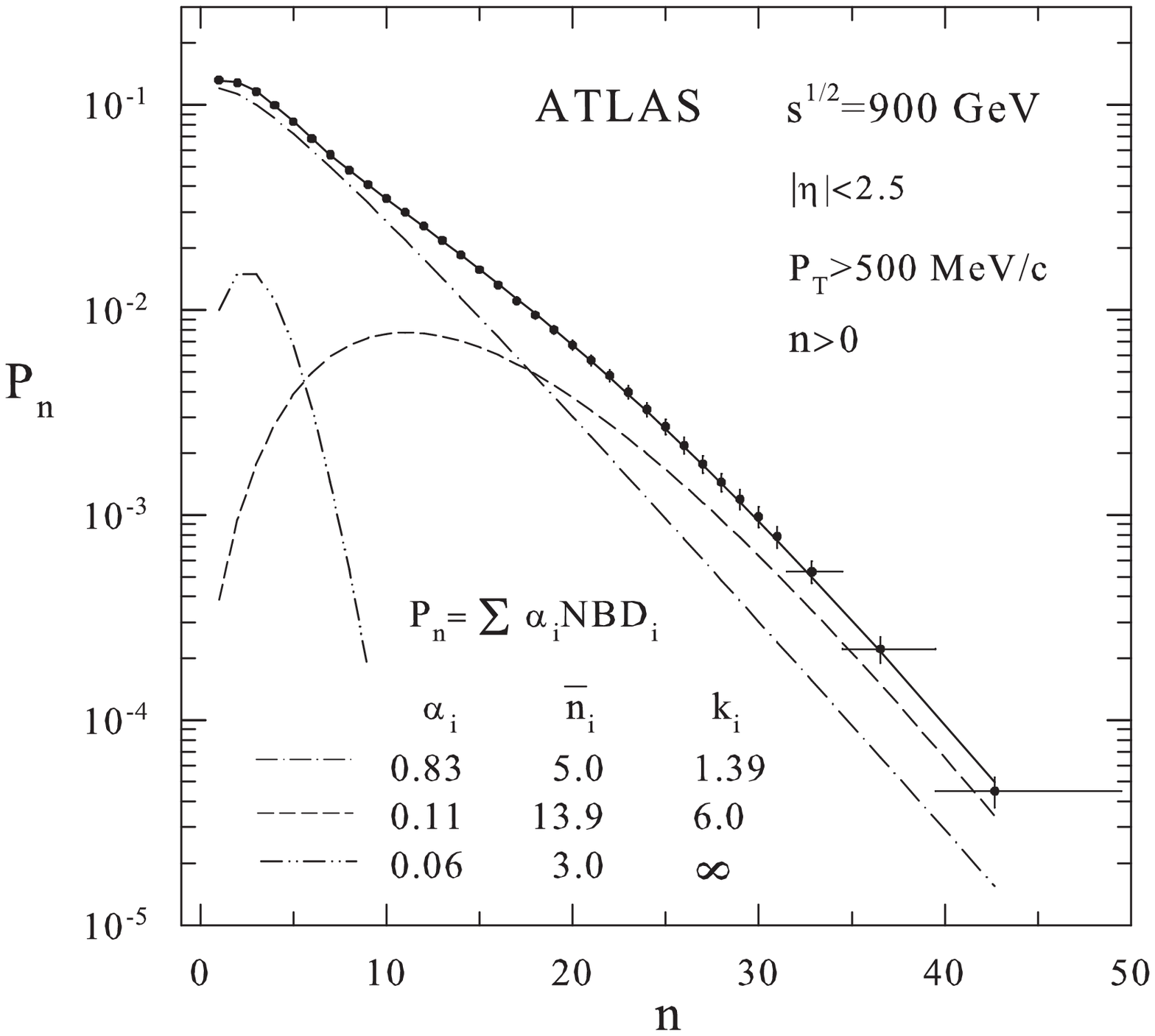}
\vskip -10mm
\hspace{1.cm}    (a) \hspace*{70mm} (b)
\caption{\label{Fig5}
MD of charged particles
measured by the ATLAS Collaboration \protect\cite{ATLAS}
in the pseudorapidity interval $|\eta|<2.5$ for $p_T > 500$~MeV/c, $n>0$
at (a) $\sqrt s=7$~TeV and (b) $\sqrt s=0.9$~TeV.
The solid lines represent fits to the data by three-component
superposition of NBDs. The dash-dot, dash and dash-dot-dot lines show
single components corresponding to the indicated parameters.  }
\end{center}
\end{figure}

We have studied the peaky structure of MD using the ATLAS data in the separate phase-space region 
defined by the conditions $p_T > 500$~MeV/c and $n>0$. Figure \ref{Fig5} shows the data 
measured in the window $|\eta|<2.5$ at $\sqrt s=7$~TeV and 0.9~TeV. 
The symbols and the lines have the same meaning as in figure \ref{Fig3}.
The corresponding parameters and the respective values of $\chi^2/dof$ are given in table~\ref{tab2}.
The parameter $k_3$ takes values larger than 100 at $\sqrt s=0.9$~TeV with error  
comparable with infinity. In such a case we have fixed 
$k_3=\infty$ and considered the third NBD as a Poisson distribution.     
\begin{table}[b]
\lineup 
\caption{\label{tab2}
The parameters of superposition of three  and two NBDs  
obtained from fits to  data \protect\cite{ATLAS} on MD measured by the ATLAS Collaboration
in the window $|\eta|<2.5$ with the cut $p_T>500$~MeV/c, $n>0$ at 
$ \sqrt{s}=7$ and 0.9 TeV.  }
\begin{indented}
\item[]
\begin{tabular}{@{}p{0.7cm}@{}lll@{}p{0.5cm}@{}lll@{}}
\br
  & \multicolumn{3}{c}{$\sqrt{s}$ = 7 TeV} & & \multicolumn{3}{c}{$\sqrt{s}$ = 0.9 TeV} \\ 
\cline{2-4}\cline{6-8} \\[-0.30cm]
{\it i}  & $\alpha_i$ & $\bar{n}_i$ & $k_i$ &   & $\alpha_i$ & $\bar{n}_i$ & $k_i$  \\
\cline{1-4}\cline{6-8} \\[-0.25cm]
1 &  0.61$^{+0.19 }_{-0.30 }$  &   11.7$^{+1.3  }_{-1.9  }$   &   1.66$^{+1.60 }_{-0.50 }$  &           
  &  0.83$^{+0.07 }_{-0.14 }$  &  \05.0$^{+0.5  }_{-1.0  }$   &   1.39$^{+0.24 }_{-0.10 }$  \\[0.15cm]  
2 &  0.11$^{+0.16 }_{-0.05 }$  &   30.5$^{+3.0  }_{-5.9  }$   &  5.7\0$^{+1.7  }_{-1.7  }$  &           
  &  0.11$^{+0.13 }_{-0.05 }$  &   13.9$^{+1.4  }_{-2.0  }$   &  6.0\0$^{+2.6  }_{-1.6  }$  \\[0.15cm]  
3 &  0.28$^{+0.14 }_{-0.14 }$  &  \02.9$^{+0.3  }_{-0.1  }$   &  2.5\0$^{+2.4  }_{-0.4  }$  &           
  &  0.06$^{+0.01 }_{-0.02 }$  &  \03.0$^{+0.3  }_{-0.3  }$   &       $\infty$              \\[0.08cm]  
\cline{2-4}\cline{6-8}\\[-0.35cm]
  & \multicolumn{3}{c}{$\chi^2/dof$ = 9.7/(39-8)}  & &  \multicolumn{3}{c}{$\chi^2/dof$ = 3.2/(34-7)} \\ 
\cline{1-4}\cline{6-8} \\[-0.35cm] 
1 &  0.446$\pm$0.017   &  \03.55$\pm$0.09  &    1.60$\pm$0.07    &    
  &  0.637$\pm$0.075   &  \03.31$\pm$0.30  &    2.13$\pm$0.21   \\    
2 &  0.554$\pm$0.017   &   16.68$\pm$0.32  &    2.04$\pm$0.08    &    
  &  0.363$\pm$0.075   &   10.60$\pm$0.93  &    3.68$\pm$0.64   \\    
\cline{2-4}\cline{6-8} \\[-0.32cm]
  & \multicolumn{3}{c}{$\chi^2/dof$ = 70.4/(39-5)} & & \multicolumn{3}{c}{$\chi^2/dof$ = 7.0/(34-5)} \\ 
\br
\end{tabular}
\end{indented}
\end{table}

One can see some similarities when comparing the description of the most inclusive and 
the $p_T$-cut biased data shown in figures \ref{Fig3} and \ref{Fig5}, respectively.   
The component with the largest probability $\alpha_1$ has the smallest parameter $k_1$.
Its contribution $ \alpha_1\bar{n}_1 $ to the total average multiplicity is dominant at both energies.
The parameters $k_i$ increase as the probabilities $\alpha_i$ decrease.
The average multiplicities $\bar{n}_1$ and  $\bar{n}_2$
increase with energy resulting in a broader total distribution at $\sqrt s=7$~TeV.
The average multiplicity $\bar{n}_3 \simeq 3$ of the third component  
is nearly energy independent similarly to the most inclusive data sample.
On the other hand, there are differences between the data shown in figures \ref{Fig3} and \ref{Fig5}, respectively.
With the imposed $p_T$ cut, the value of $\bar{n}_3$ becomes significantly smaller 
relative to the average multiplicities of the first and second component.  
The probability $\alpha_3$ of the third component 
increases with $\sqrt{s}$ at the expense of $\alpha_1$
in the $p_T$-cut data sample.
The probability $\alpha_2$ of the second component remains nearly energy independent.
The increase of $\alpha_3$
at $\sqrt s=7$~TeV is accompanied by considerable decrease of the parameter $k_3$.  
Unlike the most inclusive sample, the third component becomes more significant
($\alpha_3\simeq 0.28$) and wider ($k_3\simeq 2.5$) at  $\sqrt s=7$~TeV.
These observations are supported by the CMS measurements \cite{CMS} in the  
window $ |\eta|<2.4$ with the same condition $ p_T>500$~MeV/c (see figure \ref{Fig12} below).

\begin{figure}[b]
\begin{center}
\vskip 0cm
\hspace*{0mm}
\includegraphics[width=78mm,height=78mm]{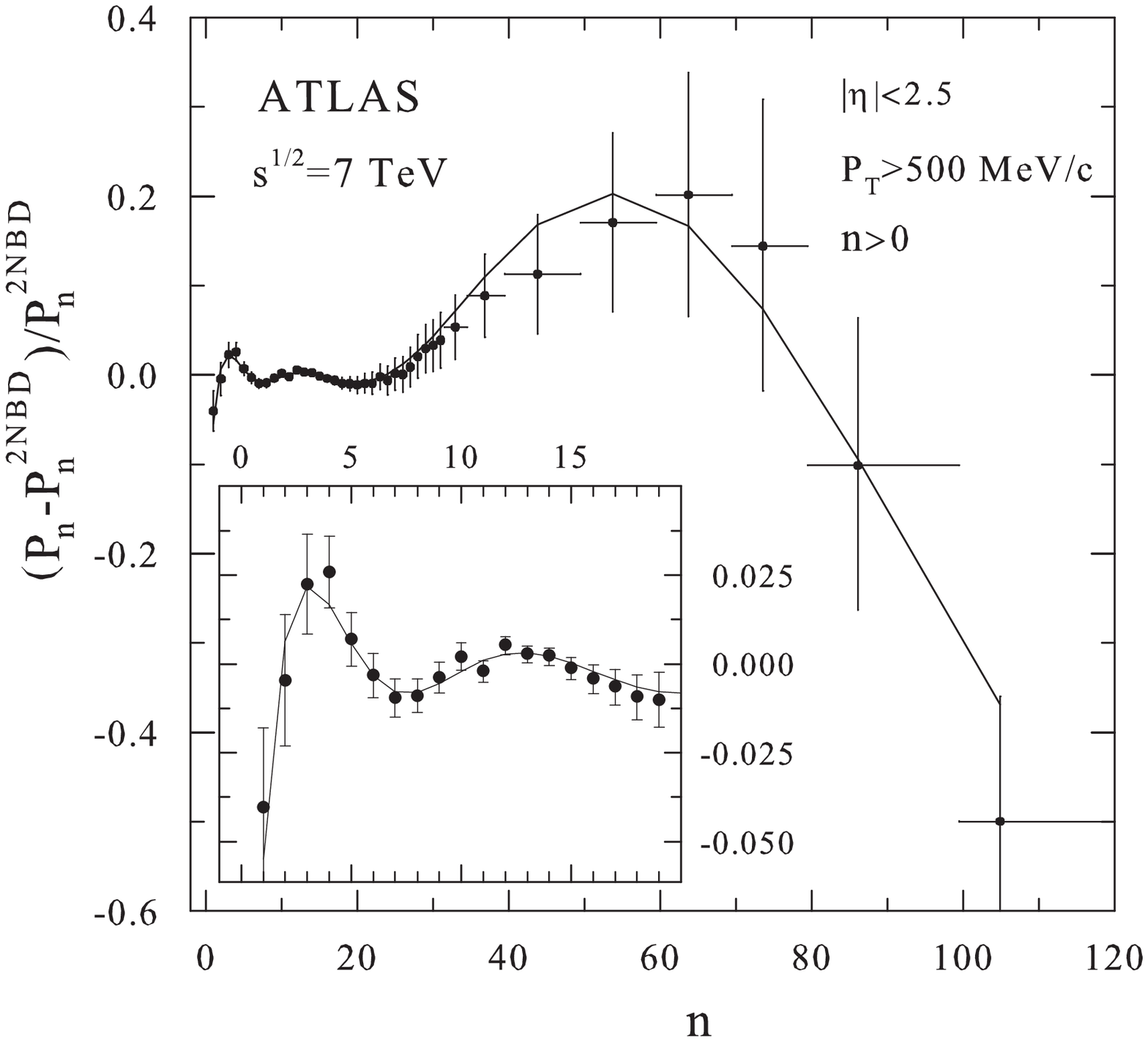}
\hspace{-0.5cm}
\includegraphics[width=78mm,height=78mm]{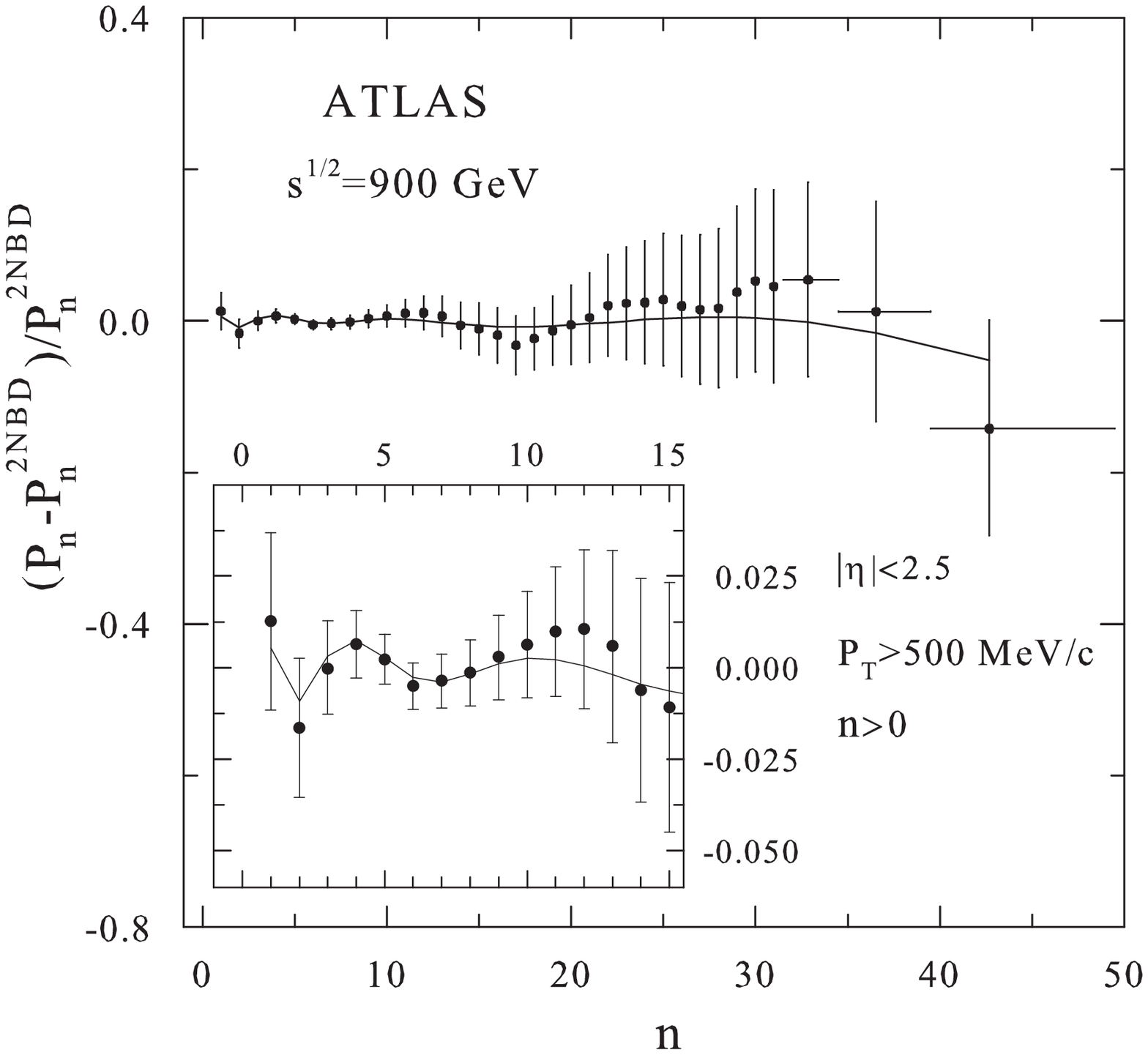}
\vskip -10mm
\hspace{1.cm}    (a) \hspace*{70mm} (b)
\caption{\label{Fig6}
Normalized residues of MD relative to the weighted superposition of two NBDs $(\mathrm{P_n^{2NBD}})$
with the parameters listed in table~\ref{tab2}.
The points correspond to data on MD  measured by the ATLAS Collaboration \protect\cite{ATLAS}
in the pseudorapidity interval $|\eta|<2.5$ for $p_T > 500$~MeV/c, $n>0$
at (a) $\sqrt s=7$~TeV and (b) $\sqrt s=0.9$~TeV.
The lines correspond to fits to the data by superposition of three NBDs. 
The insets show details of the residues at low $n$.  }
\end{center}
\end{figure}

The normalized residues with respect 
to the two-NBD parametrization (see table~\ref{tab2}) 
of the ATLAS data measured in the pseudorapidity window  $|\eta|<2.5$ 
in the phase-space region $p_T > 500$~MeV/c,
$n>0$ at the energies $\sqrt s=7$~TeV and 0.9~TeV
are depicted in figures \ref{Fig6}(a) and (b), respectively.
The data at $\sqrt s=7$~TeV show sizeable discrepancies relative to the weighted superposition
of two NBDs. 
The corresponding value of $\chi^2/dof=70.4/34$ for the two NBD fit is too large.
As seen from figure \ref{Fig6}(a), a clear peak is visible at low multiplicities around $n\simeq 3$.
The description of the peak was obtained by the third negative binomial component 
with $\bar{n}_3 \simeq 3$ as indicated by the dash-dot-dot line in figure \ref{Fig5}(a).
The residues at the energy $ \sqrt{s}=900$~GeV shown in figure \ref{Fig6}(b) are nearly flat though 
some indication of a peak around $ n\sim 3$ is visible.   

The experimental analysis \cite{ATLAS} of the ATLAS data was complemented by higher cuts
on multiplicity requiring $n_{ch}\ge 20$ with $p_T>100$~MeV/c and $n_{ch}\ge 6$ with $p_T>500$~MeV/c, respectively.
In the first sample, the high-multiplicity cut does not allow to study the third low-multiplicity component 
within superposition of three NBDs. This is evident from figure \ref{Fig3} where the dash-dot-dot line showing the third component is negligible for $n>20$.
The second data sample with the cut $n_{ch}\ge 6$ allows a check of consistency of the three-component description, though leaving all parameters free is problematic here as well.
Consequently, we have fixed some parameters related to the low-multiplicity component in this case. 
At $\sqrt{s}=7$~TeV, we have kept  
the probability of the third component 
at the value  $\alpha_3=0.28$ obtained for $n>0$ (see figure \ref{Fig5}(a)).

\begin{figure}[b]
\begin{center}
\vskip 0cm
\hspace*{0mm}
\includegraphics[width=78mm,height=78mm]{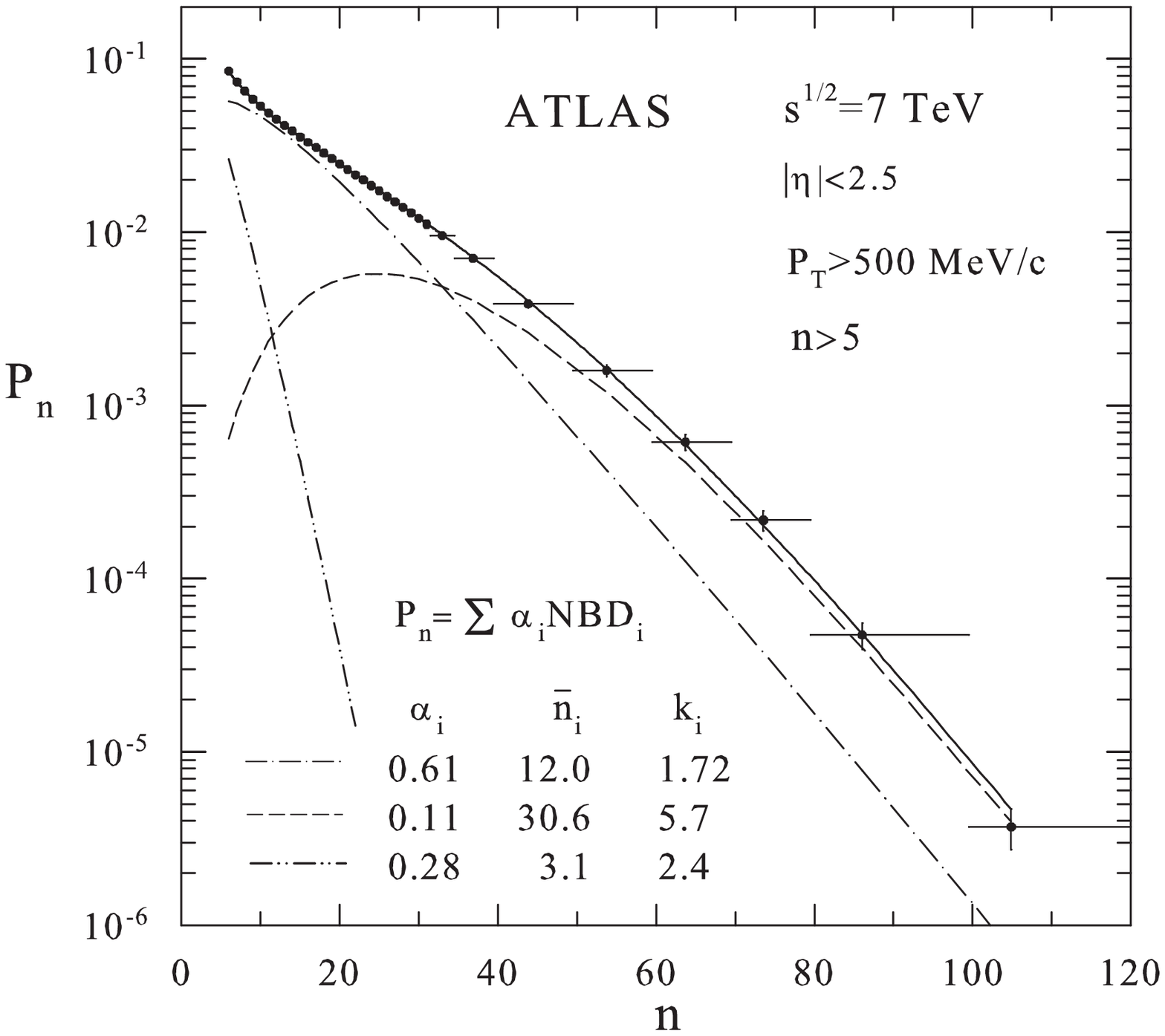}
\hspace{-0.5cm}
\includegraphics[width=78mm,height=78mm]{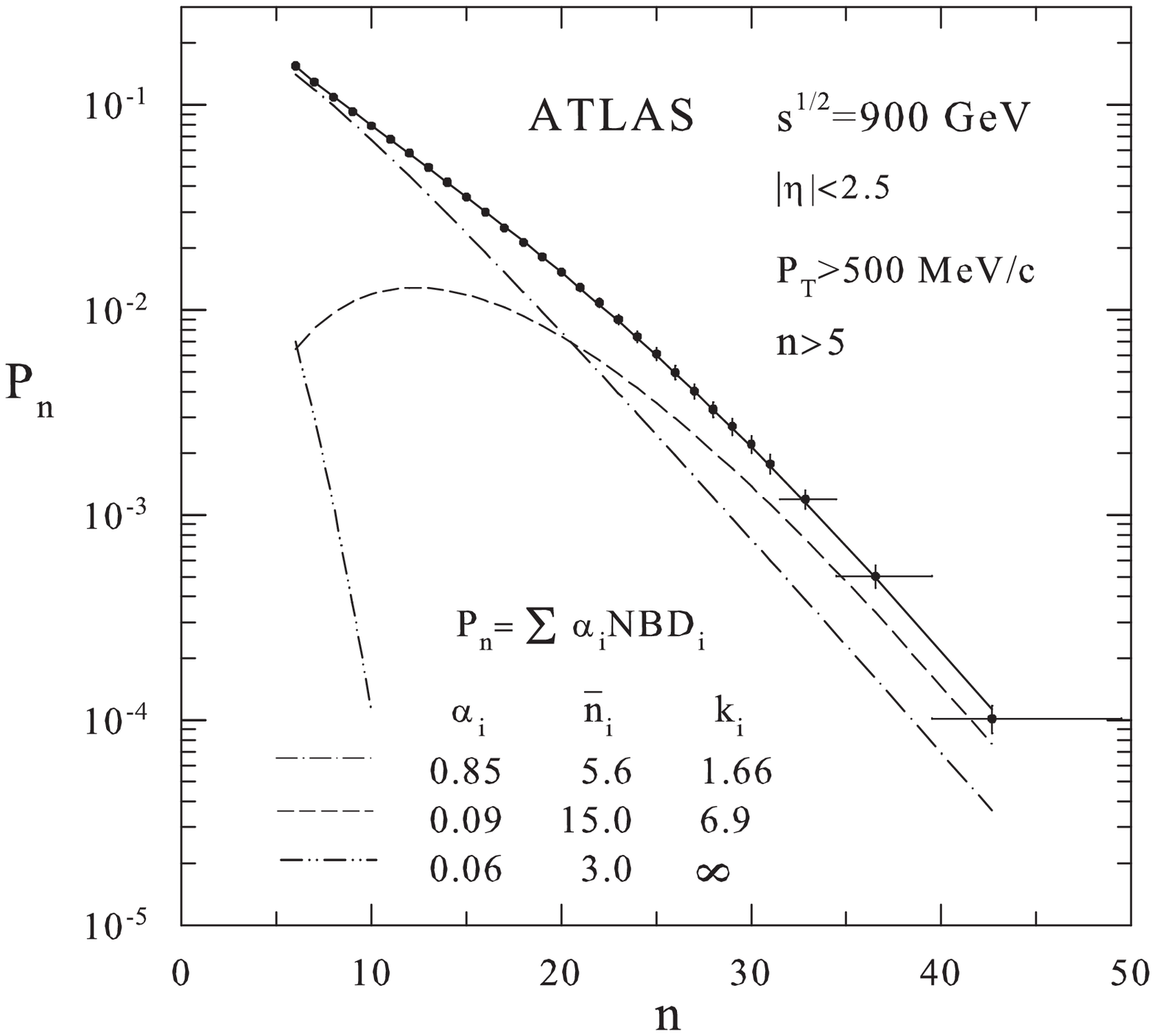}
\vskip -10mm
\hspace{1.cm}    (a) \hspace*{70mm} (b)
\caption{\label{Fig7}
MD of charged particles measured by the ATLAS Collaboration \protect\cite{ATLAS}
in the pseudorapidity interval $|\eta|<2.5$ for $p_T > 500$~MeV/c, $n>5$
at (a) $\sqrt{s}=7$~TeV and (b) $\sqrt{s}=0.9$~TeV.
The solid lines represent fits to the data with the restrictions quoted in table~\ref{tab3}.
The dash-dot, dash and dash-dot-dot lines show
single components corresponding to the indicated parameters.  }
\end{center}
\end{figure}

Figure \ref{Fig7}(a) shows data on MD measured by the ATLAS Collaboration at $\sqrt{s}=7$~TeV
in the pseudorapidity window $|\eta|<2.5$ in the phase space region $p_T > 500$~MeV/c and $n>5$.
The data points and the lines are shown in a similar way to figure \ref{Fig5}.
The multiplicity cut $n>5$ has an impact on the disappearance of the peak at maximum, though
its remnant is still visible in the low multiplicity region.
This is manifested by the increase of $P_n$ as $ n\rightarrow 5$.
The experimental data shown in figure \ref{Fig7}(a) allow us to determine all free parameters 
provided $\alpha_3$ is fixed.
The fitted parameters are quoted in figure \ref{Fig7}(a) and table~\ref{tab3}. 
\begin{table}[t]
\lineup
\caption{\label{tab3}
The parameters of superposition of three NBDs  
obtained from fits to  data \protect\cite{ATLAS} on MD measured by the ATLAS Collaboration
in the window $|\eta|<2.5$ with the cut $p_T>500$~MeV/c, $n>5$ at 
$ \sqrt{s}=7$ and 0.9 TeV.  }
\begin{indented}
\item[]
\begin{tabular}{@{}p{0.7cm}@{}lll@{}p{0.5cm}@{}lll@{}}
\br
  & \multicolumn{3}{c}{$\sqrt{s}$ = 7 TeV} & & \multicolumn{3}{c}{$\sqrt{s}$ = 0.9 TeV} \\ 
\cline{2-4}\cline{6-8} \\[-0.30cm]
{\it i}  & $\alpha_i$ & $\bar{n}_i$ & $k_i$ &   & $\alpha_i$ & $\bar{n}_i$ & $k_i$  \\
\cline{1-4}\cline{6-8} \\[-0.22cm] 
1 &  0.61$^{+0.05 }_{-0.15 }$  &  12.0$^{+1.6  }_{-2.4  }$   &   1.72$^{+0.75 }_{-0.40 }$  &           
  &  0.85$^{+0.05 }_{-0.17 }$  & \05.6$^{+0.6  }_{-1.0  }$   &   1.66$^{+0.72 }_{-0.29 }$  \\[0.15cm]  
2 &  0.11$^{+0.15 }_{-0.05 }$  &  30.6$^{+3.3  }_{-5.0  }$   &  5.7\0$^{+1.8  }_{-1.5  }$  &           
  &  0.09$^{+0.17 }_{-0.05 }$  &  15.0$^{+2.0  }_{-2.6  }$   &  6.9\0$^{+4.1  }_{-2.2  }$  \\[0.15cm]  
3 &  0.28-fixed                & \03.1$^{+0.8  }_{-1.0  }$   &  2.4\0$^{+1.6  }_{-0.8  }$  &           
  &  0.06-fixed                & \03.0-fixed                 &       $\infty$              \\[0.08cm]  
\cline{2-4}\cline{6-8}\\[-0.32cm]
  & \multicolumn{3}{c}{$\chi^2/dof$ = 8.5/(34-7)}  & &  \multicolumn{3}{c}{$\chi^2/dof$ = 4.7/(29-5)} \\ 
\br
\end{tabular}
\end{indented}
\end{table}
They have nearly the same values as obtained with the condition $n>0$ (see figure \ref{Fig5}(a) and table~\ref{tab2}). 
This result represents a consistency check of the description of data samples with different multiplicity cuts.
The ATLAS data measured in the window $|\eta|<2.5$ in the phase-space region $p_T > 500$~MeV/c, $n>5$ 
at the energy $\sqrt{s}=900$~GeV are depicted in figure \ref{Fig7}(b).
The multiplicity cut $n>5$  does not allow to determine any parameter of the low-multiplicity component at this energy.
Therefore, we have fixed $\alpha_3=0.06$, $\bar{n}_3=3.0$ and $k_3=\infty$ 
at the values obtained for $n>0$ (figure \ref{Fig5}(b)) leaving the remaining parameters free. 
The result of the fit is shown in figure \ref{Fig7}(b) and table~\ref{tab3}. Within the errors indicated,
the fitted parameters are more or less consistent for both cases with and without the $n>5$ cut.

\subsection{Pseudorapidity dependence of the third component and CMS data}

The CMS Collaboration provided the results of systematic measurements \cite{CMS} of MD of charged particles
produced in proton-proton collisions in the central interaction region. 
The data were accumulated in five pseudorapidity ranges from $|\eta|<0.5$ to
$|\eta|<2.4$ at the energies $\sqrt{s}=900$, 2360 and 7000 GeV.
The measurements refer to the inelastic NSD interactions.
The experimental analysis was based on about 132~000, 12~000 and 442~000 events
at $\sqrt{s}=900$, 2360 and 7000 GeV, respectively.
The data on MD in the central pseudorapidity windows
allow us to study the behavior of the fitted parameters in dependence
on the size of the phase-space regions.
We have analyzed data recorded by the CMS experiment at $\sqrt{s}=7$ and 0.9 TeV.
The reason is that the analysis in the framework of the superposition of three
NBDs is very sensitive to the statistics of data which is sufficiently high at both these energies.
The statistics of 12~000 events are too low to study a three-component description 
of the CMS data at the energy $\sqrt{s}=2360$~GeV.
Though systematic errors of the measurements are relatively large especially near the maximum
of the distributions, the values of $P_n$, when based on high statistics, allow us
to extract information on the fitted parameters in most cases sufficiently reliably.
The large systematic errors of the CMS measurements, as compared with the ATLAS data,
give relatively small values of $\chi^2$ even for the two-NBD hypothesis.
Following the objectives of this combined analysis, 
significantly smaller $\chi^2$ can be obtained in larger pseudorapidity windows $ |\eta|<\eta_c$
when superposition of three NBDs is considered.
The results of the fits in different pseudorapidity intervals are shown in table~\ref{tab4}.
Here some guidance was taken from the description of the ATLAS data  
in the window $|\eta|<2.5$ concerning 
the values of the parameter $k_3$ listed in table~~\ref{tab1}.
Within the errors quoted therein, $k_3\sim 24$  was found to be finite 
at $\sqrt{s}=7$~TeV.
Consequently, we have fitted the CMS data 
using $k_3$ as one of the adjustable parameters
in the largest intervals $|\eta|<2.4$ and $|\eta|<2.0$ at this energy.
The fittings give the corresponding finite values of $k_3$ in this region, though   
uncertainties of their determination from the CMS measurements admit description 
with $k_3=\infty$.
In smaller pseudorapidity windows, the mean values of $k_3$ turned out to be  
even larger than 100 with error intervals of infinity. 
Therefore, we have fixed $k_3\equiv\infty$ for
$\eta_c=1.5$, 1.0 and 0.5. 
\begin{table}[b]
\lineup
\caption{\label{tab4}
The parameters of superposition of three NBDs  
obtained from fits to data \protect\cite{CMS} on MD measured by the CMS Collaboration
in the pseudorapidity intervals $|\eta|<\eta_c$, $ \eta_c=2.4$, 2.0, 1.5, 1.0, 0.5 
at $\sqrt{s}=7$ and 0.9 TeV. 
}
\begin{indented}
\item[]
\begin{tabular}{@{}lp{0.7cm}@{}lll@{}p{0.5cm}@{}lll@{}}
\br
 &  & \multicolumn{3}{c}{$\sqrt{s}$ = 7 TeV} & & \multicolumn{3}{c}{$\sqrt{s}$ = 0.9 TeV} \\ 
\cline{3-5}\cline{7-9} \\[-0.30cm]
$\eta_c$ & {\it i}  & $\alpha_i$ & $\bar{n}_i$ & $k_i$ &   & $\alpha_i$ & $\bar{n}_i$ & $k_i$  \\
\cline{1-5}\cline{7-9} \\[-0.35cm]
2.4 & 1 & 0.824$^{+0.079}_{-0.238}$ & 28.4$^{+2.6 }_{-7.4  }$     &\01.66$^{+0.27 }_{-0.09 }$  &         
        &  0.74$^{+0.14 }_{-0.28 }$ & 15.7$^{+1.7 }_{-4.1  }$     &  2.09$^{+0.27 }_{-0.24 }$ \\[0.15cm] 
    & 2 & 0.107$^{+0.202}_{-0.053}$ & 67.\0$^{+5. }_{\!\!\!-11.}$ &\06.6\0$^{+3.3  }_{-2.2  }$ &         
        &  0.19$^{+0.25 }_{-0.11 }$ & 32.3$^{+4.7 }_{-4.8  }$     & 6.7\0$^{+4.9  }_{-2.0  }$ \\[0.15cm] 
    & 3 & 0.069$^{+0.036}_{-0.026}$ & 13.0$^{+0.9 }_{-0.9  }$     &40.\0$^{\,+\infty}_{\,-29. }$   &     
        &  0.07$^{+0.03 }_{-0.03 }$ & 11.7$^{+0.9 }_{-1.0  }$     &      $\infty$           \\[0.08cm]   
\cline{3-5}\cline{7-9}\\[-0.32cm]
 & & \multicolumn{3}{c}{$\chi^2/dof$ = 2.5/(127-8-1)}  & &  \multicolumn{3}{c}{$\chi^2/dof$ = 4.4/(68-7-1)} \\ 
\cline{1-5}\cline{7-9} \\[-0.35cm]
2.0 & 1 & 0.795$^{+0.071}_{-0.145}$ & 22.1$^{+2.2 }_{-3.0  }$  &\01.69$^{+0.18 }_{-0.10 }$  &         
        &  0.81$^{+0.12 }_{-0.42 }$ & 13.6$^{+1.4 }_{-5.1  }$  &  2.07$^{+0.50 }_{-0.24 }$\\[0.15cm]  
    & 2 & 0.145$^{+0.095}_{-0.051}$ & 55.2$^{+3.7 }_{-4.7  }$  &\06.2\0$^{+1.4  }_{-1.1  }$ &         
        &  0.14$^{+0.38 }_{-0.09 }$ & 28.6$^{+4.8 }_{-6.7  }$  & 7.0\0$^{+8.6  }_{-3.0  }$\\[0.15cm]  
    & 3 & 0.060$^{+0.050}_{-0.020}$ & 10.6$^{+1.0 }_{-1.0  }$  & 72.\0$^{\,+\infty}_{\,-63. }$ &      
         & 0.05$^{+0.04 }_{-0.03 }$ &\09.4$^{+1.2 }_{-1.3  }$  &      $\infty$            \\[0.08cm]  
\cline{3-5}\cline{7-9}\\[-0.32cm]
 & & \multicolumn{3}{c}{$\chi^2/dof$ = 3.8/(115-8-1)}  & &  \multicolumn{3}{c}{$\chi^2/dof$ = 2.4/(62-7-1)} \\  
\cline{1-5}\cline{7-9} \\[-0.35cm]
1.5 & 1 & 0.796$^{+0.069}_{-0.116}$ &  16.2$^{+1.4 }_{-2.2  }$  &\01.65$^{+0.20 }_{-0.11 }$  &         
        &  0.81$^{+0.12 }_{-0.28 }$ & \09.8$^{+1.1 }_{-2.5  }$  &  2.17$^{+0.65 }_{-0.25 }$ \\[0.15cm] 
    & 2 & 0.149$^{+0.094}_{-0.050}$ &  41.5$^{+2.8 }_{-3.6  }$  &\06.0\0$^{+1.3 }_{-1.0  }$  &         
        &  0.15$^{+0.24 }_{-0.08 }$ &  22.8$^{+3.1 }_{-4.3  }$  & 7.6\0$^{+4.9  }_{-2.7  }$ \\[0.15cm] 
    & 3 & 0.055$^{+0.022}_{-0.019}$ & \07.8$^{+1.0 }_{-0.8  }$  &    \0$\infty$              &         
        &  0.04$^{+0.04 }_{-0.04 }$ & \06.8$^{+2.3 }_{-2.3  }$  &      $\infty$             \\[0.08cm] 
\cline{3-5}\cline{7-9}\\[-0.32cm]
 & & \multicolumn{3}{c}{$\chi^2/dof$ = 2.9/(95-7-1)}  & &  \multicolumn{3}{c}{$\chi^2/dof$ = 2.0/(52-7-1)} \\   
\cline{1-5}\cline{7-9} \\[-0.35cm]
1.0 & 1 & 0.806$^{+0.092}_{-0.180}$ &  10.8$^{+1.3 }_{-2.2  }$  &\01.58$^{+0.37 }_{-0.14 }$ &          
        &  0.81$^{+0.13 }_{-0.35 }$ & \06.5$^{+1.0 }_{-2.2  }$  &  2.15$^{+1.54 }_{-0.41 }$\\[0.15cm]  
    & 2 & 0.136$^{+0.143}_{-0.060}$ &  28.5$^{+2.8 }_{-3.8  }$  &\06.1\0$^{+2.3 }_{-1.5  }$ &          
        &  0.15$^{+0.29 }_{-0.09 }$ &  15.6$^{+2.7 }_{-3.6  }$  & 7.3\0$^{+5.2  }_{-2.8  }$\\[0.15cm]  
    & 3 & 0.058$^{+0.037}_{-0.032}$ & \05.0$^{+1.0 }_{-0.9  }$  &    \0$\infty$             &          
        &  0.04$^{+0.06 }_{-0.04 }$ & \04.5$^{+1.8 }_{-1.7  }$  &      $\infty$             \\[0.08cm] 
\cline{3-5}\cline{7-9}\\[-0.32cm]
 & & \multicolumn{3}{c}{$\chi^2/dof$ = 0.9/(70-7-1)}  & &  \multicolumn{3}{c}{$\chi^2/dof$ = 4.3/(40-7-1)} \\   
\cline{1-5}\cline{7-9} \\[-0.35cm]
0.5 & 1 & 0.792$^{+0.149}_{-0.307}$ & \05.1$^{+1.3 }_{-2.0  }$  &\01.49$^{+1.31 }_{-0.24 }$ &          
        &  0.79$^{+0.06 }_{-0.13 }$ & \03.3$^{+1.9 }_{-0.6  }$  &  1.36$^{+0.89 }_{-0.53 }$\\[0.15cm]  
    & 2 & 0.148$^{+0.213}_{-0.089}$ &  14.4$^{+2.6 }_{-2.9  }$  &\05.9\0$^{+4.1 }_{-2.0  }$ &          
        &  0.15-fixed               & \06.6$^{+1.6 }_{-1.8  }$  & 5.1\0$^{+5.4  }_{-2.7  }$\\[0.15cm]  
    & 3 & 0.060$^{+0.094}_{-0.060}$ & \02.7$^{+1.4 }_{-1.3  }$  &    \0$\infty$             &          
        &  0.06$^{+0.13 }_{-0.06 }$ & \03.3$^{+1.6 }_{-1.5  }$  &      $\infty$             \\[0.08cm]  
\cline{3-5}\cline{7-9}\\[-0.32cm]
 & & \multicolumn{3}{c}{$\chi^2/dof$ = 2.3/(41-7-1)}  & &  \multicolumn{3}{c}{$\chi^2/dof$ = 0.8/(23-6-1)} \\     
\br
\end{tabular}
\end{indented}
\end{table}
\begin{figure}
\begin{center}
\vskip 0cm
\hspace*{0mm}
\includegraphics[width=78mm,height=78mm]{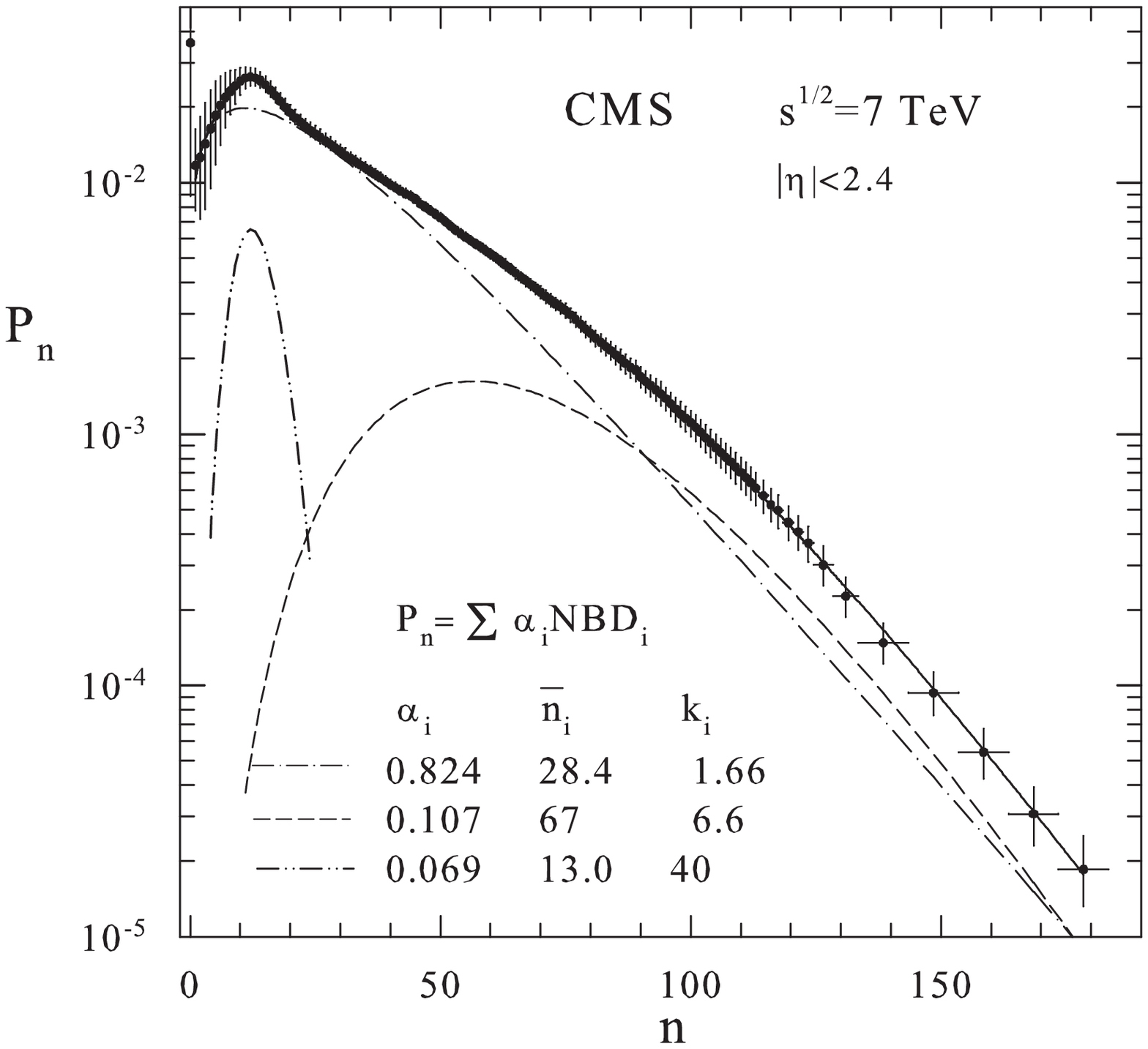}
\hspace{-0.5cm}
\includegraphics[width=78mm,height=78mm]{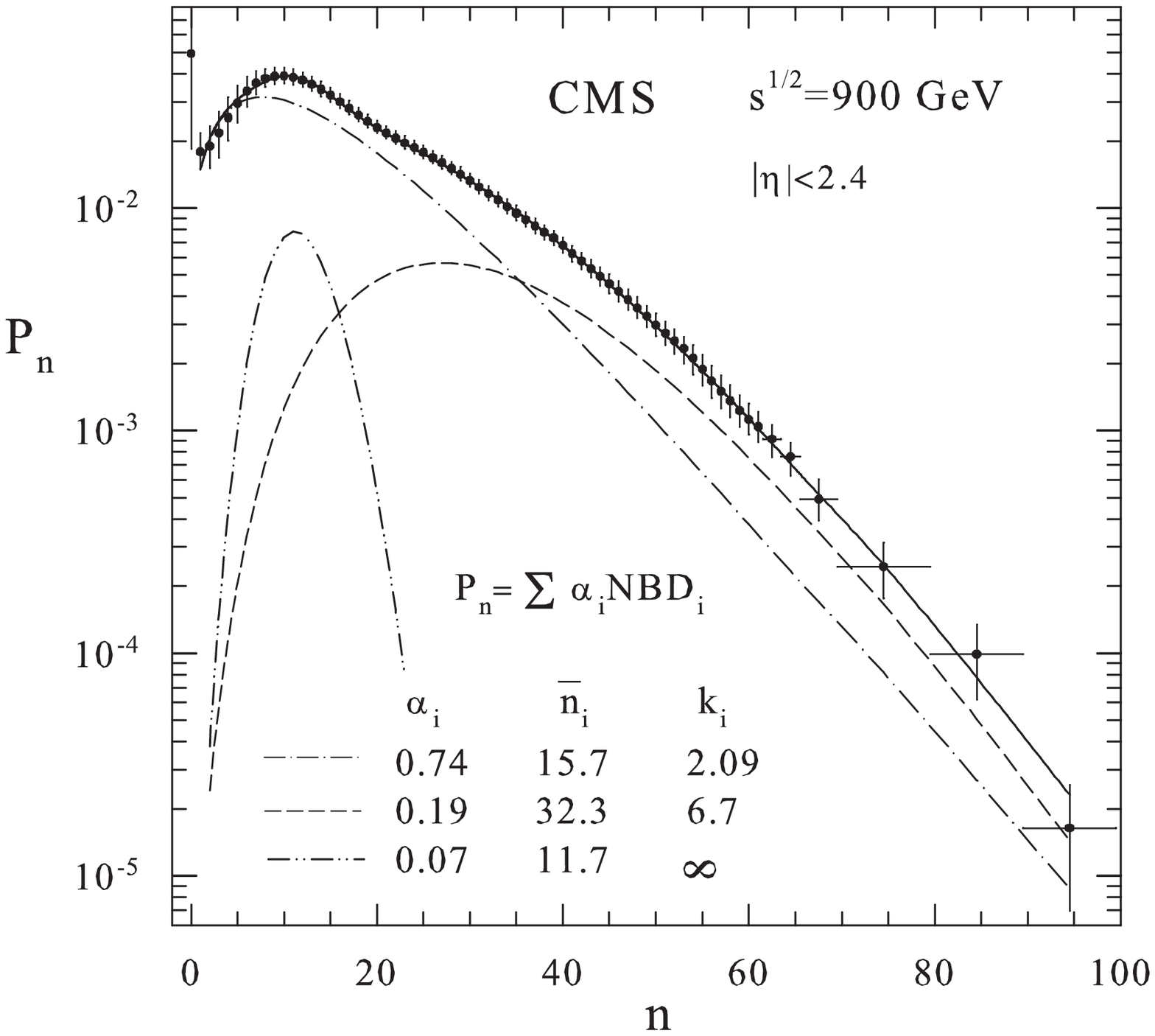}
\vskip -10mm
\hspace*{0mm}
\includegraphics[width=78mm,height=78mm]{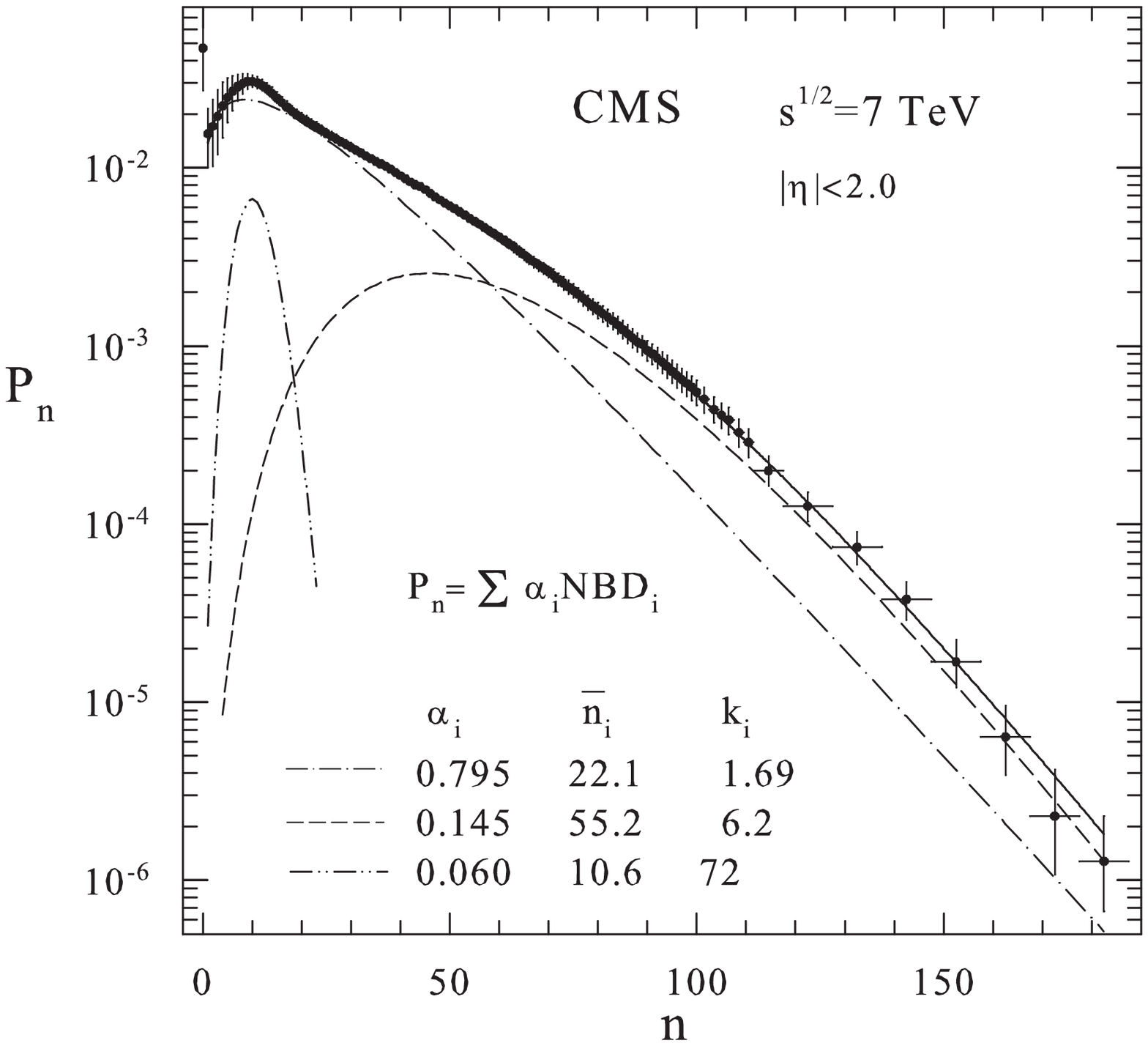}
\hspace{-0.5cm}
\includegraphics[width=78mm,height=78mm]{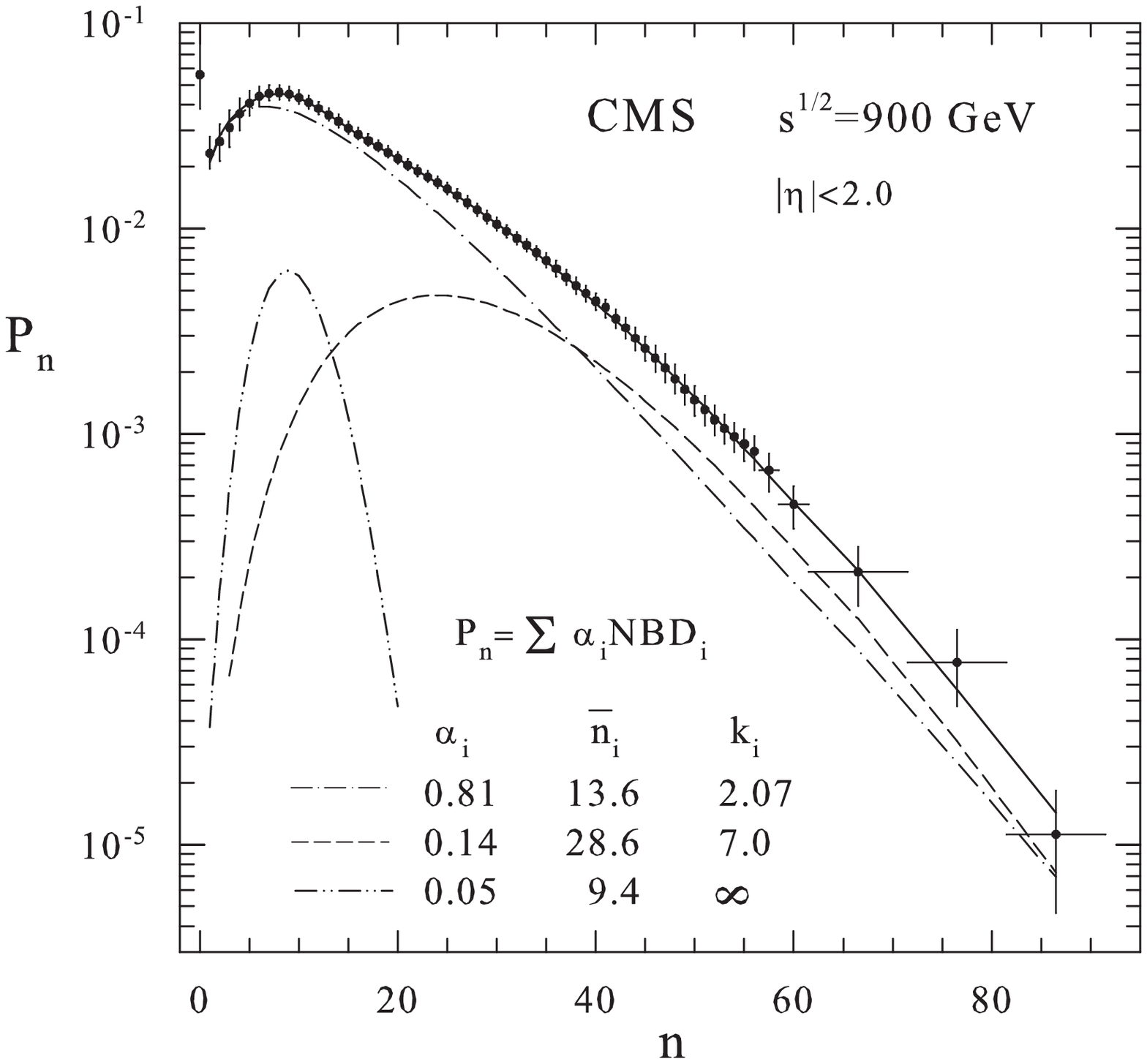}
\vskip -10mm
\hspace*{0mm}
\includegraphics[width=78mm,height=78mm]{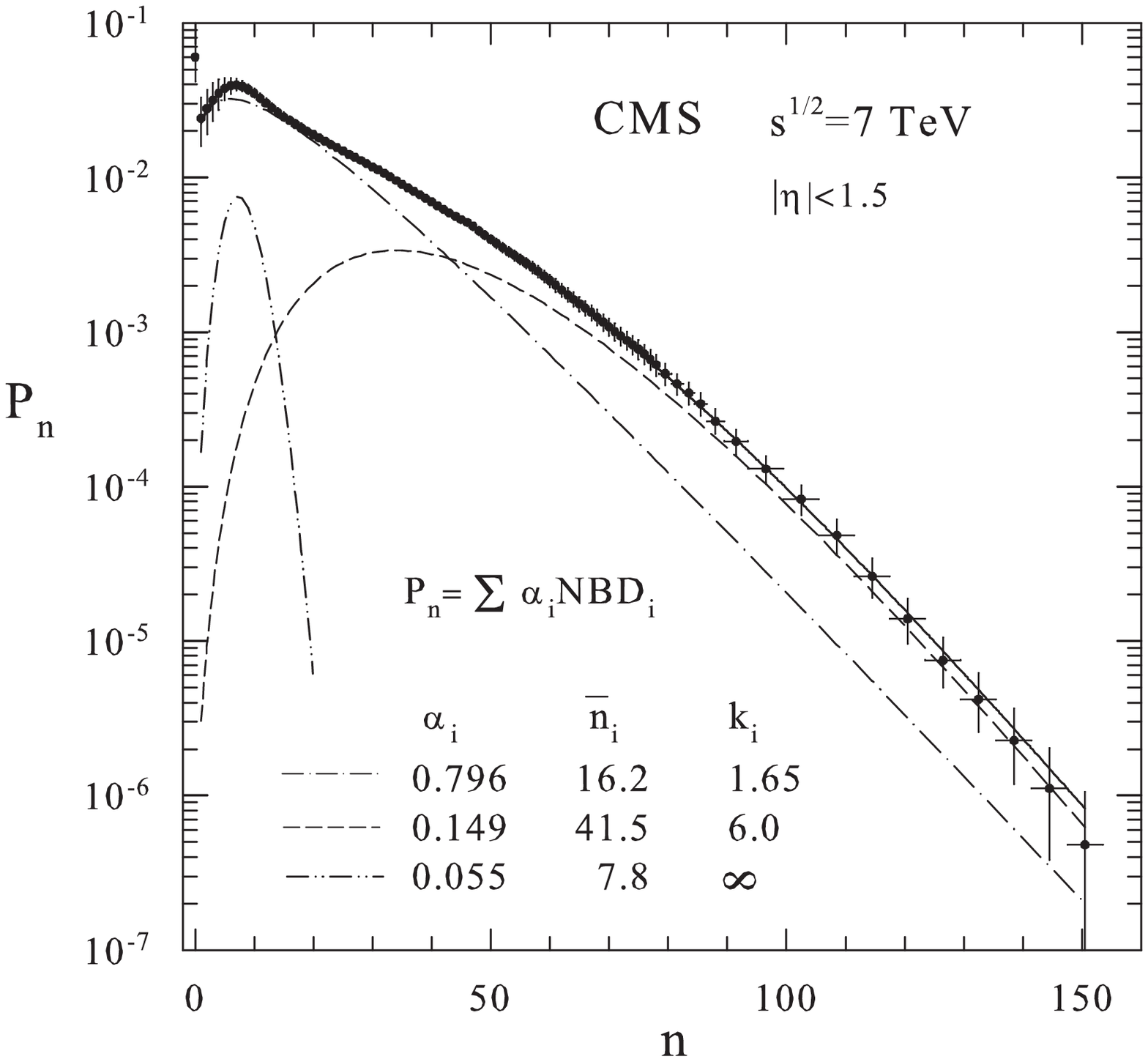}
\hspace{-0.5cm}
\includegraphics[width=78mm,height=78mm]{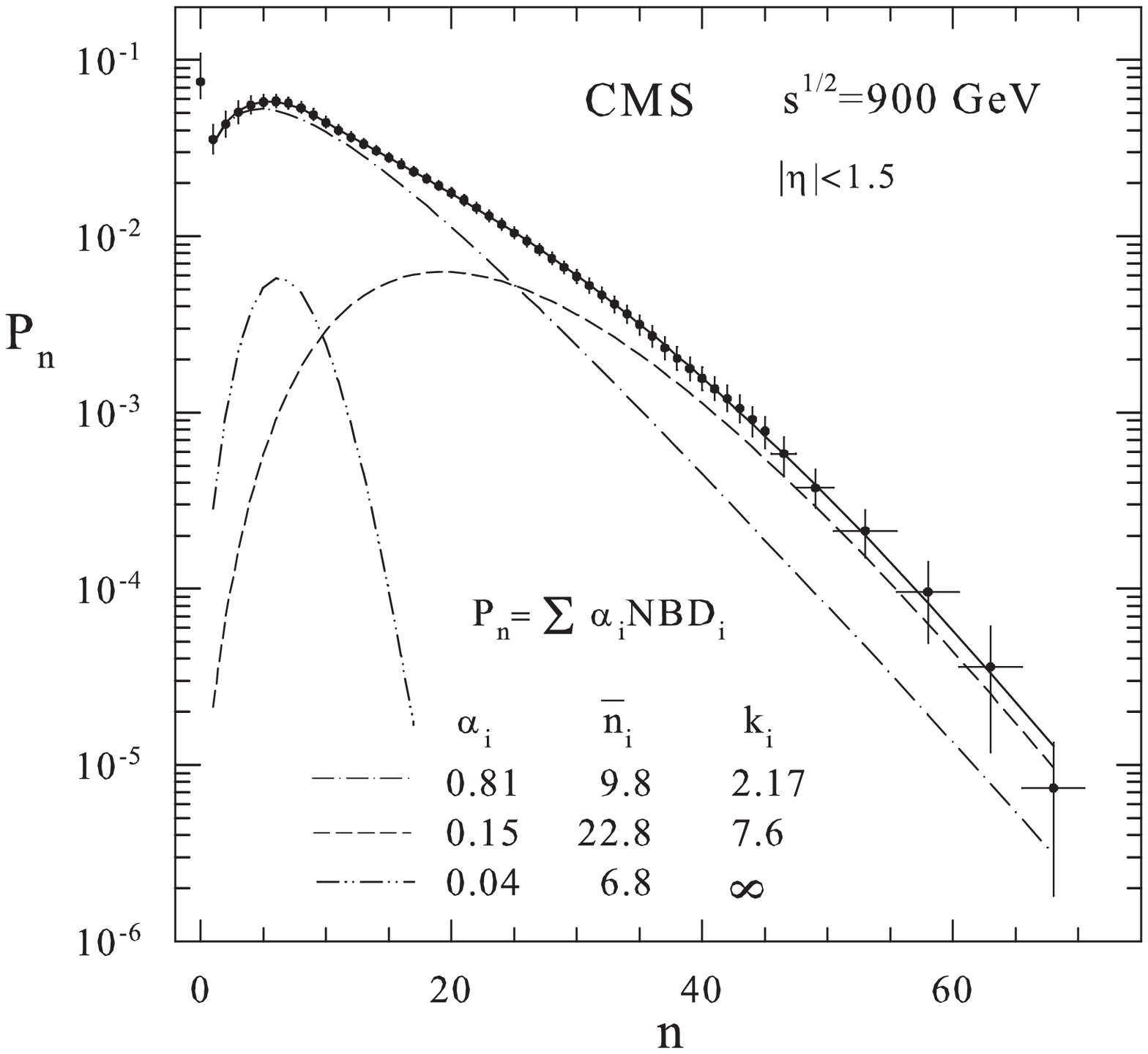}
\vskip -10mm
\hspace{1.cm}    (a) \hspace*{70mm} (b)
\caption{\label{Fig8}
MD of charged particles measured by the CMS Collaboration \protect\cite{CMS}
in the pseudorapidity windows $|\eta|<\eta_c$, $ \eta_c=2.4$, 2.0 and  1.5
at (a) $\sqrt{s}=7$~TeV and (b) $\sqrt{s}=0.9$~TeV.
The solid lines represent fits to the data by the three-component
superposition of NBDs. The dash-dot, dash and dash-dot-dot lines show
single components corresponding to the indicated parameters.  }
\end{center}
\end{figure} 
\begin{figure}
\begin{center}
\vskip 0cm
\hspace*{0mm}
\includegraphics[width=78mm,height=78mm]{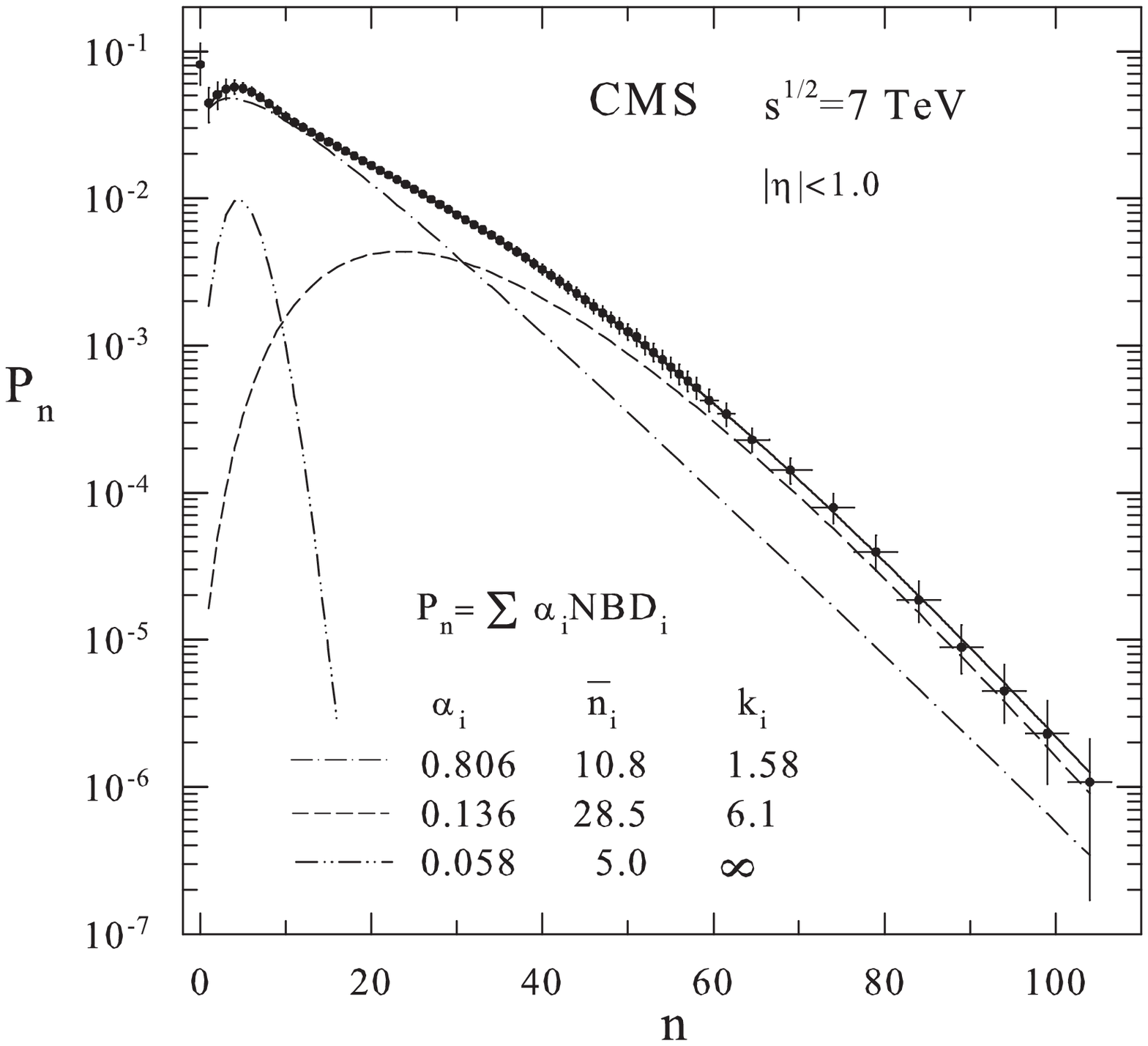}
\hspace{-0.5cm}
\includegraphics[width=78mm,height=78mm]{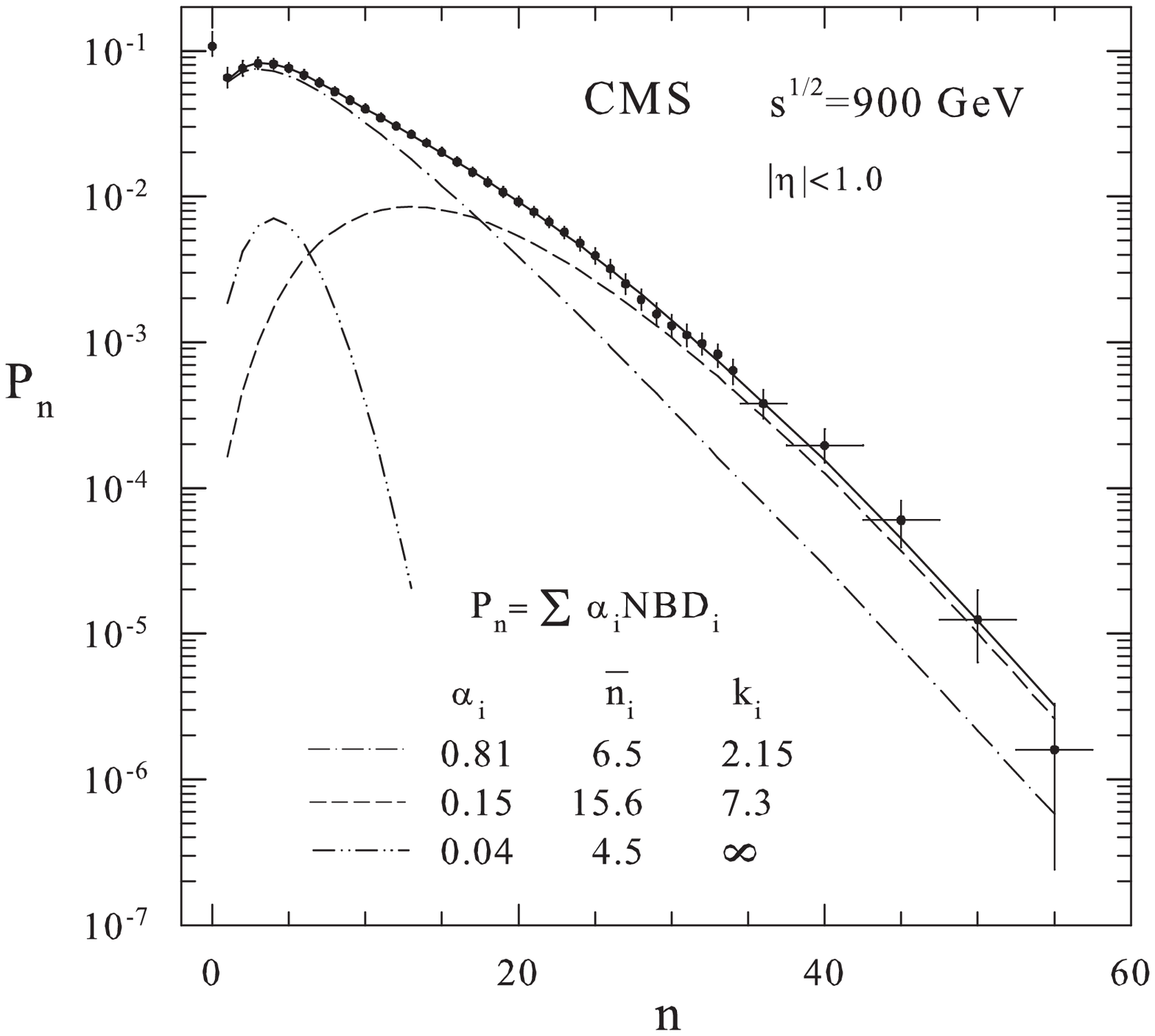}
\vskip -10mm
\hspace*{0mm}
\includegraphics[width=78mm,height=78mm]{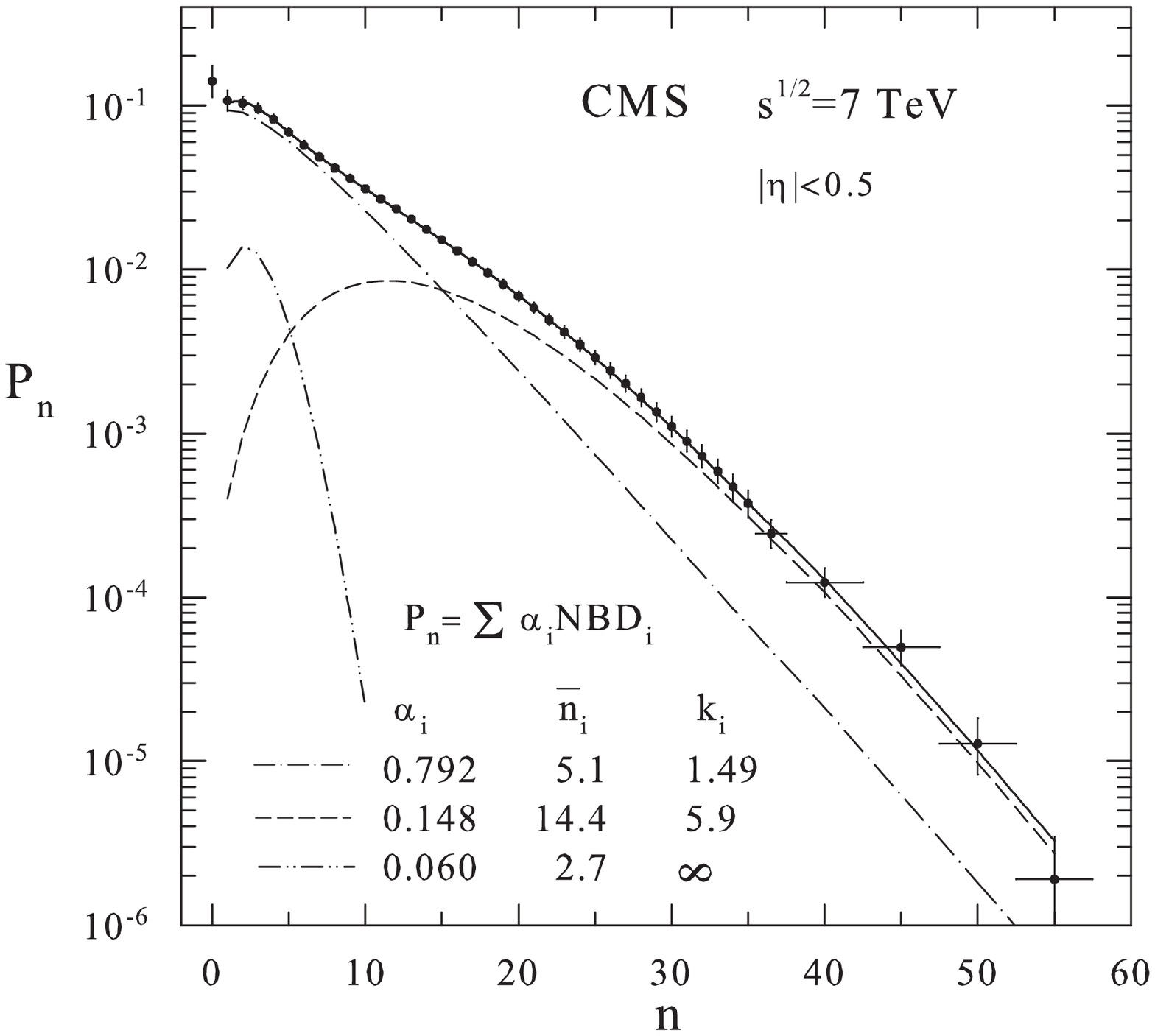}
\hspace{-0.5cm}
\includegraphics[width=78mm,height=78mm]{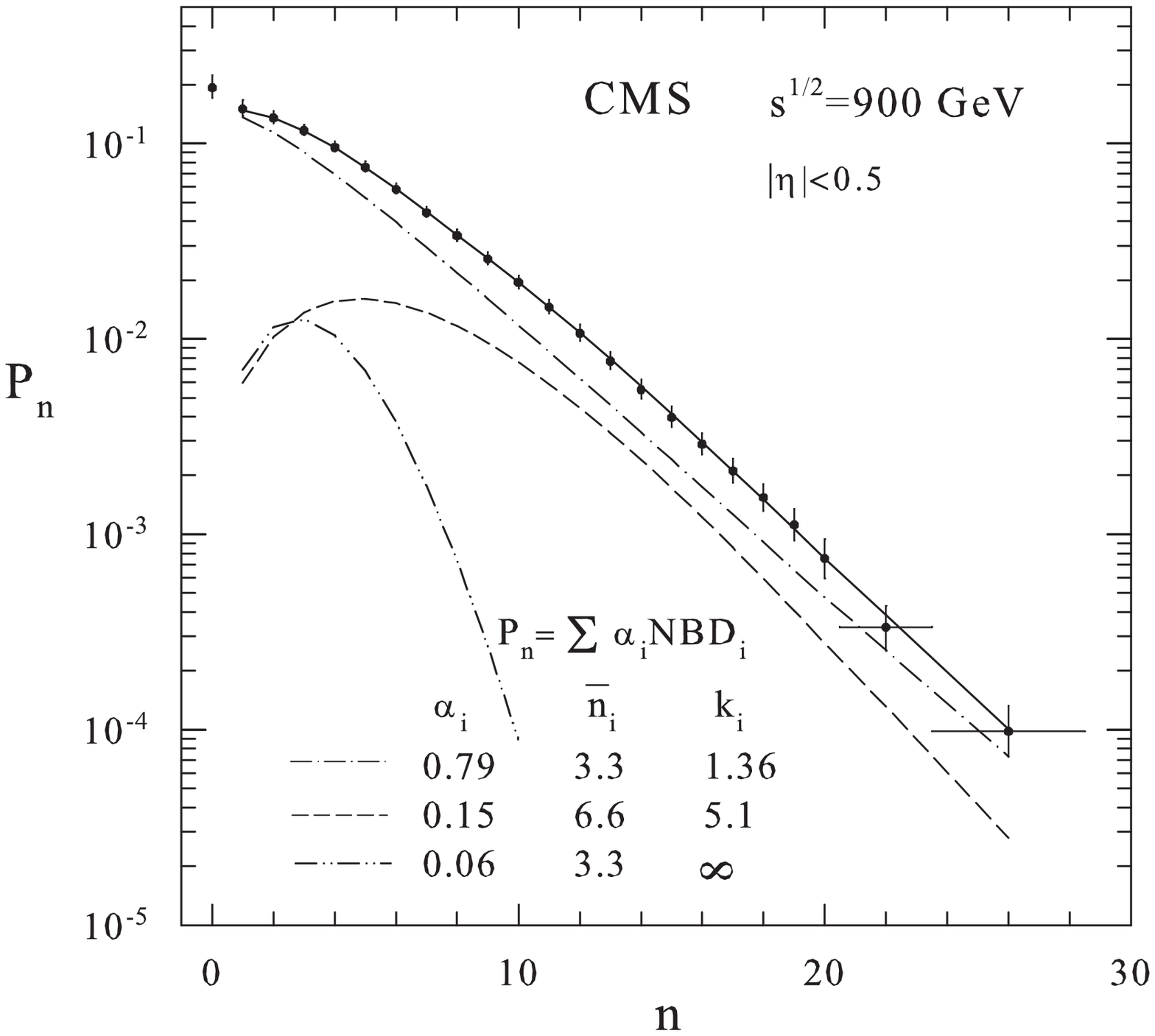}
\vskip -10mm
\hspace{1.cm}    (a) \hspace*{70mm} (b)
\caption{\label{Fig9}
MD of charged particles measured by the CMS Collaboration \protect\cite{CMS}
in the small windows $|\eta|<1.0$ and $|\eta|<0.5$
at (a) $\sqrt{s}=7$~TeV and (b) $\sqrt{s}=0.9$~TeV.
The solid lines represent fits to the data by three-component
superposition of NBDs. The dash-dot, dash and dash-dot-dot lines show
single components corresponding to the indicated parameters.  }
\end{center}
\end{figure} 
As follows from table~\ref{tab1}, the errors of all parameters 
are larger at smaller energy.
Within the errors, the value of $k_3$ is consistent with infinity in the window $|\eta|<2.5$ 
for the ATLAS data at $\sqrt{s}=900$~GeV.
In view of the large systematic uncertainties of the CMS measurements,  this 
substantiates the choice $k_3=\infty$ in all pseudorapidity intervals when fitting the CMS data  
at $\sqrt{s}=900$~GeV.

Figure \ref{Fig8} shows the MD of charged particles measured in the pseudorapidity windows $|\eta|<2.4$,
$|\eta|<2.0$ and  $|\eta|<1.5$ at the energies $\sqrt{s}=7$~TeV and $\sqrt{s}=0.9$~TeV.
The experimental data are depicted by symbols.
The solid lines represent fits corresponding to
weighted superposition of three NBDs.
The single components are shown by the dash-dot, dash and dash-dot-dot curves, respectively.
The zero point $P_0$ was not included in the fits.
Figure \ref{Fig9} shows the MD measured in the small windows $|\eta|<1.0$ and $|\eta|<0.5$. 
The data at the energy $ \sqrt{s}=900$~GeV do not allow to extract reliable information on 
superposition of three NBDs in the narrowest window $|\eta|<0.5$.
Here we have fixed the probability of the second component at the value $\alpha_2=0.15$
obtained in the same window at $\sqrt{s}=7$~TeV and also 
in the larger window $|\eta|<1.0$ at $\sqrt{s}=0.9$~TeV.

One can see from figures \ref{Fig8} and \ref{Fig9} that the three-component description of the CMS data in different 
pseudorapidity ranges manifests analogous trends as observed from the 
analysis of the most inclusive sample of the ATLAS measurements shown in figure \ref{Fig3}.
Moreover, values of the fitted parameters in the intervals $|\eta|<2.4$ and $|\eta|<2.0$ 
are in a sense similar to those obtained from the ATLAS data. 
The multiplicity component with largest probability $\alpha_1$ has smallest parameter $k_1$ in all windows.
Its contributions $ \alpha_1\bar{n}_1 $ to the total average multiplicities are dominant at both energies.
Except for the window $|\eta|<0.5$ at $ \sqrt{s}=900$~GeV one can see that the 
other two components contribute to the high and low multiplicity region, respectively.
The values of the parameters $k_i$ increase with the decreasing probabilities $\alpha_i$.
The average multiplicities of the first and the second component, $\bar{n}_1$ and  $\bar{n}_2$,
increase with energy. 
The average multiplicity $\bar{n}_3$ of the third component is within errors 
nearly energy independent for all $\eta_c$.
The approximate energy independence is valid for the probabilities $ \alpha_i $ as well. 

\begin{figure}
\begin{center}
\vskip 0cm
\hspace*{0mm}
\includegraphics[width=78mm,height=78mm]{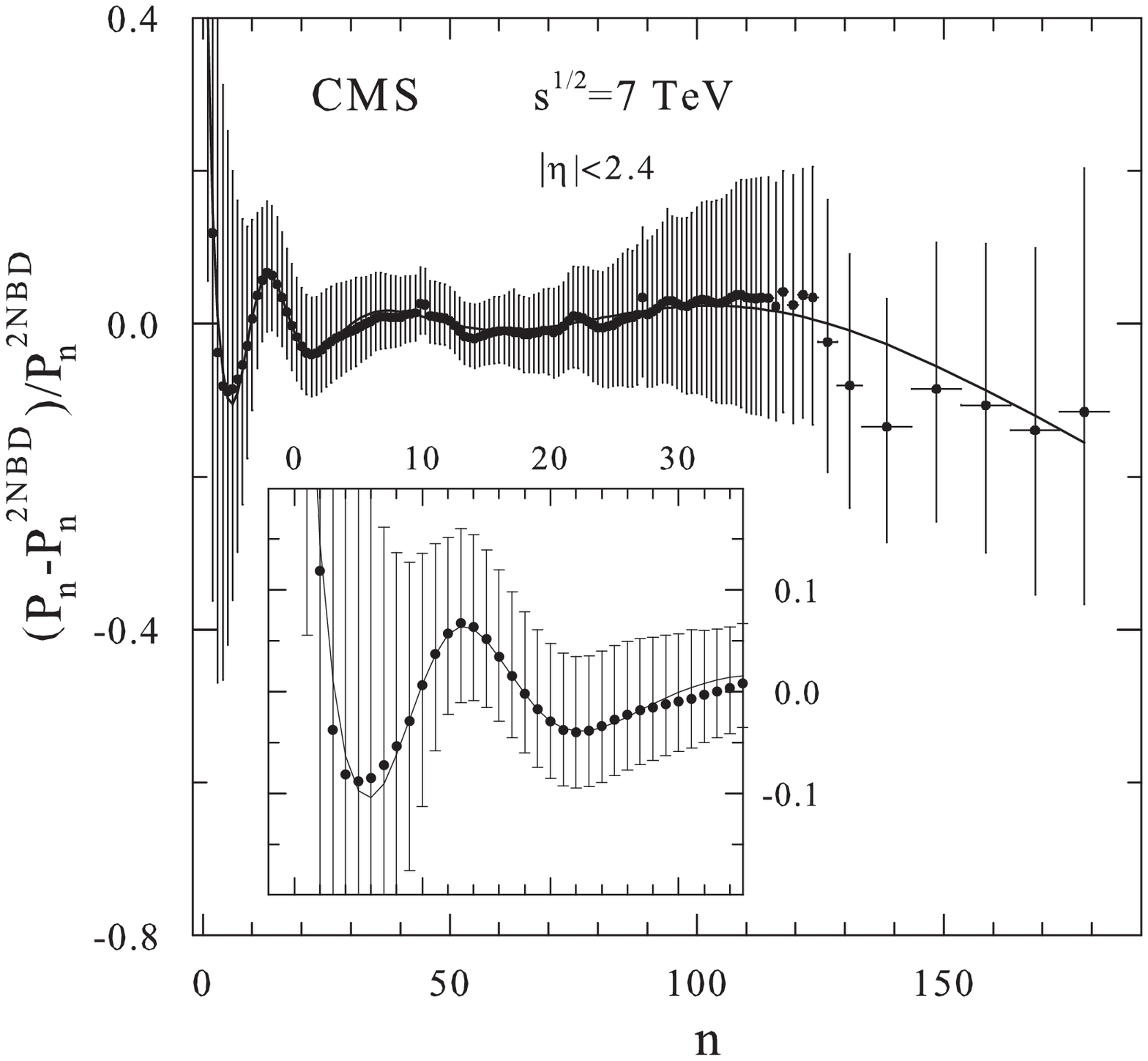}
\hspace{-0.5cm}
\includegraphics[width=78mm,height=78mm]{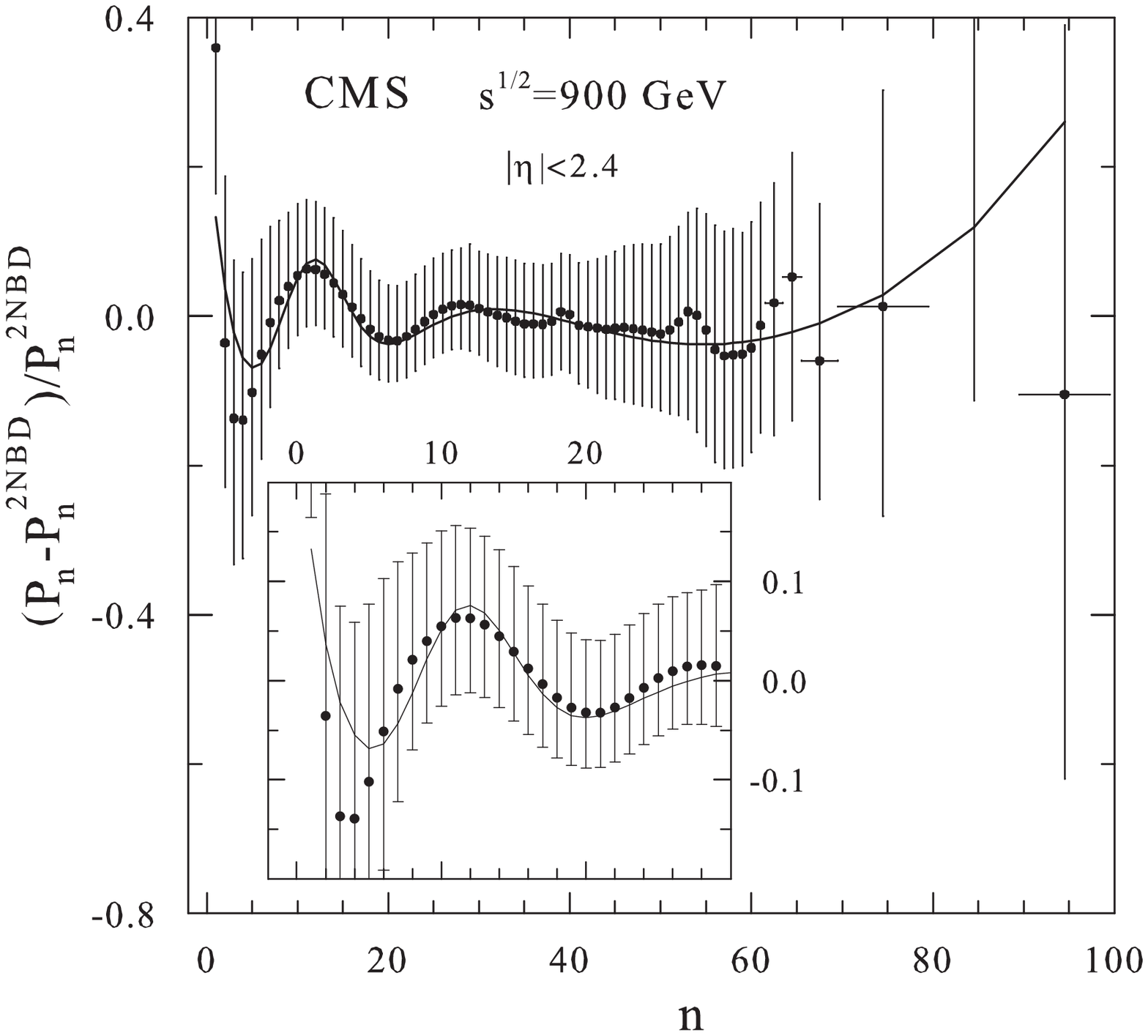}
\vskip -10mm
\hspace*{0mm}
\includegraphics[width=78mm,height=78mm]{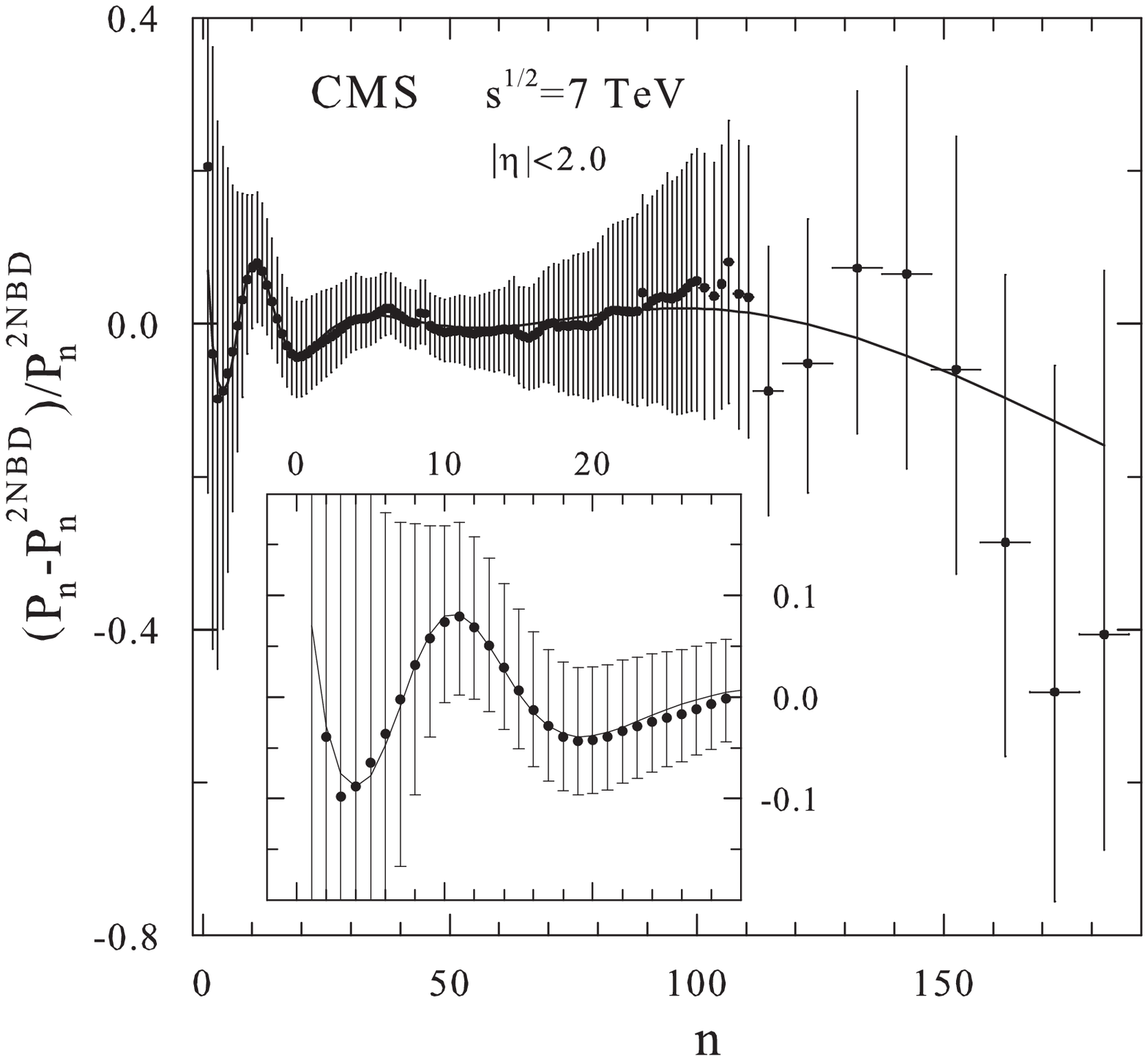}
\hspace{-0.5cm}
\includegraphics[width=78mm,height=78mm]{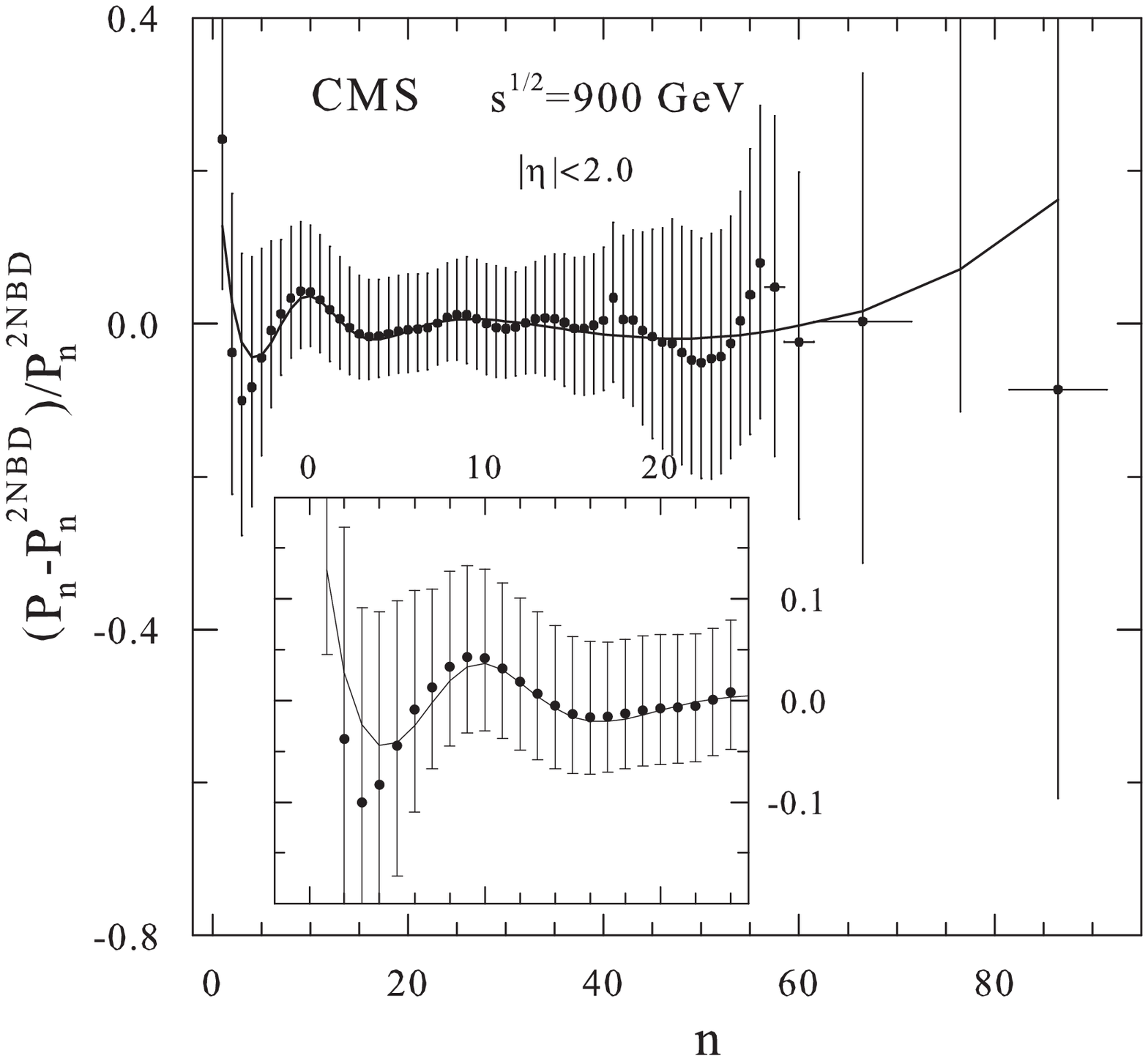}
\vskip -10mm
\hspace*{0mm}
\includegraphics[width=78mm,height=78mm]{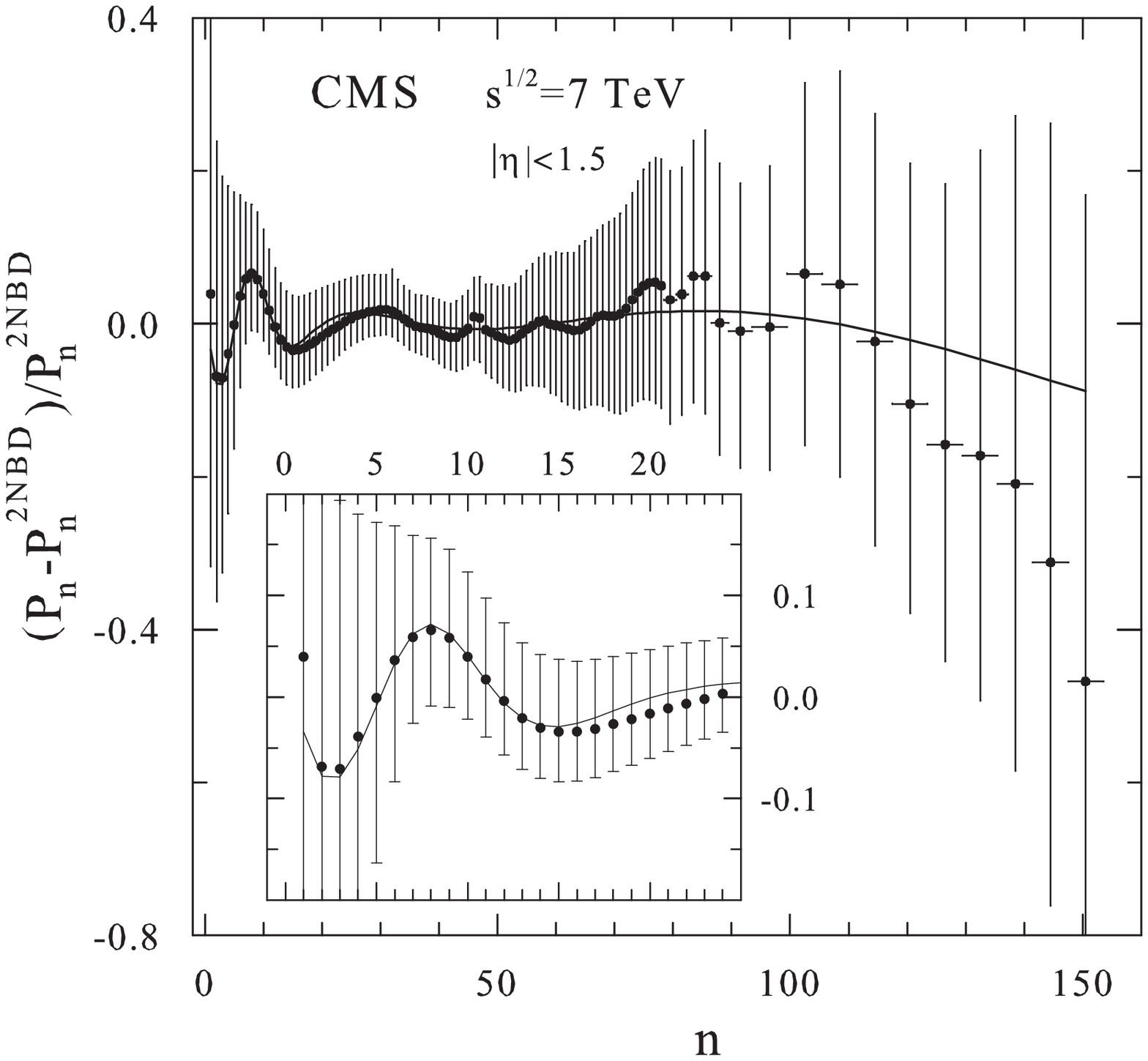}
\hspace{-0.5cm}
\includegraphics[width=78mm,height=78mm]{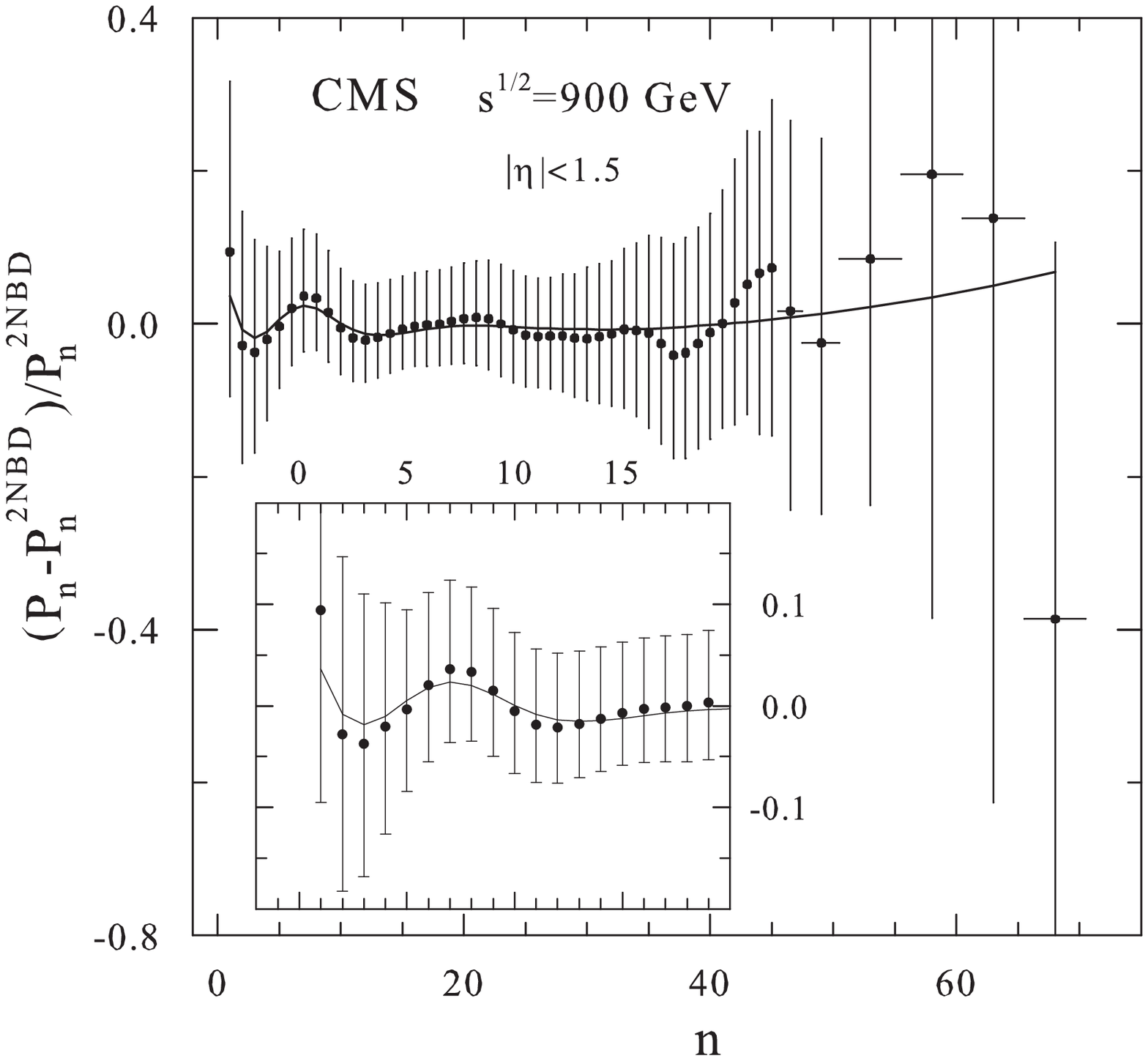}
\vskip -10mm
\hspace{1.cm}    (a) \hspace*{70mm} (b)
\caption{\label{Fig10}
Normalized residues of MD relative to the weighted superposition of two NBDs $(\mathrm{P_n^{2NBD}})$
with the parameters listed in table~\ref{tab5}.
The symbols correspond to data on MD measured by the CMS Collaboration \protect\cite{CMS}
in the pseudorapidity intervals $|\eta|<\eta_c$, $ \eta_c=2.4$, 2.0 and 1.5 at 
(a) $\sqrt s=7$~TeV and (b) $\sqrt s=0.9$~TeV.
The lines correspond to fits to the data by three-component 
superposition of NBDs. 
The insets show the detailed behavior of the residues at  
low $n$.
 }
\end{center}
\end{figure}

We have fitted the CMS data in different pseudorapidity intervals with weighted superposition of two NBDs.
The obtained values of parameters are listed in table~\ref{tab5}.
Figure \ref{Fig10} shows 
the normalized residues of MD relative to the two-NBD fits 
\begin{table}[b]
\lineup
\caption{\label{tab5}
The parameters of superposition of two NBDs  
obtained from fits to  data \protect\cite{CMS} on MD measured by the CMS Collaboration
in the pseudorapidity  windows $|\eta|<\eta_c$, $ \eta_c=2.4$, 2.0, 1.5,1.0, 0.5 at 
$\sqrt{s}=7$ and 0.9 TeV. 
}
\begin{indented}
\item[]
\begin{tabular}{@{}lp{0.7cm}@{}lll@{}p{0.5cm}@{}ll@{}l@{}}
\br
 &  & \multicolumn{3}{c}{$\sqrt{s}$ = 7 TeV} & & \multicolumn{3}{c}{$\sqrt{s}$ = 0.9 TeV} \\ 
\cline{3-5}\cline{7-9} \\[-0.30cm]
$\eta_c$ & {\it i}  & $\alpha_i$ & $\bar{n}_i$ & $k_i$ &   & $\alpha_i$ & $\bar{n}_i$ & $k_i$  \\
\cline{1-5}\cline{7-9} \\[-0.35cm]
2.4 & 1 &  0.470$\pm$0.071   &   15.1$\pm$1.3    &    2.57$\pm$0.31  &             
        &  0.82$\pm$0.11     &   15.0$\pm$1.9    &    2.48$\pm$0.32  \\[0.00cm]    
    & 2 &  0.530$\pm$0.071   &   46.2$\pm$2.9    &    3.15$\pm$0.42  &             
        &   0.18$\pm$0.11    &   35.4$\pm$3.8    &   8.4\0$\pm$2.5\0   \\[0.00cm]  
\cline{3-5}\cline{7-9}\\[-0.32cm]
 & & \multicolumn{3}{c}{$\chi^2/dof$ = 13.4/(127-5-1)}  & &  \multicolumn{3}{c}{$\chi^2/dof$ = 10.7/(68-5-1)} \\     
\cline{1-5}\cline{7-9} \\[-0.35cm]
2.0 & 1 &  0.615$\pm$0.081   &   14.9$\pm$1.6    &    2.16$\pm$0.23  &             
        &   0.75$\pm$0.17    &   11.5$\pm$2.2    &    2.60$\pm$0.46  \\[0.00cm]    
    & 2 &  0.385$\pm$0.081   &   44.3$\pm$3.5    &    4.14$\pm$0.64  &             
        &  0.25$\pm$0.17     &   27.8$\pm$4.5    &   6.6\0$\pm$2.3\0  \\[0.00cm]   
\cline{3-5}\cline{7-9}\\[-0.32cm]
 & & \multicolumn{3}{c}{$\chi^2/dof$ = 14.2/(115-5-1)}  & &  \multicolumn{3}{c}{$\chi^2/dof$ = 5.2/(62-5-1)} \\      
\cline{1-5}\cline{7-9} \\[-0.35cm]
1.5 & 1 &  0.666$\pm$0.084   &   11.7$\pm$1.4    &    1.99$\pm$0.24  &             
        &   0.77$\pm$0.17    &  \08.7$\pm$1.8    &    2.55$\pm$0.56  \\[0.00cm]    
    & 2 &  0.334$\pm$0.084   &   35.0$\pm$3.0    &    4.47$\pm$0.69  &             
        &  0.23$\pm$0.17     &   21.7$\pm$3.8    &   7.0\0$\pm$2.6\0  \\[0.00cm]   
\cline{3-5}\cline{7-9}\\[-0.32cm]
 & & \multicolumn{3}{c}{$\chi^2/dof$ = 9.7/(95-5-1)}  & &  \multicolumn{3}{c}{$\chi^2/dof$ = 3.0/(52-5-1)} \\   
\cline{1-5}\cline{7-9} \\[-0.35cm]
1.0 & 1 & 0.70\0$\pm$0.11\0  &  \08.0$\pm$1.2    &    1.86$\pm$0.30  &             
        &   0.78$\pm$0.22    &  \05.8$\pm$1.5    &    2.45$\pm$0.76  \\[0.00cm]    
    & 2 & 0.30\0$\pm$0.11\0  &   24.2$\pm$2.7    &    4.56$\pm$0.94  &             
        &   0.22$\pm$0.22    &   14.9$\pm$3.4    &   6.8\0$\pm$3.0\0  \\[0.00cm]   
\cline{3-5}\cline{7-9}\\[-0.32cm]
 & & \multicolumn{3}{c}{$\chi^2/dof$ = 3.8/(70-5-1)}  & &  \multicolumn{3}{c}{$\chi^2/dof$ = 4.5/(40-5-1)} \\   
\cline{1-5}\cline{7-9} \\[-0.35cm]
0.5 & 1 & 0.72\0$\pm$0.14\0  &  \03.9$\pm$0.8    &    1.84$\pm$0.49  &             
        &   1.0              &\03.9$\pm$0.1\0\0\0&    1.73$\pm$0.10  \\[0.00cm]    
    & 2 & 0.28\0$\pm$0.14\0  &   12.6$\pm$2.1    &    4.68$\pm$1.42  &             
        &   \0\0\-           &                   &                    \\[0.00cm]   
\cline{3-5}\cline{7-9}\\[-0.32cm]
 & & \multicolumn{3}{c}{$\chi^2/dof$ = 3.3/(41-5-1)}  & &  \multicolumn{3}{c}{$\chi^2/dof$ = 2.0/(23-2-1)} \\   
\br
\end{tabular}
\end{indented}
\end{table}
\begin{figure}[t]
\begin{center}
\vskip 0cm
\hspace*{0mm}
\includegraphics[width=78mm,height=78mm]{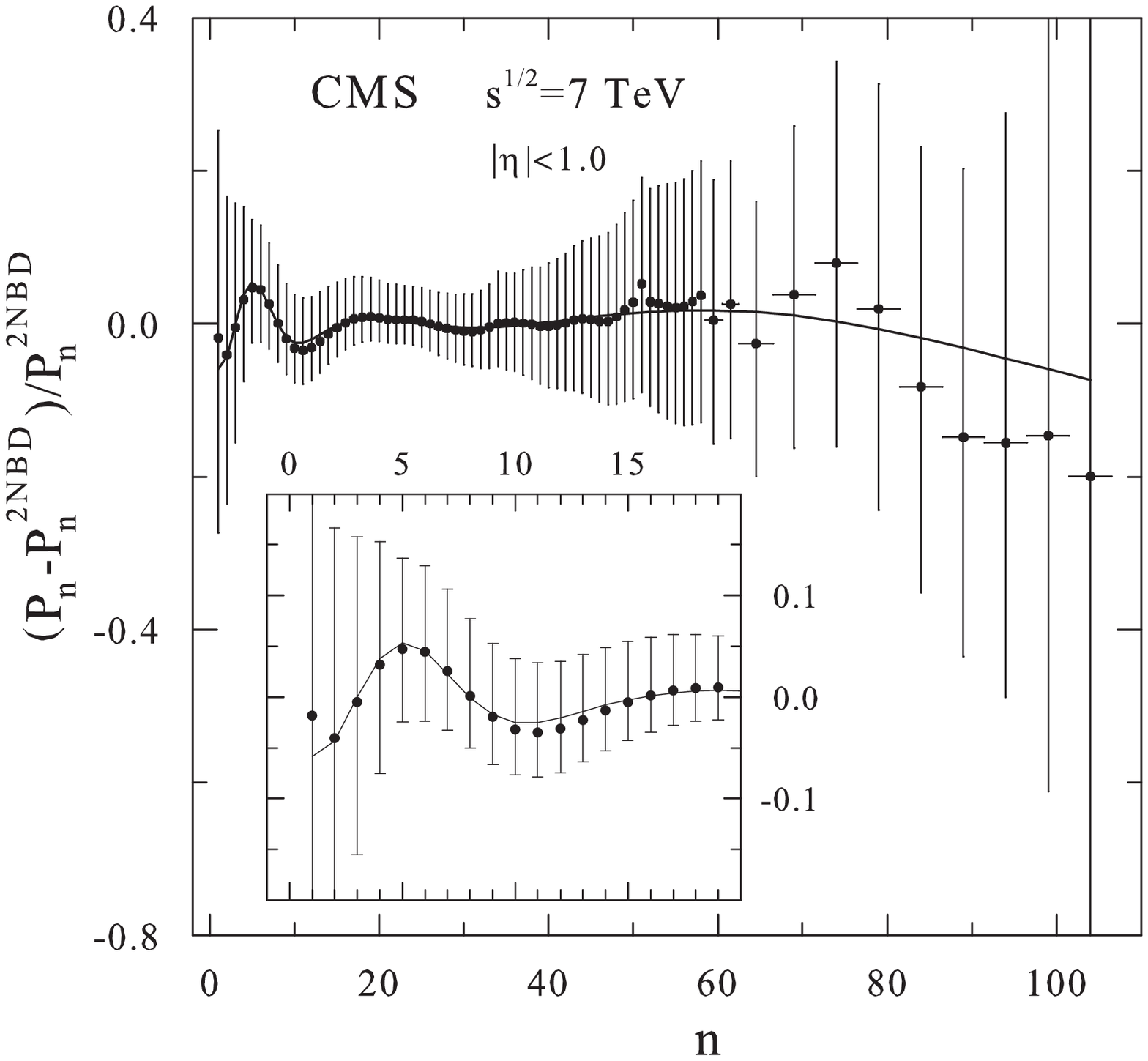}
\hspace{-0.5cm}
\includegraphics[width=78mm,height=78mm]{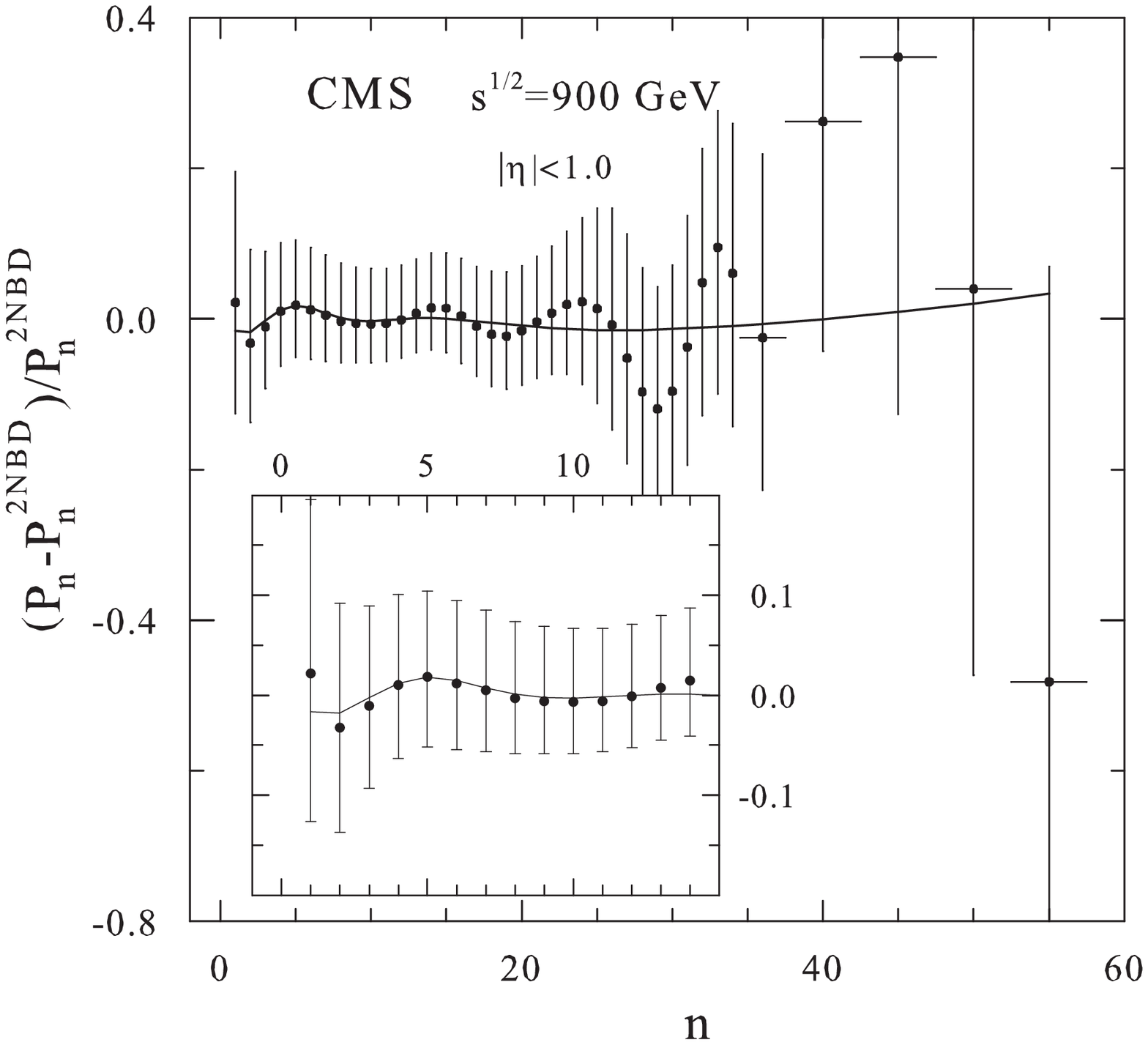}
\vskip -10mm
\vskip 0cm
\hspace*{0mm}
\includegraphics[width=78mm,height=78mm]{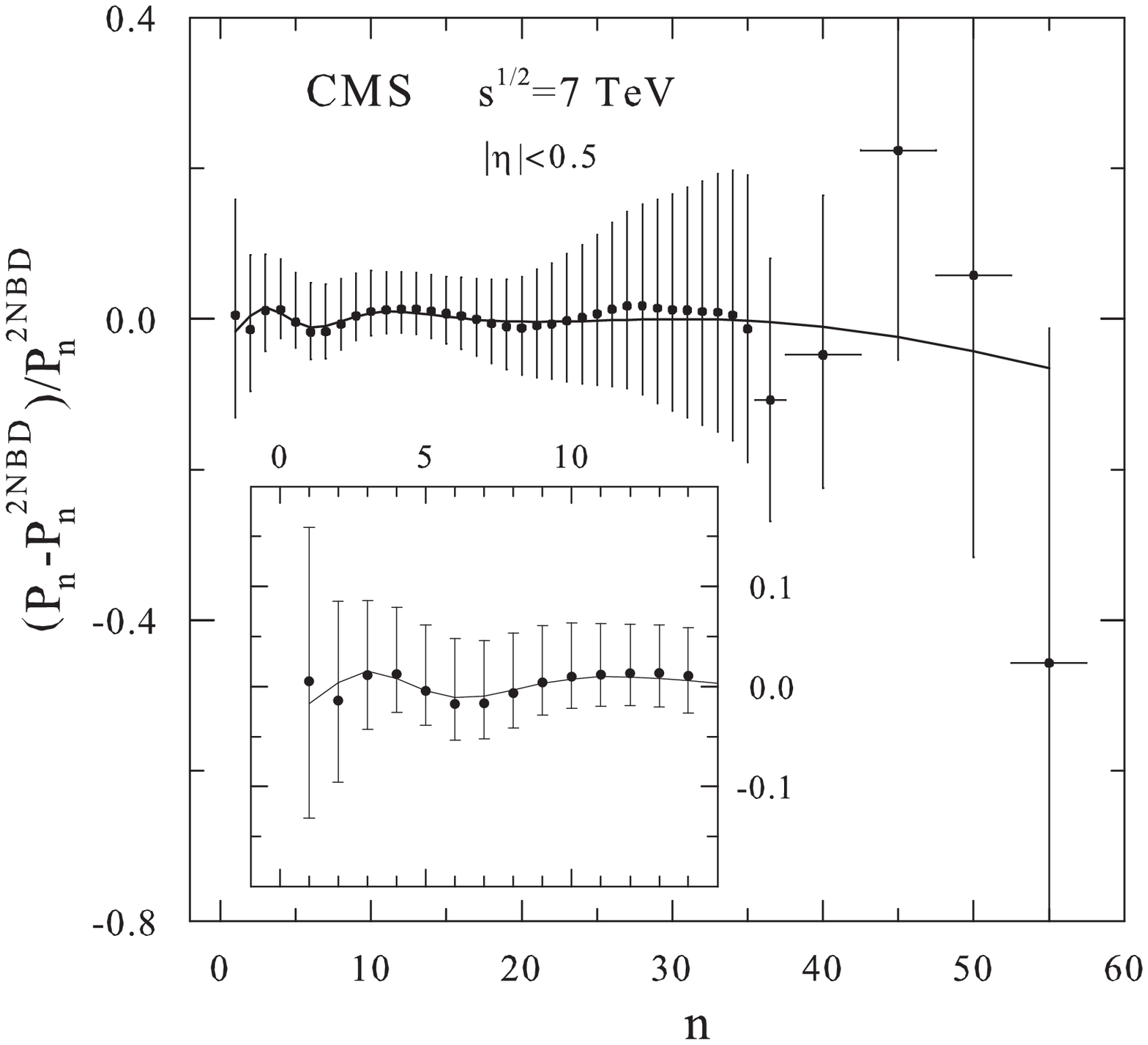}
\hspace{-0.5cm}
\includegraphics[width=78mm,height=78mm]{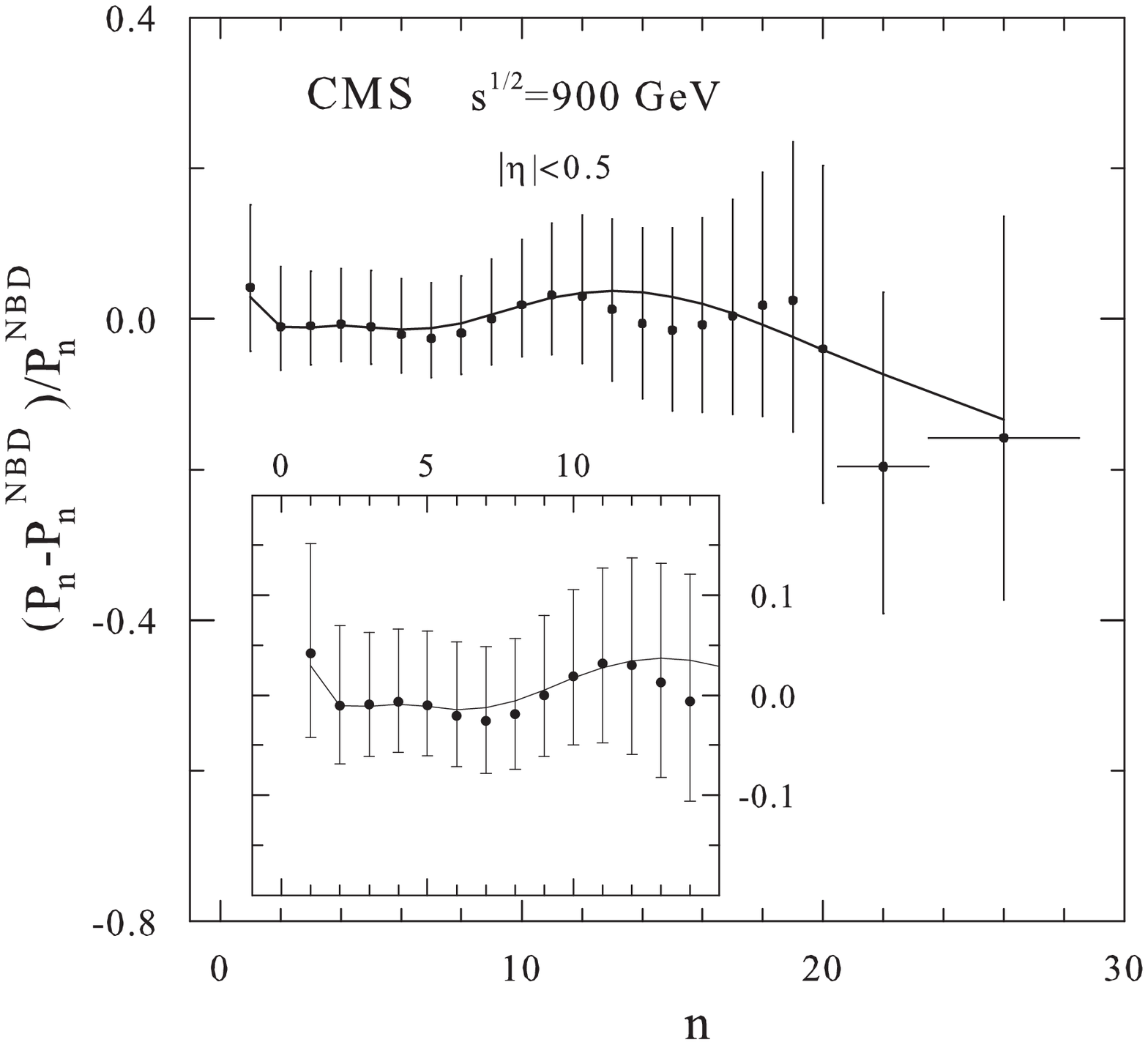}
\vskip -10mm
\hspace{1.cm}    (a) \hspace*{70mm} (b)
\caption{\label{Fig11}
Normalized residues of MD relative to the weighted superposition of two NBDs $(\mathrm{P_n^{2NBD}})$
with the parameters listed in table~\ref{tab5}.
The symbols correspond to data on MD measured by the CMS Collaboration \protect\cite{CMS}
in the small windows $|\eta|<1.0$ and $|\eta|<0.5$ at 
(a) $\sqrt s=7$~TeV and (b) $\sqrt s=0.9$~TeV.
The lines correspond to fits to the data by three-component 
superposition of NBDs. The insets show the detailed behavior of the residues 
at low $n$. 
}
\end{center}
\end{figure} 
at the energies $\sqrt s=7$ and 0.9~TeV
in the windows  $|\eta|<2.4$,  $|\eta|<2.0$ and  $|\eta|<1.5$.
The residues for small intervals $|\eta|<1.0$ and $|\eta|<0.5$
are depicted in figure \ref{Fig11}.
The insets represent the detailed structure of the residues at low multiplicity.
The solid lines correspond to the three-component description of the data 
shown in figures \ref{Fig8} and \ref{Fig9}. 
Due to large errors, the residual analysis is roughly compatible with hypothesis of description of 
the CMS data by two NBDs in all pseudorapidity intervals at both energies.  
However, in regard to a similar analysis of the ATLAS data one has to look at the residues in more detail. 
Let us note that the error bars shown in figures \ref{Fig10} and \ref{Fig11} 
form a characteristic envelope around mean values of the residues mainly
reflecting systematic uncertainties of the data.  
The envelope follows the fine structure of the residual mean values quite accurately. 
As seen from figure \ref{Fig10}, the structure indicates the systematic emergence of a peak at low multiplicities.
The peak in the residues is best visible in the large windows $|\eta|<2.4$ and $|\eta|<2.0$. 
A comparison of the corresponding insets in figures \ref{Fig4} and \ref{Fig10} shows that the peak has 
nearly the same properties as observed in the ATLAS data. 
This concerns its position as well as its size. 
The peaky structure at low $n$ diminishes with decreasing width of the pseudorapidity intervals. 
As seen from figure \ref{Fig11}, the residues in the small windows are flat. 
At $\sqrt{s}=900$~GeV, two NBD superposition (single NBD) gives a nearly perfect  
description of the data in the interval $|\eta|<1.0$ ($|\eta|<0.5$), respectively.
Similar holds at the energy $\sqrt{s}=7$~TeV for $|\eta|<0.5$ 
where two-NBD superposition approximates the data with high accuracy.

The CMS Collaboration measured the MD of charged particles in the pseudorapidity window
$|\eta|<2.4$ exploiting the transverse momentum cut $p_T>500$~MeV/c \cite{CMS}.
The complementary $p_T$ cut represents selection of harder processes which has significant impact 
on the shape of the distribution relative to unbiased data.
The measurements allow as to study the distribution in the framework of the three-component model 
and compare it with analysis of the ATLAS data obtained under the same restriction on the
transverse momentum of registered particles.
The obtained values of the fitted parameters are quoted in table~\ref{tab6}.
The  $p_T$-cut data at $\sqrt{s}=0.9$~TeV does neither  allow to extract reliable information 
on three-component parametrization of MD in the case if the third component is considered in the 
Poisson form. 
Therefore, we have fixed the probability $\alpha_2=0.11$ at the value 
found from analysis of ATLAS data at the same energy with the same $p_T$ cut  
in the interval $ |\eta|<2.5 $  (see table~\ref{tab2}).

\begin{table}[h]
\lineup
\caption{\label{tab6}
The parameters of superposition of three and two NBDs  
obtained from fits to data \protect\cite{CMS} on MD measured by the CMS Collaboration
in the window $|\eta|<2.4$ with the cut  $p_T>500$~MeV/c at 
$ \sqrt{s}=7$ and 0.9 TeV.  }
\begin{indented}
\item[]
\begin{tabular}{@{}p{0.7cm}@{}lll@{}p{0.5cm}@{}lll@{}}
\br
  & \multicolumn{3}{c}{$\sqrt{s}$ = 7 TeV} & & \multicolumn{3}{c}{$\sqrt{s}$ = 0.9 TeV} \\ 
\cline{2-4}\cline{6-8} \\[-0.30cm]
{\it i}  & $\alpha_i$ & $\bar{n}_i$ & $k_i$ &   & $\alpha_i$ & $\bar{n}_i$ & $k_i$  \\
\cline{1-4}\cline{6-8} \\[-0.22cm]
1 &  0.63$^{+0.28 }_{-0.63 }$  &  12.0$^{+3.3  }_{-2.6  }$  &   2.0$^{+15.   }_{\,\,\,-0.8}$   &       
  &  0.79$^{+0.05 }_{-0.06 }$  & \05.6$^{+0.5  }_{-0.3  }$  &  1.41$^{+0.46  }_{-0.21 }$  \\[0.15cm]   
2 &  0.12$^{+0.24 }_{-0.09 }$  &  31.1$^{+7.5  }_{-8.6  }$  &   6.1$^{\,\,\,+6.3 }_{\,\,\,-2.6}$   &   
  &  0.11-fixed                &  12.4$^{+1.5  }_{-1.0  }$  & 5.9\0$^{+1.7   }_{-1.2  }$  \\[0.15cm]   
3 &  0.25$^{+0.39 }_{-0.19 }$  & \03.5$^{+2.4  }_{-0.4  }$  &   3.7$^{\,\,\,+\infty}_{\,\,\,-1.9  }$ & 
  &  0.10$^{+0.06 }_{-0.05 }$  & \03.5$^{+0.8  }_{-0.9  }$  &      $\infty$              \\[0.08cm]    
\cline{2-4}\cline{6-8}\\[-0.32cm]
  & \multicolumn{3}{c}{$\chi^2/dof$ = 2.0/(79-8-1)}  & &  \multicolumn{3}{c}{$\chi^2/dof$ = 0.6/(37-6-1)} \\ 
\cline{1-4}\cline{6-8} \\[-0.35cm]   
1 &  0.721$\pm$0.103   &   \07.1$\pm$1.2\0 &    1.41$\pm$0.21  &    
  &   0.50$\pm$0.16    &  \03.17$\pm$0.49  &    3.24$\pm$1.34  \\   
2 &  0.279$\pm$0.103   &    24.2$\pm$2.9   &    3.88$\pm$0.82  &    
  &   0.50$\pm$0.16    &  \09.44$\pm$1.40  &    3.21$\pm$0.91  \\   
\cline{2-4}\cline{6-8} \\[-0.32cm]
  & \multicolumn{3}{c}{$\chi^2/dof$ = 9.9/(79-5-1)} & & \multicolumn{3}{c}{$\chi^2/dof$ = 0.6/(37-5-1)} \\   
\br
\end{tabular}
\end{indented}
\end{table}

\begin{figure}[t]
\begin{center}
\vskip 0cm
\hspace*{0mm}
\includegraphics[width=78mm,height=78mm]{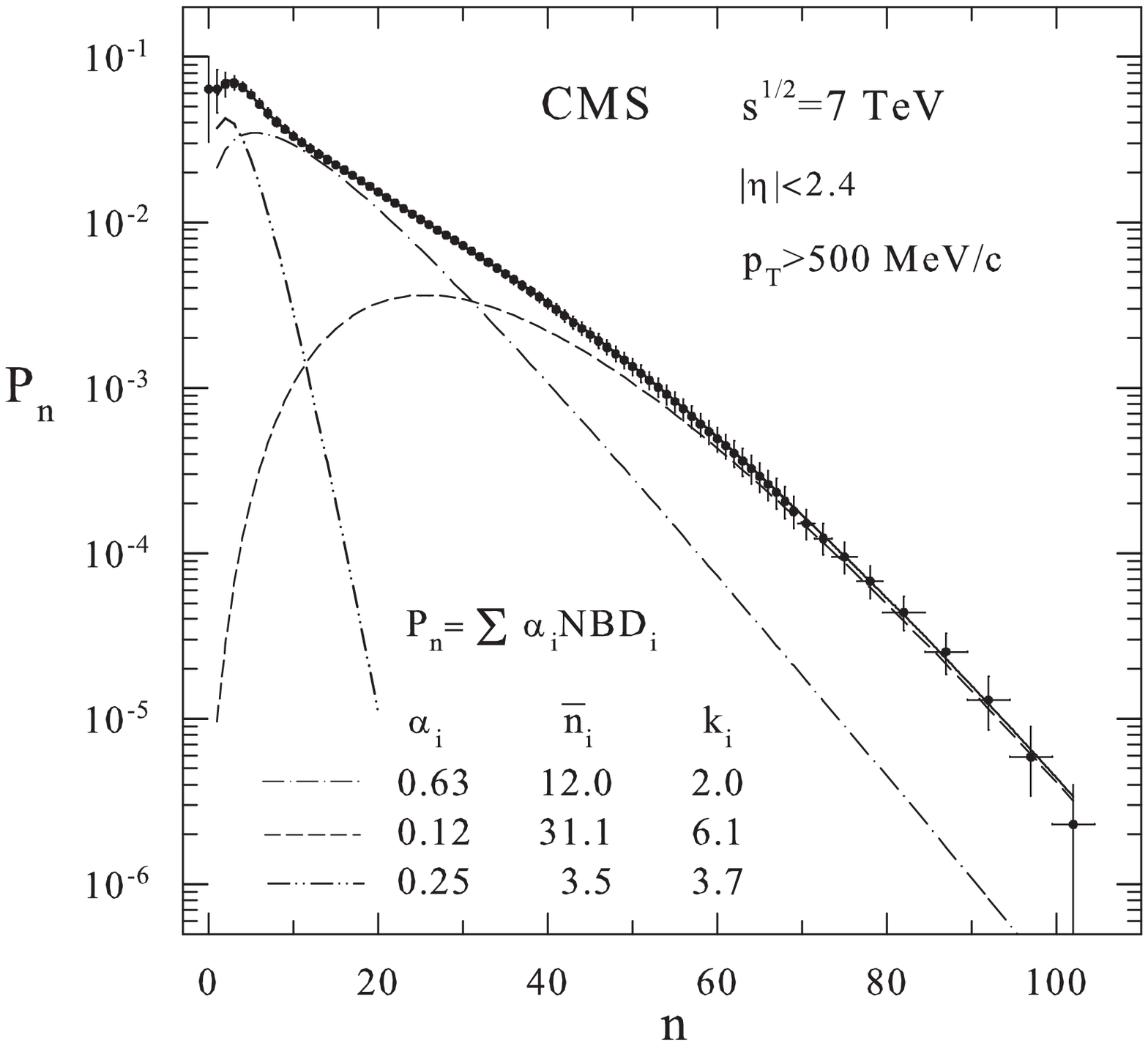}
\hspace{-0.5cm}
\includegraphics[width=78mm,height=78mm]{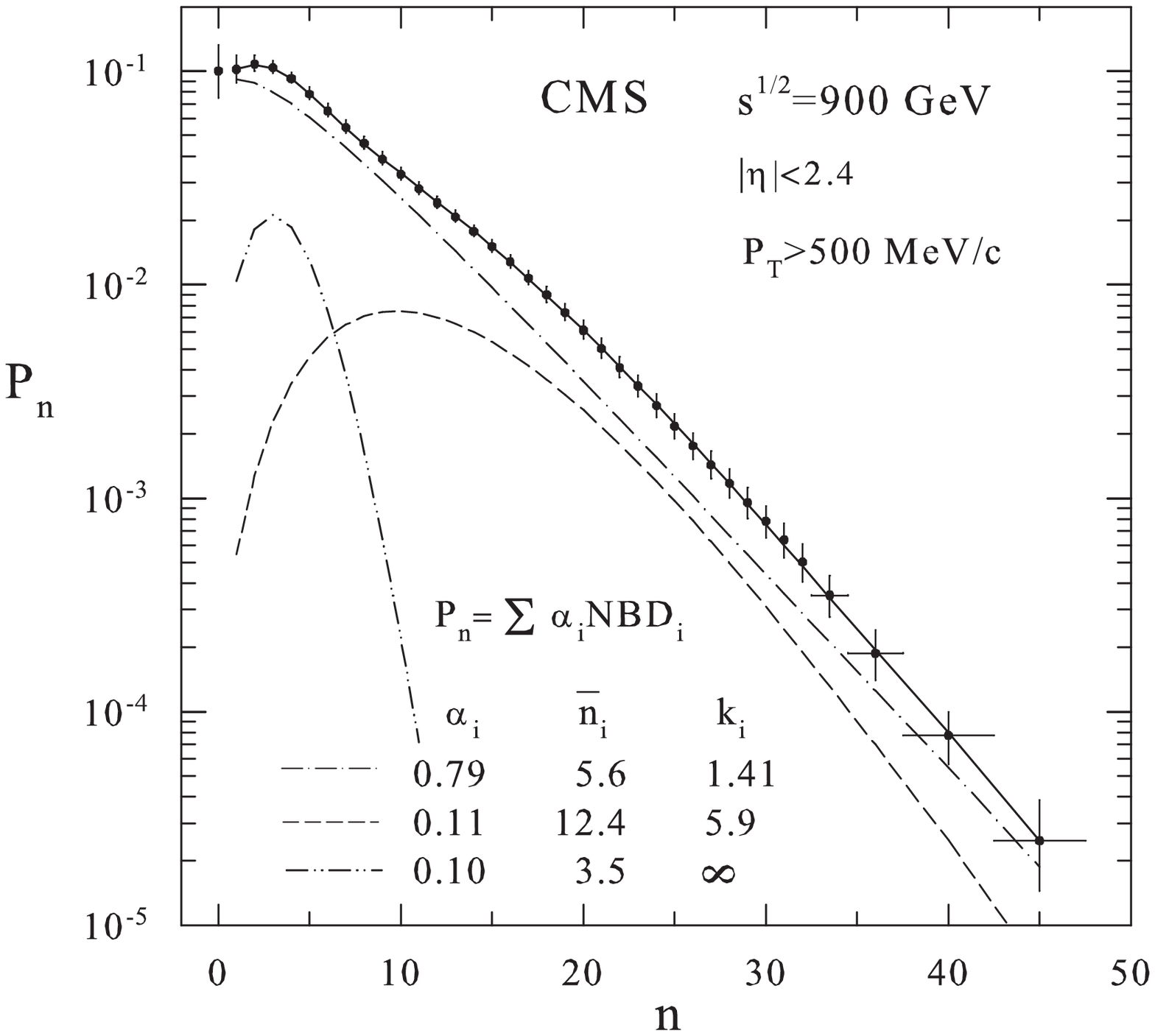}
\vskip -10mm
\hspace{1.cm}    (a) \hspace*{70mm} (b)
\caption{\label{Fig12}
MD of charged particles in the pseudorapidity window
$|\eta|<2.4$ measured  by the CMS Collaboration \protect\cite{CMS}
with the transverse momentum cut $ p_T>500$~MeV/c at
(a) $\sqrt{s}=7$~TeV and (b) $\sqrt{s}=0.9$~TeV.
The solid lines represent fits using three-component
superposition of NBDs. The dash-dot, dash and dash-dot-dot lines show
single components corresponding to the indicated parameters.  }
\end{center}
\end{figure}

Figure \ref{Fig12} shows the MD measured by the CMS Collaboration in the
pseudorapidity interval $|\eta|<2.4$  with the transverse momentum cut $p_T>500$~MeV/c
at the energies $\sqrt{s}=7$ and 0.9~TeV.
The experimental data are depicted as symbols. The fitted superposition of three NBDs
is indicated by the solid line. 
The dash-dot, dash and dash-dot-dot lines show single negative binomial components 
of the total distribution.
A comparison of the results of analysis displayed in figures \ref{Fig12} and \ref{Fig5} leads to similar
conclusions for both measurements when the cut $p_T>500$~MeV/c is applied.
The values of all parameters obtained from fits to the CMS data are compatible
with those extracted from the ATLAS data.
The performed combined analysis shows that the average multiplicity $\bar{n}_3$ of the third
component under the peak at low $n$ is approximately energy independent.
This observation does not depend on the $p_T$ cut imposed on the data.
On the other hand, the condition $p_T>500$~MeV/c influences the energy behavior 
of the parameter $\alpha_3$ significantly.
As seen from figure \ref{Fig12} and table~\ref{tab6}, the probability of the third component $\alpha_3$  
indicates an increasing tendency with energy reflecting 
growth of the peak at low multiplicities at $\sqrt{s}=7$~TeV.
The observation is in accord with analysis of the ATLAS data with 
$p_T>500$~MeV/c (see figure~\ref{Fig5} and table~\ref{tab2}).
Such a trend is nearly negligible in the minimum biased CMS data (figure \ref{Fig8})
and is totally absent in the ATLAS data with the low transverse momentum cut $p_T>100$~MeV/c 
(figure \ref{Fig3}).
The obtained value of $k_3$ from the $p_T$-cut CMS data at $ \sqrt{s}=7$~TeV   
does not contradict the decreasing tendency of this parameter which is
visible in the ATLAS data with the same restriction on transverse momentum (figure \ref{Fig5} and  table \ref{tab2}).

\begin{figure}
\begin{center}
\vskip 0cm
\hspace*{0mm}
\includegraphics[width=78mm,height=78mm]{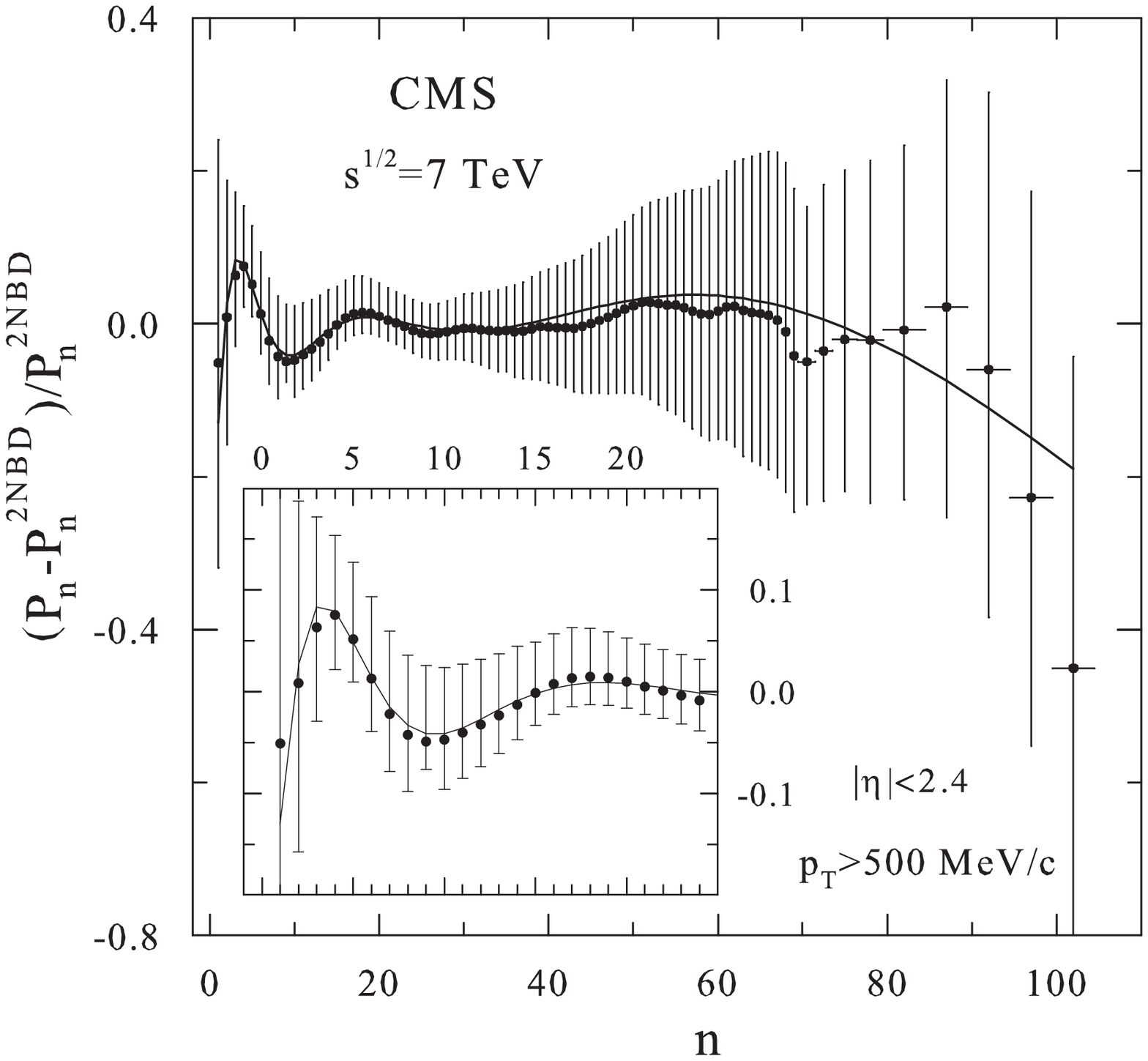}
\hspace{-0.5cm}
\includegraphics[width=78mm,height=78mm]{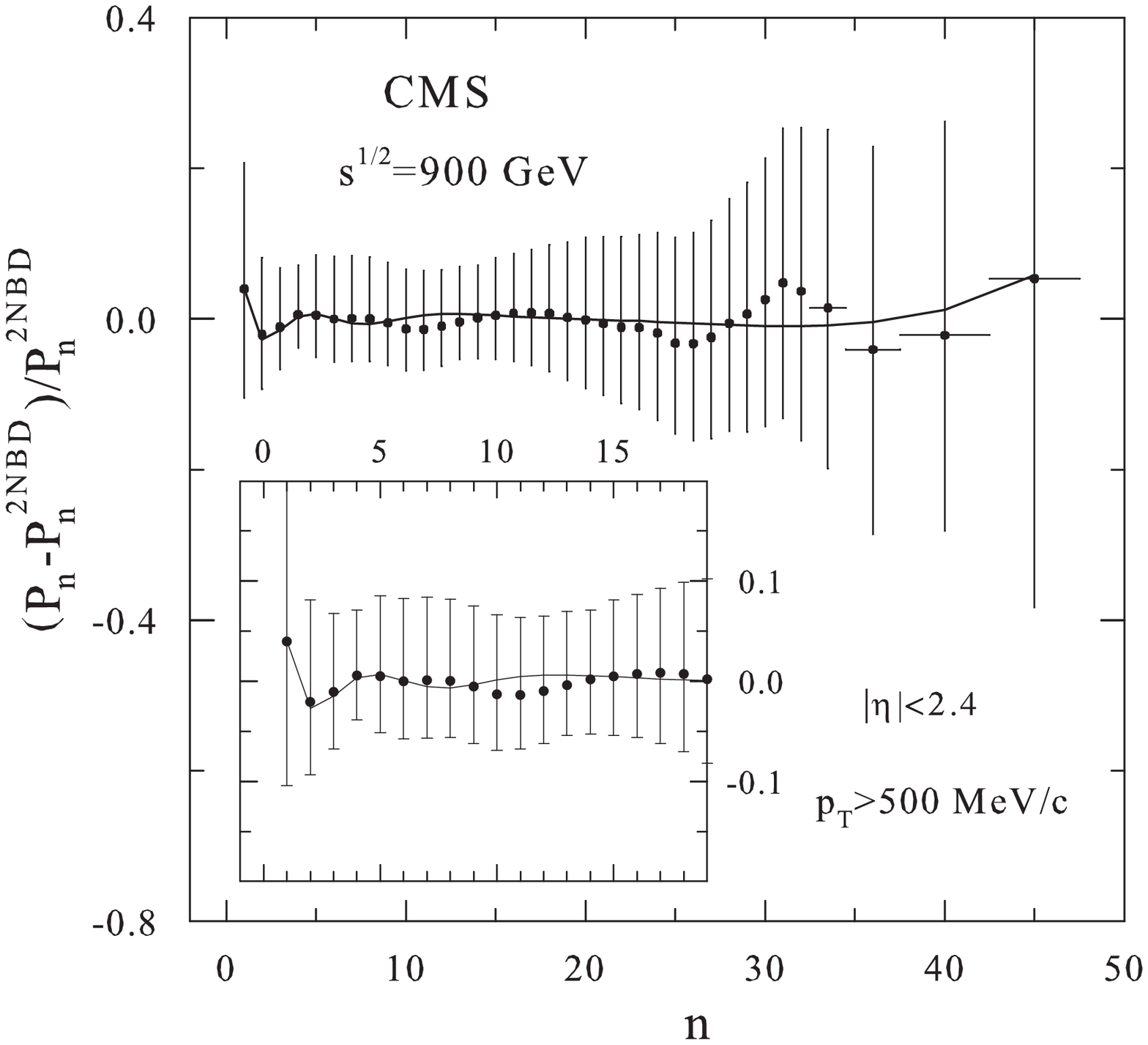}
\vskip -10mm
\hspace{1.cm}    (a) \hspace*{70mm} (b)
\caption{\label{Fig13}
Normalized residues of MD relative to the weighted superposition of two NBDs $(\mathrm{P_n^{2NBD}})$
with the parameters listed in table~\ref{tab6}.
The symbols correspond to data on MD measured by the CMS Collaboration \protect\cite{CMS}
in the pseudorapidity interval
$|\eta|<2.4$ with the cut $ p_T>500$~MeV/c at
(a) $\sqrt{s}=7$~TeV and (b) $\sqrt{s}=0.9$~TeV.
The lines correspond to fits to the data by three-component 
superposition of NBDs. The insets show the detailed behavior of the residues 
at low $n$. 
 }
\end{center}
\end{figure}

Figure \ref{Fig13} shows the normalized residues of MD with respect to the two-NBD parametrization
(see table~\ref{tab6}) of the CMS data with the cut $p_T>500$~MeV/c.   
Though the error bars dominated by systematic uncertainties are large, 
the residual mean values indicate existence
of a peaky structure near $n\sim 3-4$ at the energy $\sqrt{s}=7$~TeV.
The structure is similar to the peak clearly visible in the analysis of the ATLAS data 
shown in figure \ref{Fig6}(a).  
The full line corresponds to description of the CMS data by weighted superposition 
of three NBDs with the third component
indicated by the dash-dot-dot line in figure~\ref{Fig12}(a). 
The residues at $\sqrt{s}=900$~GeV shown in figure~\ref{Fig13}(b) are flat. 
This is in compliance with residual analysis  
of the ATLAS data for $p_T>500$~MeV/c at the same energy (figure~\ref{Fig6}(b)). 
Both $p_T$-cut data are here well-parametrized by superposition of two NBDs.

\subsection{The third component in the ALICE and LHCb data}

Some structure in the distributions of charged particles in the low multiplicity region is   
seen also in the measurements of two other experiments at the LHC.
The ALICE and LHCb Collaborations presented data which allow to search for complementary  
information on the third component in description of MD in $pp$ collisions
at high energies.

\begin{figure}[t]
\begin{center}
\vskip 0cm
\hspace*{0mm}
\includegraphics[width=78mm,height=78mm]{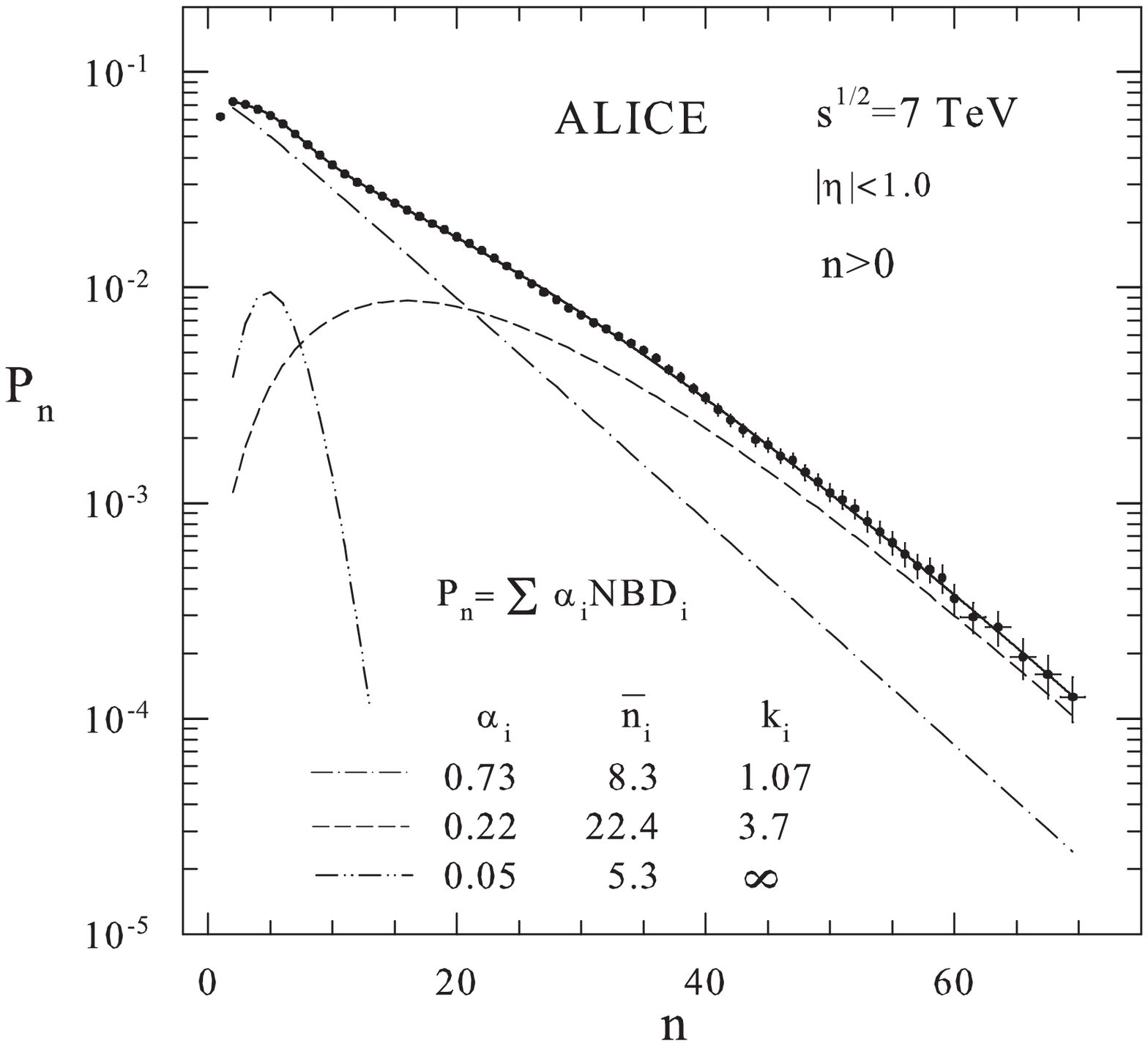}
\hspace{-0.5cm}
\includegraphics[width=78mm,height=78mm]{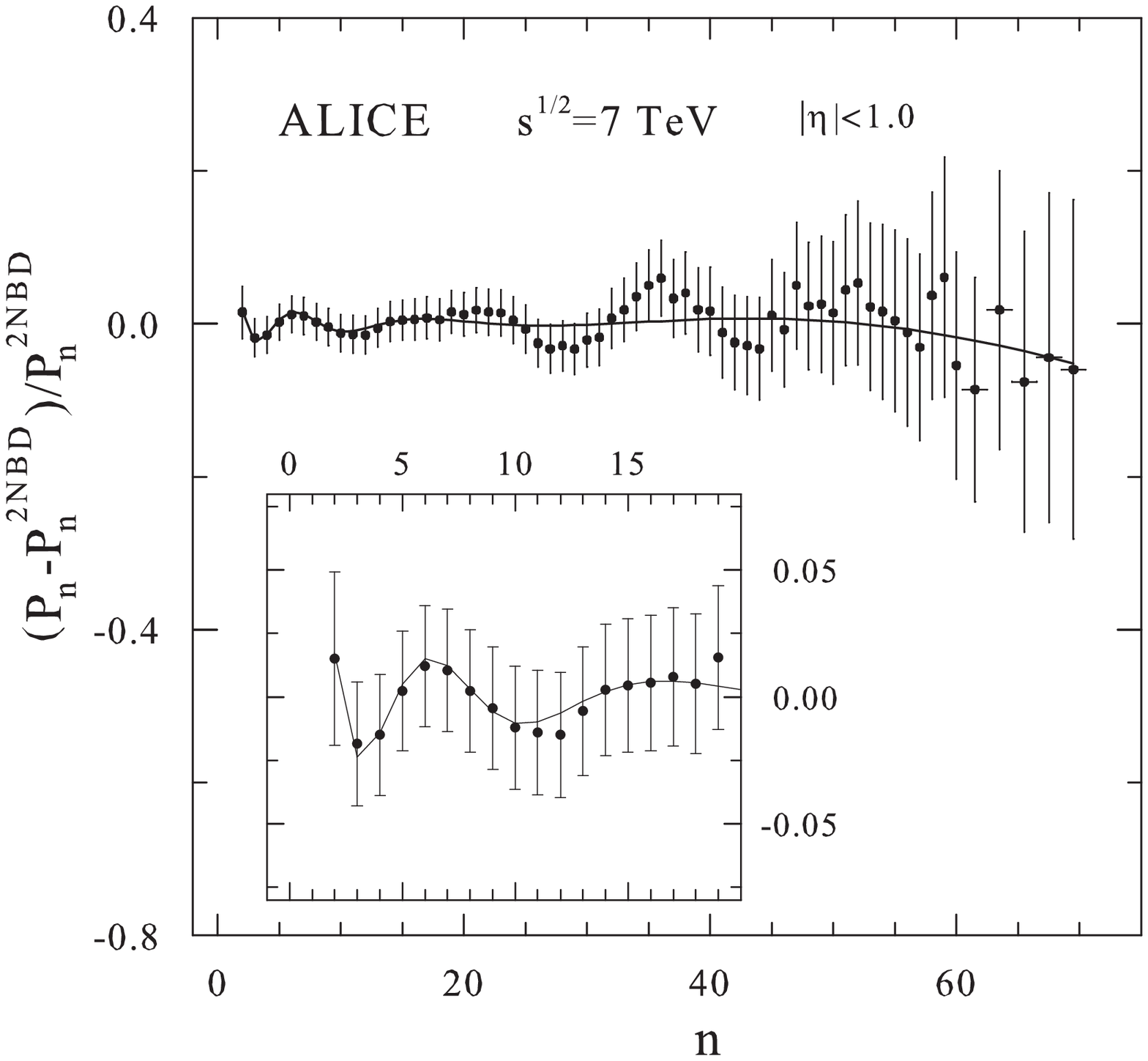}
\vskip -10mm
\hspace{1.cm}    (a) \hspace*{70mm} (b)
\caption{\label{Fig14}
(a) MD of charged particles 
measured by the ALICE Collaboration \protect\cite{ALICE2}
in the central window $|\eta|<1.0$ 
at $\sqrt{s}=7$~TeV.
(b) Normalized residues of MD relative to the weighted superposition of two NBDs $(\mathrm{P_n^{2NBD}})$
with the parameters listed in table~\ref{tab7}.
The inset shows the behavior of the residues at low $n$.
The solid lines represent fitted three-component superposition of NBDs. 
The dash-dot, dash and dash-dot-dot lines show
single components corresponding to the indicated parameters. 
 }
\end{center}
\end{figure}

The ALICE Collaboration measured \cite{ALICE2} multiplicity of charged particles produced
in the central pseudorapidity region $|\eta|<1$ at the energy $\sqrt{s}=7$ TeV.
The data sample was collected from 240~000 events
without normalization to the NSD processes.
The event class was defined by requiring at least one charged particle in the
measured pseudorapidity interval. The measurements cover the multiplicity region $ 1\leq n \leq 70$.
The ALICE data at $\sqrt{s}=7$ TeV are shown as points in figure \ref{Fig14}(a). 
The solid line represents a fitted three-component function (\ref{eq:r2}) 
with the parameters quoted in table~\ref{tab7}. 
The point $P_1$ was not included in the fit.
The parameter $k_3$ was found to be consistent with infinity, having  
a mean value much larger than 100. 
Therefore, for better stability of the fit, we set $k_3=\infty$. 
The single components of the total distribution are depicted by the 
dash-dot, dash and dash-dot-dot lines, respectively.
The values of the fitted parameters indicate a similar trend 
as follows from the analysis of the CMS data
in the pseudorapidity window $|\eta|<1$ at the energy $\sqrt{s}=7$ TeV (figure \ref{Fig9}(a)).
The average multiplicity $\bar{n}_1\sim 5$ of the third component
is approximately the same for both data sets.
Analysis of the ALICE data gives somewhat smaller values of $\bar{n}_1$ and $\bar{n}_2$ than 
obtained from the CMS measurements.
The parameters $k_i$ increase with decreasing probabilities $\alpha_i$.
Unlike the distributions measured by the CMS and ATLAS Collaborations, small wavy fluctuations
are seen in the ALICE data for $n>25$.
The fluctuations are attributed to the correlation of errors over a range comparable
to multiplicity resolution \cite{ALICE2}.
The wavy behavior does not allow to extract the parameters of the third component of the
function (\ref{eq:r2}) from ALICE data at lower energies.

We have fitted the ALICE data at $\sqrt{s}=7$ TeV with weighted superposition of two NBDs. 
The obtained parameters are listed in table~\ref{tab7}.
Figure~\ref{Fig14}(b) shows relative residues of MD with respect to the two-NBD fit to the data.
The fitted three-component superposition of NBDs is represented by the solid line.
The inset depicts the detailed behavior of the residues at low $n$.
As follows from figure~\ref{Fig14}(b), the ALICE measurement in the window $ |\eta|<1.0$ is 
approximately compatible with the hypothesis of description of MD by two NBDs. 
However, in regard to the residual analysis of the ATLAS data   
complemented by the study of $\eta_c$-dependence of the residual structures with the CMS data, 
one needs to look at 
figure~\ref{Fig14}(b) in more detail. 
There are some oscillations of the residues for $n>25$ which reflect the fluctuations in data 
mentioned above. Besides this, there is an indication of a little peak around $n\sim 6$.
We consider that the peak is of the same origin as the residual structure at low multiplicities emerging  
in the CMS data in the window $|\eta|<1$ at $\sqrt{s}=7$ TeV (see figure~\ref{Fig11}(a)).
An indirect support for this claim can serve values of the parameters $\alpha_3$ and $\bar{n}_3$
which are within errors mutually comparable  for both data sets
(see corresponding numbers in tables~\ref{tab4} and \ref{tab7}). 

\begin{table}[h]
\lineup
\caption{\label{tab7}
The parameters of superposition of three and two NBDs  
obtained from fits to the ALICE data \protect\cite{ALICE2}
in the central region $|\eta|<1.0$ and the minimum biased LHCb data \protect\cite{LHCb} 
in the forward region $2.0<\eta<4.5$, all at $\sqrt{s}=7$~TeV.  }
\begin{indented}
\item[]
\begin{tabular}{@{}p{0.7cm}@{}lll@{}p{0.5cm}@{}ll@{}l@{}}
\br
 & \multicolumn{3}{c}{ALICE,  $|\eta|<1.0$} & & \multicolumn{3}{c}{LHCb,  $2.0<\eta<4.5$} \\ 
\cline{2-4}\cline{6-8} \\[-0.30cm]
{\it i}  & $\alpha_i$ & $\bar{n}_i$ & $k_i$ &   & $\alpha_i$ & $\bar{n}_i$ & $k_i$  \\
\cline{1-4}\cline{6-8} \\[-0.22cm]
1 & 0.73$^{+0.15 }_{-0.32 }$  & \08.3$^{+1.9  }_{-3.7  }$  &    1.07$^{+0.42 }_{-0.14 }$   &            
  & 0.88$^{+0.06 }_{-0.19 }$  &  12.8$^{+1.1  }_{-3.0  }$  &  \01.43$^{+0.21 }_{-0.08 }$  \\[0.15cm]    
2 & 0.22$^{+0.31 }_{-0.13 }$  &  22.4$^{+2.5  }_{-4.9  }$  &   3.7\0$^{+1.7  }_{-1.2  }$   &            
  & 0.07$^{+0.18 }_{-0.05 }$  &  31.5$^{+1.9  }_{-5.0  }$  &     10.$^{\0+\,26.}_{\0\0-5.}$  \\[0.15cm] 
3 & 0.05$^{+0.01 }_{-0.02 }$  & \05.3$^{+0.4  }_{-0.4  }$  &       $\infty$                &            
  & 0.05$^{+0.01 }_{-0.01 }$  & \06.2$^{+0.4  }_{-0.2  }$  &       $\infty$               \\[0.08cm]    
\cline{2-4}\cline{6-8}\\[-0.32cm]
  & \multicolumn{3}{c}{$\chi^2/dof$ = 11.6/(65-7-1)}  & &  \multicolumn{3}{c}{$\chi^2/dof$ = 2.7/(39-7)} \\ 
\cline{1-4}\cline{6-8} \\[-0.35cm]   
1 & 0.433$\pm$0.058   &  \04.68$\pm$0.32   &    2.13$\pm$0.31  &    
  & 0.627$\pm$0.087   &  \07.67$\pm$0.85\0 &  \02.18$\pm$0.16  \\   
2 & 0.567$\pm$0.058   &  17.2\0$\pm$0.9\0  &    2.42$\pm$0.25  &    
  & 0.373$\pm$0.087   &  24.1\0$\pm$2.3\0  & \04.5\0$\pm$1.2\0 \\   
\cline{2-4}\cline{6-8} \\[-0.32cm]
  & \multicolumn{3}{c}{$\chi^2/dof$ = 15.7/(65-5-1)} & & \multicolumn{3}{c}{$\chi^2/dof$ = 16.3/(39-5)} \\     
\br
\end{tabular}
\end{indented}
\end{table}

\begin{figure}[t]
\begin{center}
\vskip 0cm
\hspace*{0mm}
\includegraphics[width=78mm,height=78mm]{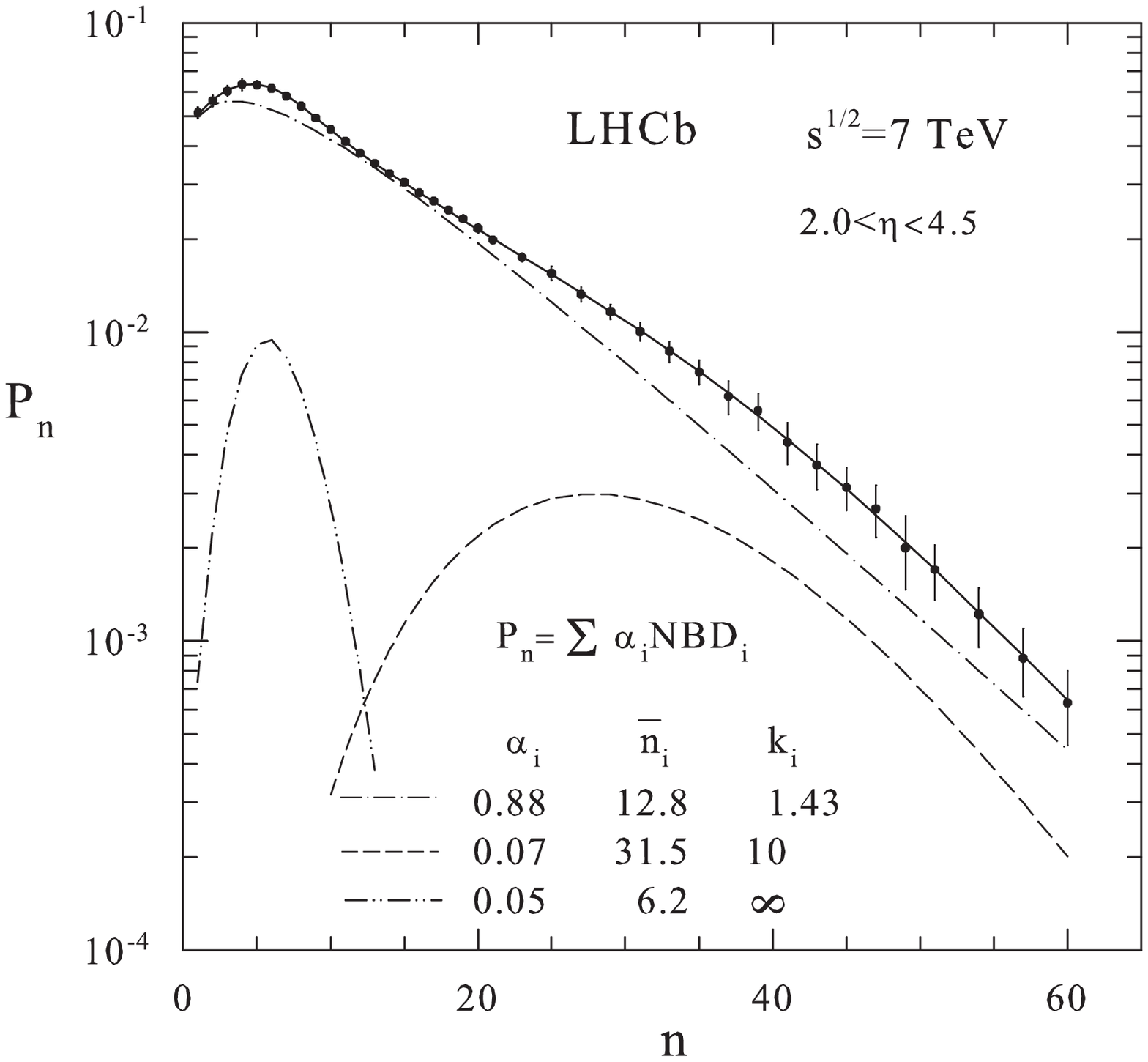}
\hspace{-0.5cm}
\includegraphics[width=78mm,height=78mm]{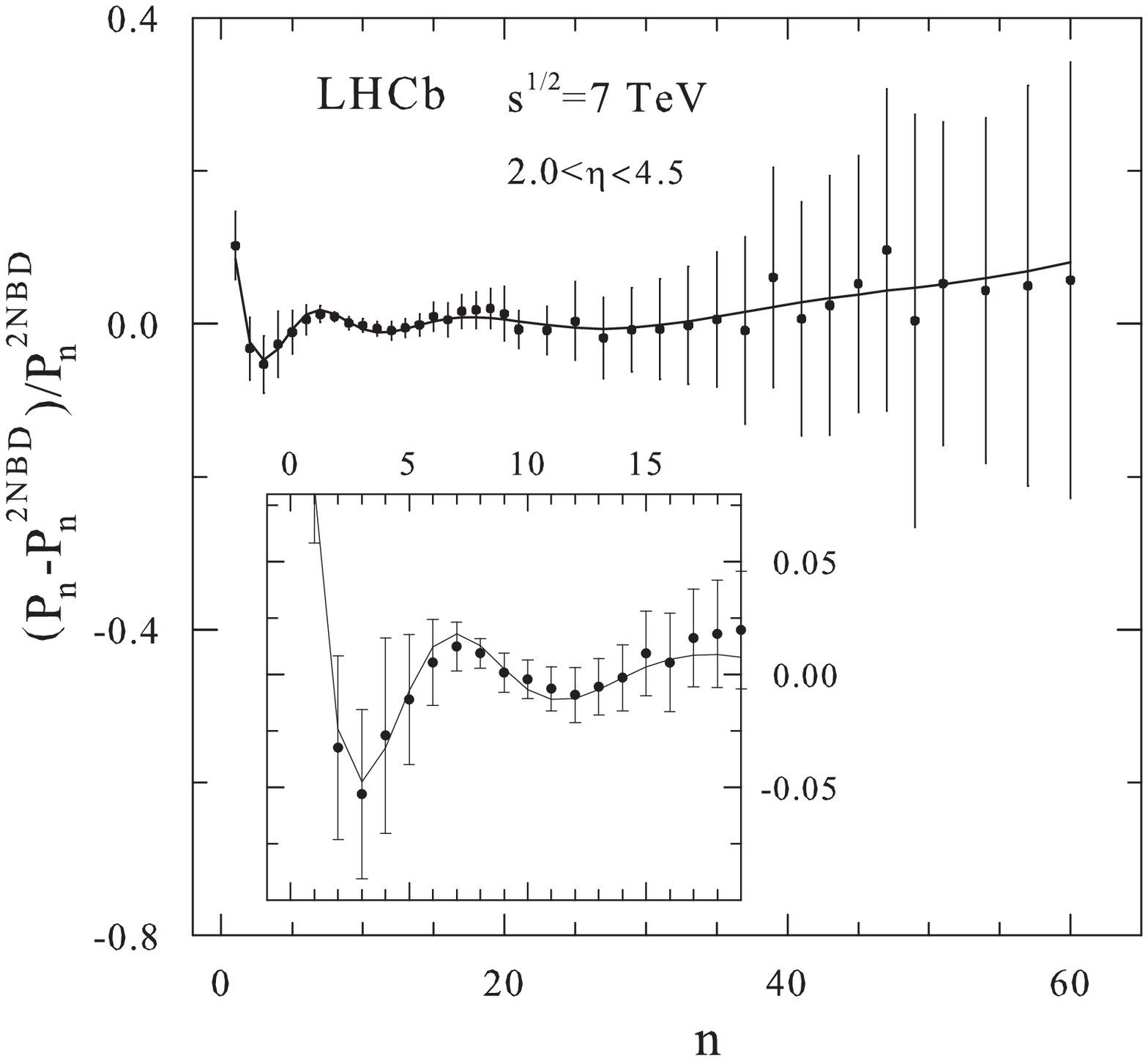}
\vskip -10mm
\hspace{1.cm}    (a) \hspace*{70mm} (b)
\caption{\label{Fig15}
(a) MD of charged particles 
measured by the LHCb Collaboration \protect\cite{LHCb}
in the forward region $2.0<\eta<4.5$ at $\sqrt{s}=7$~TeV. 
(b) Normalized residues of MD relative to the weighted superposition of two NBDs $(\mathrm{P_n^{2NBD}})$
with the parameters listed in table~\ref{tab7}.
The inset shows the behavior of the residues at low $n$.
The solid lines represent fitted three-component superposition of NBDs. 
The dash-dot, dash and dash-dot-dot lines show
single components corresponding to the indicated parameters. 
 }
\end{center}
\vskip -5mm
\end{figure}

The LHCb Collaboration measured the MD of charged particles  at the
energy $\sqrt{s}=7$ TeV in the forward region \cite{LHCb}.
The experimental analysis was based on a high statistic sample of 3M events. In this paper we
study the data taken with the minimum bias trigger defined by requiring at least one
reconstructed track in the vertex detector.
The LHCb data obtained in the forward pseudorapidity region $ 2.0<\eta<4.5$ are depicted as points
in figure \ref{Fig15}(a).
The solid line represents the best fit to the data by superposition of three NBDs. 
The single components are shown as dash-dot, dash and dash-dot-dot lines, respectively.
The parameters of the distribution are stated in table~\ref{tab7}.
The analysis gives a large value of $k_3>100$.
For the sake of better stability of the fit, the parameter was fixed at infinity, $k_3=\infty$.
All parameters $k_i$ increase with the decreasing probabilities $\alpha_i$ as
in the central interaction region.
The average multiplicities $\bar{n}_i$ of single components 
match with the values obtained from the analysis of the CMS data
in the regions with nearly the same widths of the pseudorapidity windows.
The span of the pseudorapidity interval of the LHCb measurements, $ \Delta\eta=2.5$,
corresponds to the size of the central windows between $|\eta|<1.0$ and $|\eta|<1.5$. 
All three parameters $\bar{n}_i$ shown in
figure \ref{Fig15}(a) are found between the corresponding average 
multiplicities obtained from analysis of the CMS data measured 
in these windows (see tables~\ref{tab4} and  \ref{tab7}). 
The observation complies with the extensive
character of the average multiplicity for each component separately.

The shapes of the MD in the forward and central region are similar. 
One can see from figure \ref{Fig15}(a) that the minimum biased data 
measured by the LHCb Collaboration in the forward region
manifest existence of a narrow maximum and a broad shoulder at high multiplicities.
It is worth mentioning that a slight decrease of the probability $\alpha_2$  
accompanies an increasing tendency of the parameter $k_2$.
This is connected with a receding role of the second component in the tail 
of the total distribution which becomes narrower and suppressed with respect to the dominant one 
(compare with figure \ref{Fig14}(a)).  
On the other hand, the probability $\alpha_3$ of the third component under the peak at low $n$ is nearly 
the same as in the central region (see figures \ref{Fig9}(a) and \ref{Fig14}(a)). 

We have fitted the minimum biased LHCb data with weighted
superposition of two NBDs. The obtained values of the corresponding parameters are
listed in table~\ref{tab7}. The normalized residues of MD relative to the two-NBD fit
of the data are shown in figure~\ref{Fig15}(b).
The inset represents the detailed structure of the residues at low multiplicity. 
The solid lines correspond to the three-component description of the LHCb data with the parameters listed in table~\ref{tab7}. 
One can see a certain structure at low multiplicities with a local maximum around $n\sim 7$. 
The maximum is described by the third component of the total MD shown by the dash-dot-dot 
line in figure~\ref{Fig15}(a).

Although the residual structures of MD in the ALICE and LHCb measurements are not very significant, the data
show signs of a small peak at low multiplicities which supports analogous observation in the CMS data 
in the interval $|\eta|<1.0$ at $\sqrt{s}=7$ TeV. 
The evolution of the peak with the increasing width of the pseudorapidity window  in the CMS data
and its comparability in the largest windows with the peaky structure clearly visible 
in the ATLAS data represents a sequence of observations indicating emergence of a new signal in multiparticle 
production in hadron collisions at TeV energies.

\section{Invariant Properties of MDs}

The studies of the MD of charged particles in $pp/p\bar{p}$ collisions showed
a complex structure which was attributed by many authors to
soft and semihard components in the data.
This led to the natural conclusion that the soft component prevails  
at lower collision energies while
the semihard component should play an increasing role at higher energies.
The important experimental observation concerns the energy independence of MD of the
soft events \cite{GiovUgo1,CDF}. 
The energy invariance of the parameters of the soft component was confirmed by the recent analysis of the CMS data
in the framework of the two-NBD model \cite{Ghosh}.
The results of this study show the stability of the description of the distinct peak visible
in the region where the soft production processes dominate. 
The peaky structure can be well-described by a separate component
within  the three-component parametrization of data on multiplicities measured at the LHC. 
The component under the peak at low $n$ reveals remarkable properties.
Its average multiplicity $ \bar{n}_3$ is nearly the same at both energies $\sqrt{s}=7$ and 0.9~TeV.   
For minimum biased data, 
the third component is narrow, well-approximated by the Poisson distribution (or by NBD with large $k_3$). 
Description of the experimental data by superposition of three NBDs manifests some invariant properties
in the studied kinematic region.
These are characterized by approximate independence of the corresponding parameters 
on the collision energy $ \sqrt{s}$ and the size of the pseudorapidity window $|\eta|<\eta_c$.   

\begin{figure}[b]
\begin{center}
\vskip 0cm
\hspace*{0mm}
\includegraphics[width=78mm,height=78mm]{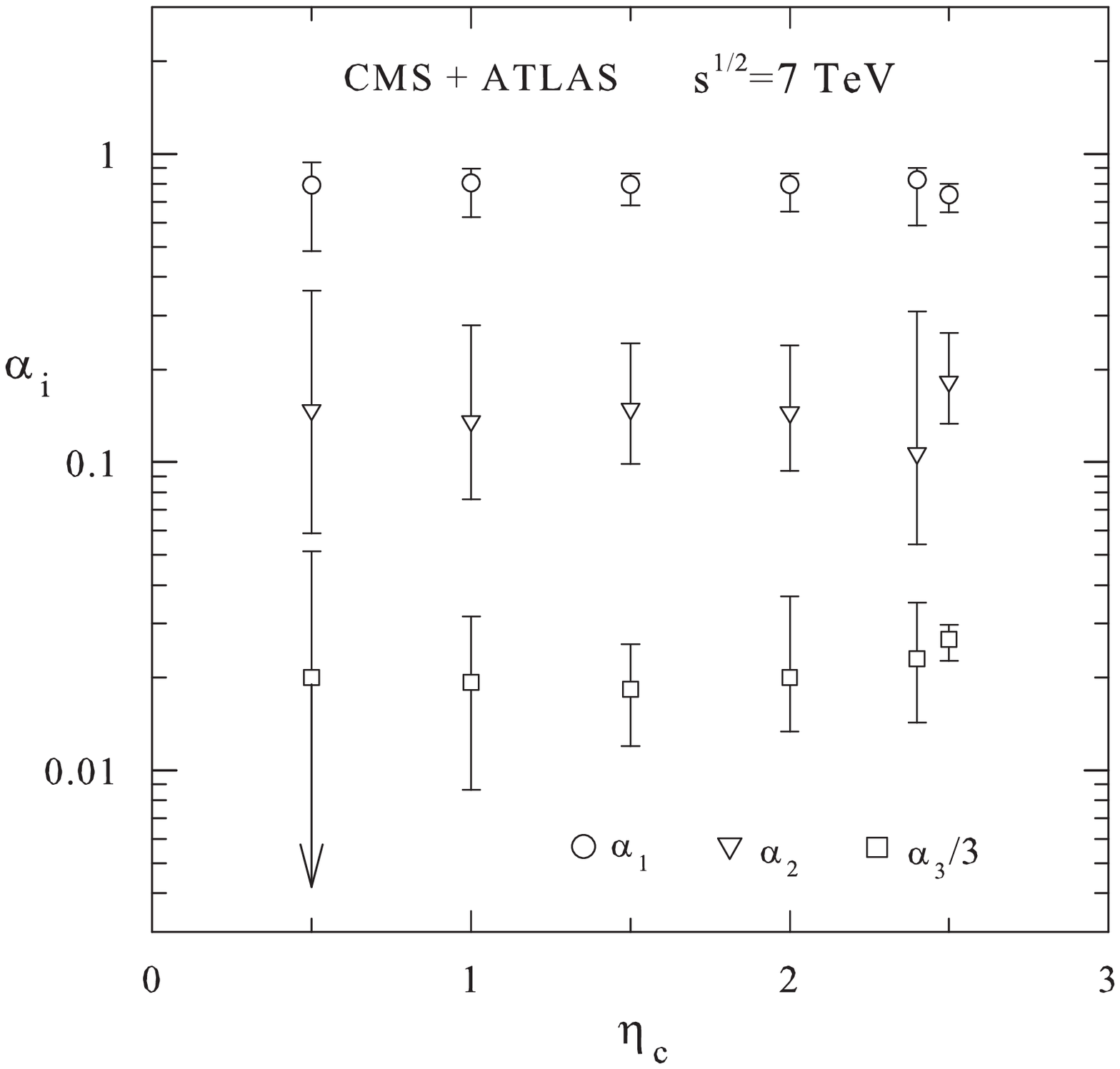}
\hspace{-0.5cm}
\includegraphics[width=78mm,height=78mm]{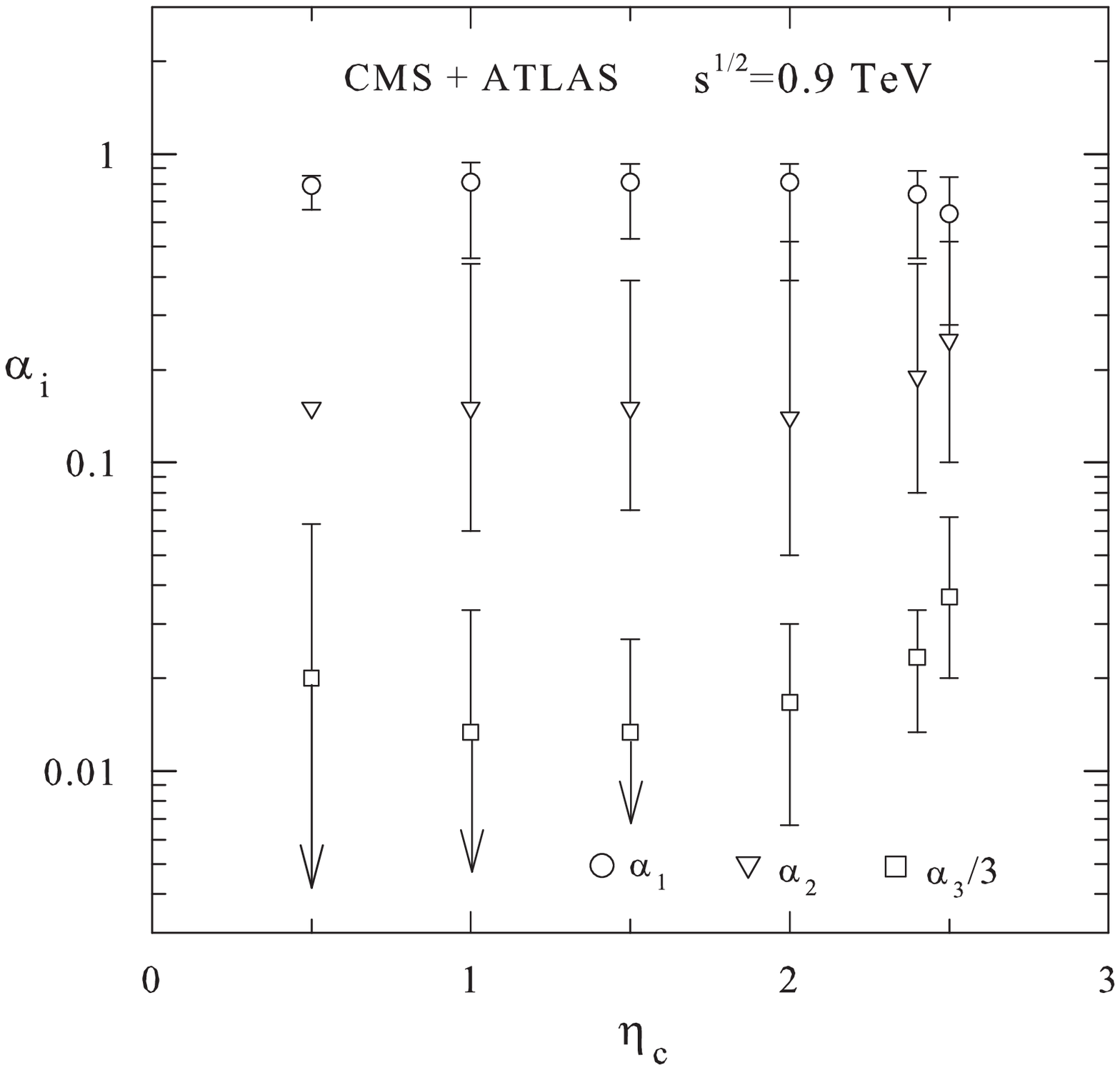}
\vskip -10mm
\hspace{1.cm}    (a) \hspace*{70mm} (b)
\caption{\label{Fig16}
The $\eta_c$ dependence of the probabilities $\alpha_i$ of single components of 
weighted superposition of three NBDs fitted to the 
NSD CMS data \protect\cite{CMS} and the most inclusive sample of the ATLAS data \protect\cite{ATLAS}
at (a) $\sqrt{s}=7$~TeV and (b) $\sqrt{s}=0.9$~TeV. 
The values of $\alpha_3$ are multiplied by a factor of one third. 
}
\end{center}
\end{figure} 
Figure \ref{Fig16} shows the probabilities $\alpha_i$ of single negative binomial components at the energies  
$\sqrt{s}=7$ and 0.9~TeV as function of $\eta_c$. 
The symbols correspond to the results of analysis of the minimum biased CMS data 
in five windows from $\eta_c=0.5$ to $\eta_c=2.4$ 
and the most inclusive sample of ATLAS data in the interval $|\eta|<2.5$.
Despite considerable errors,
there are  clear trends visible in the behavior of the probabilities. 
All $\alpha_i$ reveal signs of independence on the window size at $\sqrt{s}=7$~TeV
in the pseudorapidity ranges from $\eta_c=0.5$ up to $\eta_c=2.5$.
The same property seems to be valid at $\sqrt{s}=0.9$~TeV regardless of problems with determination of 
the parameters in the narrowest window at this energy.   
Moreover, from comparison of figures \ref{Fig16}(a) and (b) one can deduce 
that the probabilities $\alpha_i$ are approximately energy independent.

\begin{figure}[b]
\begin{center}
\vskip 0cm
\hspace*{0mm}
\includegraphics[width=78mm,height=78mm]{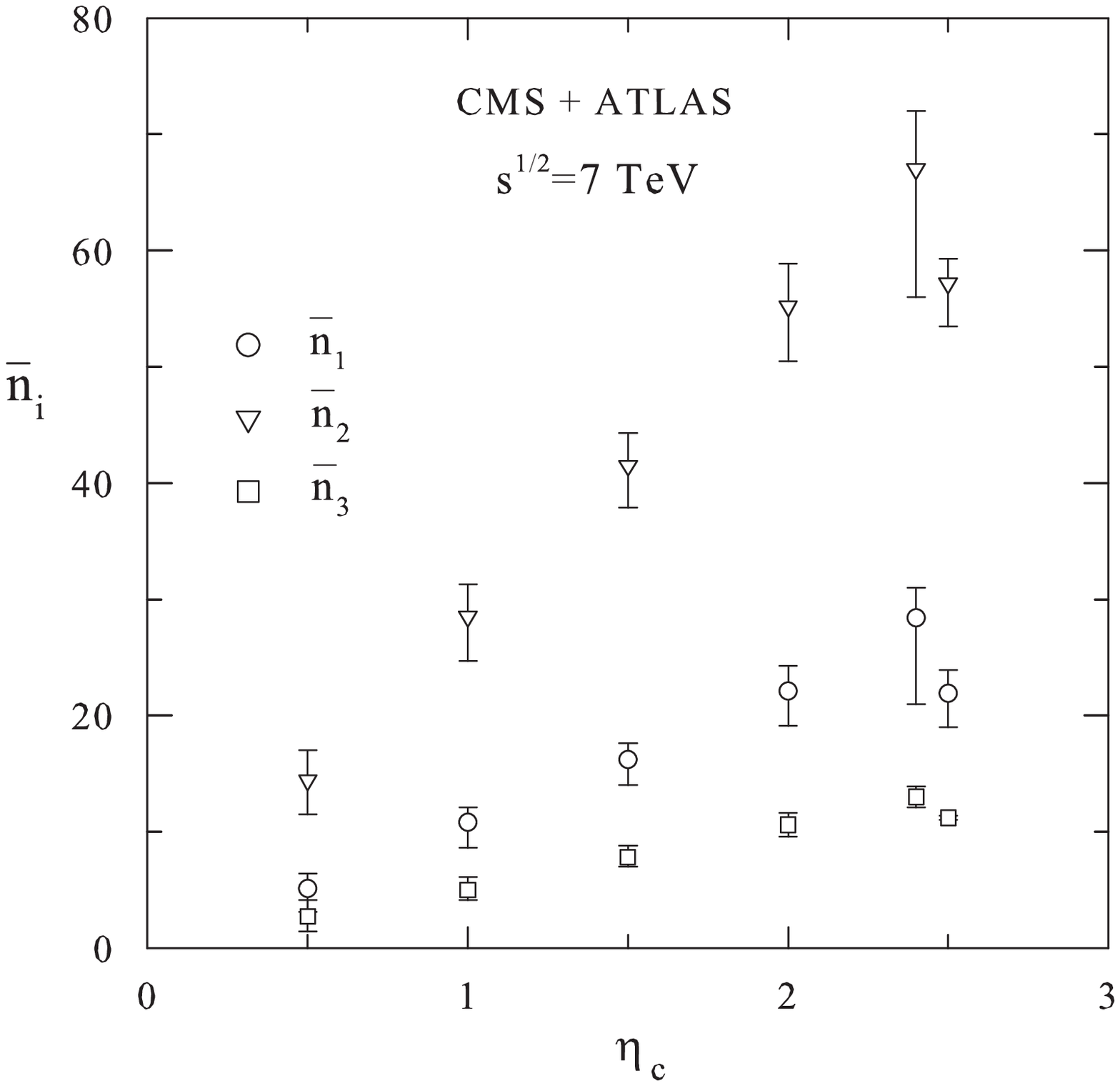}
\hspace{-0.5cm}
\includegraphics[width=78mm,height=78mm]{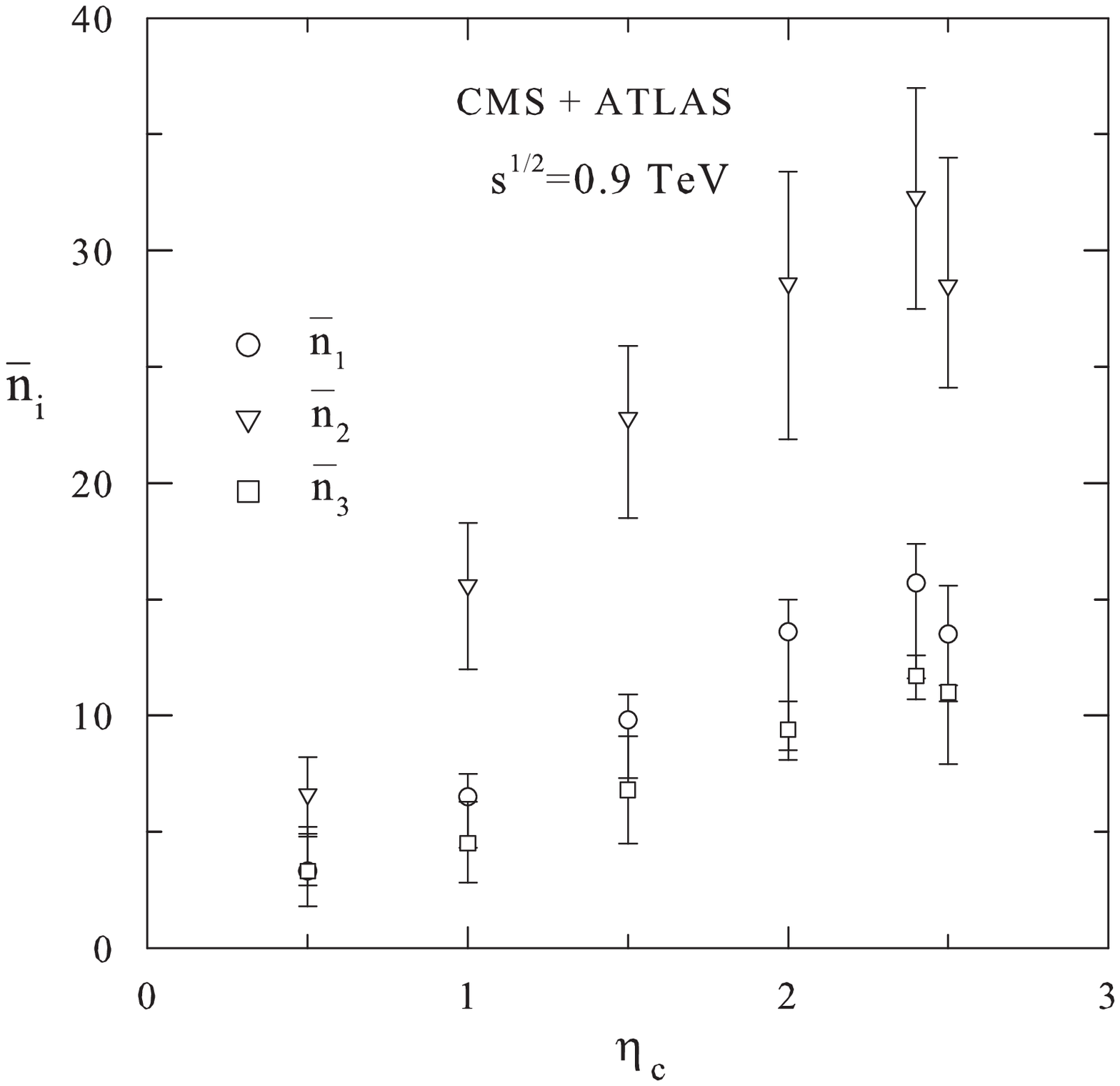}
\vskip -10mm
\hspace{1.cm}    (a) \hspace*{70mm} (b)
\caption{\label{Fig17}
The $\eta_c$ dependence of the average multiplicities $\bar{n}_i$
of single components of weighted superposition of three NBDs fitted to the 
NSD CMS data \protect\cite{CMS} and the most inclusive sample of the ATLAS data \protect\cite{ATLAS}
at (a) $\sqrt{s}=7$~TeV and (b) $\sqrt{s}=0.9$~TeV. 
}
\end{center}
\end{figure} 

Figure \ref{Fig17} shows the average multiplicities $\bar{n}_i$ of 
the negative binomial components of the total distribution 
in dependence on the size of the window $|\eta|<\eta_c$ 
at the energies $\sqrt{s}=7$~TeV and 0.9~TeV. 
The values of $\bar{n}_i$ extracted from fits to the minimum bias CMS data  
are depicted for five consecutive intervals in pseudorapidity from $\eta_c=0.5$ up to $\eta_c=2.4$. 
The symbols at the point
$\eta_c=2.5$ correspond to the parameters obtained from analysis of the most inclusive ATLAS data with  
$p_T>100$~MeV/c and $n_{ch}>1$. The average multiplicities $\bar{n}_i$ demonstrate an approximate 
linear increase with the window size $\eta_c$. Within the errors indicated, the linear dependencies can be extrapolated to the point $ \bar{n}_i=0$ at $ \eta_c=0$. 
These properties reflect the extensive character of 
the average multiplicities for all three components of the total distribution.   
The energy dependences of single $\bar{n}_i$ are mutually different. 
While the average multiplicities of the first
and the second component, $\bar{n}_1$ and $\bar{n}_2$, increase with $\sqrt{s}$, 
the average multiplicity $\bar{n}_3$ of the third component 
is within errors approximately energy independent for all $\eta_c$. 
The last statement is an extrapolation
in the pseudorapidity interval $|\eta|<0.5$   
because the CMS data at $\sqrt{s}=0.9$~TeV does not allow to extract reliable 
information on the third component in the smallest window.

\begin{figure}[t]
\begin{center}
\vskip 0cm
\hspace*{0mm}
\includegraphics[width=78mm,height=78mm]{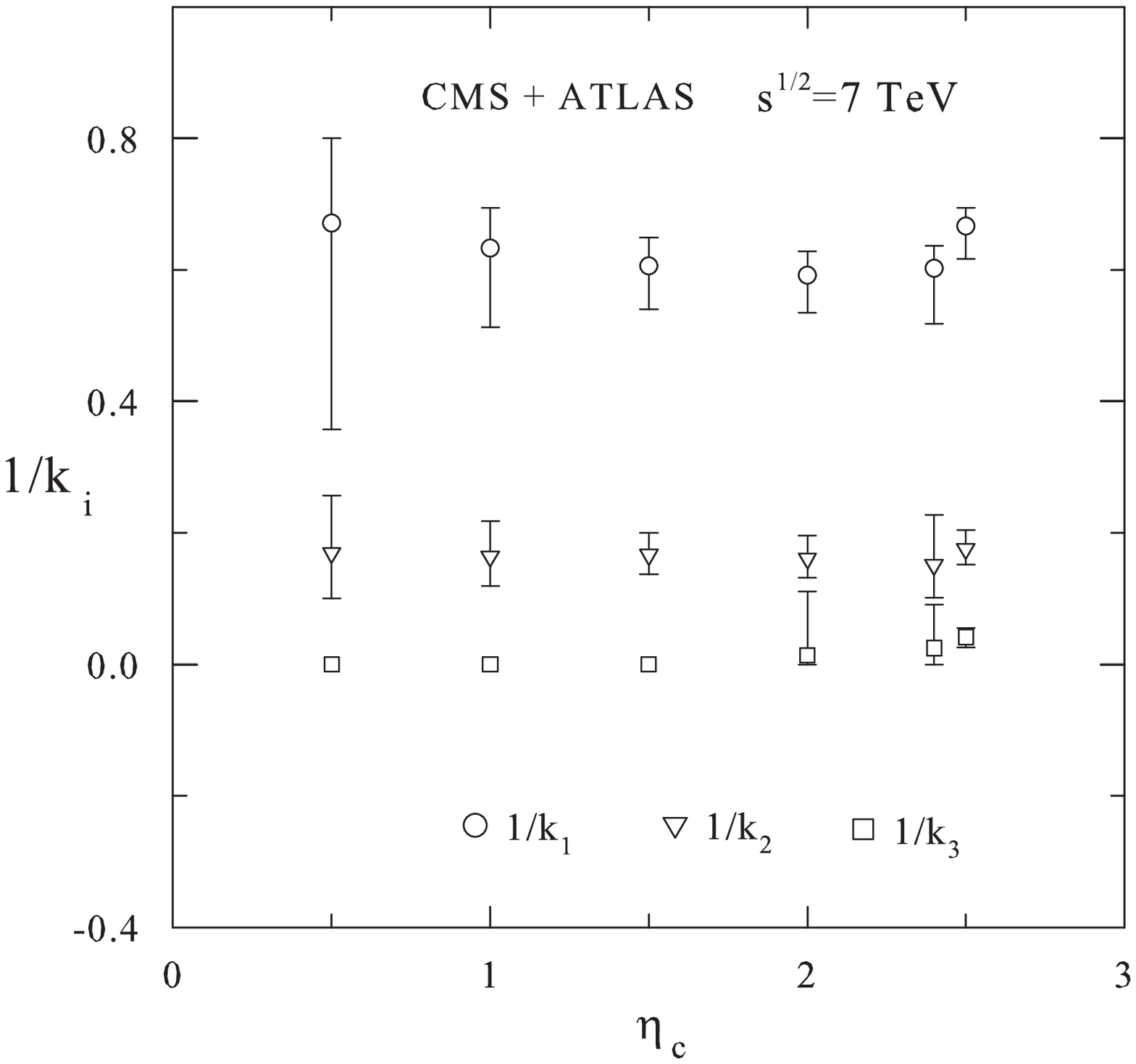}
\hspace{-0.5cm}
\includegraphics[width=78mm,height=78mm]{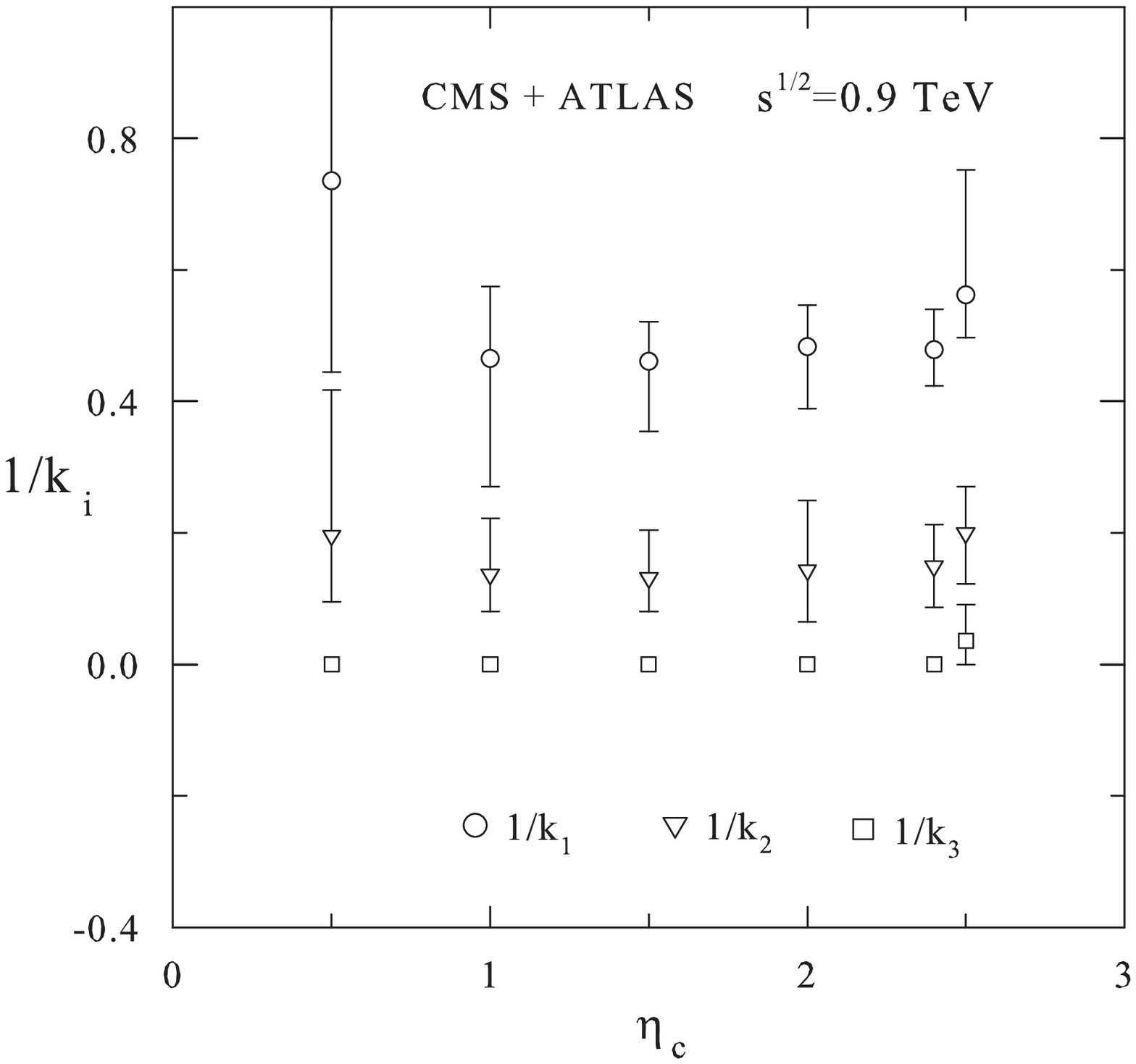}
\vskip -10mm
\hspace{1.cm}    (a) \hspace*{70mm} (b)
\caption{\label{Fig18}
The $\eta_c$ dependence of the inverse values of the parameters $k_i$ 
of single components of weighted superposition of three NBDs fitted to the 
NSD CMS data \protect\cite{CMS} and the most inclusive sample of the ATLAS data \protect\cite{ATLAS}
at (a) $\sqrt{s}=7$~TeV and (b) $\sqrt{s}=0.9$~TeV. 
}
\end{center}
\end{figure} 

Figure \ref{Fig18} shows the inverse values of the parameters $k_i$ of weighted superposition of three NBDs 
at the energies $\sqrt{s}=7$ and 0.9~TeV as a function of $\eta_c$.
The parameters were found from analysis of the same data as in the case of 
figures \ref{Fig16} and \ref{Fig17}.
Within the errors indicated, the values of $k_i$ do not depend on $\eta_c$ at both energies.
An exception is the window $|\eta|<0.5$ at $\sqrt{s}=0.9$~TeV where 
$k_1$ and $k_2$ show a tendency to decrease.   
This however should be taken with caution because 
application of three NBDs is  problematic 
in the narrowest window at this energy and 
the parameters were obtained here with fixed value of $\alpha_2$ only. 
As concerns the energy dependence, the parameter $k_1$ demonstrates some decrease with $\sqrt{s}$. 
This means that the dominant component of the total distribution reveals broadening 
as the energy increases.  
At the same time $k_2$ and $k_3$ are approximately energy independent for the minimum biased data. 
Therefore, the energy dependence of the NBD components in the tail and under the peak of MD   
is given entirely in terms of the respective average multiplicities $\bar{n}_2$ and $\bar{n}_3$.
Because the latter does not depend on energy, 
the peak at the maximum of the distribution reveals invariant properties. 
The imposed $p_T$ cut influences the width of the peak
at low $n$ but leaves the property of the energy independence of $\bar{n}_3$ preserved.

\section{Summary}

The MDs of charged particles have been studied in limited intervals of pseudorapidity
using the high statistic data obtained in proton-proton collisions at the LHC. 
The analysis concerns the symmetric pseudorapidity windows 
with the center at $\eta=0$. 
The minimum biased data measured at the LHC in the forward region was analyzed as well. 
The performed combined analysis of data obtained by the ATLAS, CMS, ALICE  and
LHCb Collaborations shows that besides the shoulder-like
structure at high multiplicities, MD manifests a distinct peak in the low multiplicity region. 
The peak is best visible in the ATLAS measurements which are based 
on recording the high statistics more than 10M events.
The description of the data within two-component superposition of NBDs is unsatisfactory.
The ATLAS data can be well parametrized by a weighted sum of three NBDs.
The main features of the three-component description 
are confirmed by the analysis of the CMS data in different pseudorapidity windows.
The obtained parametrization of MD reveals some invariant properties.
There are indications, at least at the energy $\sqrt{s}=7$~TeV, that the parameters $k_i$ of 
single NBDs do not depend on the window size. 
The analysis supports an idea that the pseudorapidity invariance concerns  
the probability $\alpha_i$ of each component as well. 
The  component in the tail of the distribution and the component under the peak at low $n$  
have nearly energy independent shapes for the minimum biased data samples.

The third component of the total distribution reflects the character of the peaky 
structure of MD at maximum.
Its probability $\alpha_3$ is at the level of a few per cent.
The  component is narrow, well-described by the Poisson distribution (or by a NBD with a large value of $k_3$). 
The corresponding average multiplicity $\bar{n}_3$ is approximately energy independent. 
The low-multiplicity component does not diminish when a cut on the transverse momentum is imposed.
Its probability $\alpha_3$ increases with the $p_T$ cut and the 
respective NBD becomes wider with a smaller value of the parameter $k_3$. 
The third component is best visible in the larger pseudorapidity windows.
A sign of the component is seen also in the minimum biased data measured 
by the LHCb Collaboration in the forward region.
The analysis of the LHC data on MD excludes interpretation
of this Poisson-like component in terms of hard processes which could lead
to \lq elbow-like\rq\ behavior at the tail of the distributions considered in \cite{GiovUgo3}.

Within the performed analysis and on the basis of the studied material we conclude that, 
besides the shoulder-like structure at high multiplicities, the high statistic data on MD 
measured in $pp$ collisions at the LHC manifest existence of a distinct peak visible at low $n$.
The total distribution can be well-described by a weighted sum of three NBDs. 
The parametrization of data reveals remarkable 
properties concerning the energy and pseudorapidity invariance of some characteristics 
of single components of the total distribution.  
Whether the third multiplicity component corresponds to a specific class of events or 
appearance of the characteristic peak in MD is a manifestation of new phenomena in multiparticle production 
remains an open question. 

We have shown that the ATLAS data on multiplicities 
measured by recording high statistics defy description within the model of IPPIs 
even if the probabilities of single interactions are considered as free parameters. 
Though the IPPI model takes somehow into account correlations of particles via assumed NBDs 
and their convolutions in each binary collision, further correlations between the interacting pairs of 
partons are probably needed to capture the enlarge role of collective effects at LHC energies.
A special question is whether the observed shift of the data at maximum of MD relative to 
the distributions shown in figure~\ref{Fig1} could be interpreted as a signature of a phase transition 
in hadron collisions at high energies.
Investigations in that direction are however beyond reach of this paper.
Physical interpretation of the obtained results represents a challenging problem and requires further
experimental and theoretical study.

\ack{ 
The investigations have been supported by RVO61389005 and 
by the Ministry of Education,
Youth and Sports of the Czech Republic grant LA08002.}
 
\appendix 
\section*{Appendix}
\setcounter{section}{1}

The results of our analysis are based on the computation and minimization of $\chi^2$
for various data sets presented in different forms as concerns the binning in multiplicity and error estimation.
Let us summarize details of the least-squares criteria and the conditions used in the minimization procedure.
The value of $\chi^2$ was computed as  
\begin{equation}
\chi^2=\sum\limits_{m=m_1}^{m_2}
\frac{(P^{ex}_m-P^{ph}_m)^2}{\sigma^2_m}
\label{eq:a1}
\end{equation} 
over the finite range $<m_1,m_2>$ of the multiplicity bins. 
The total number of bins considered in experiment is denoted by $m_2$. The value of $m_1=1$ or $2$ depends 
on the condition whether the first bin with the lowest multiplicity is included in the fit or not.
The bins are defined using the information accessible in the following form:
the probability $P^{ex}_m$ in the point $n_m$ and the bin interval $(a_m,b_m)$. 
In general, the values of $n_m$ are not given as integers.
In the case where the $m$th bin consists of a single multiplicity $n$, 
$n_m=n$ and $P^{ph}_m = P(n)$ with $P(n)$ given by (2).
For the bins containing more multiplicities, such as in the tails of the distributions,  
we distinguish two cases. 
If the point $n_m$ lies in the middle of the bin, $n_m=([b_m]+[a_m]+1)/2$ (CMS data), 
we use the average value  
\begin{equation}
P^{ph}_m=([b_m]-[a_m])^{-1}\sum\limits_{n=[a_m]+1}^{[b_m]} P(n).   
\label{eq:a2}
\end{equation} 
If $n_m\neq ([b_m]+[a_m]+1)/2$ as in the ATLAS data, we define   
\begin{equation}
P^{ph}_m= (1-\Delta)P([n_m]) + \Delta P([n_m]+1), 
\ \ \ \
\Delta\equiv n_m-[n_m].
\label{eq:a3}
\end{equation} 
All values of  $P^{ph}_m$ are then renormalized to fulfil the condition
\begin{equation}
\sum\limits_{m=m_1}^{m_2} P^{ph}_m =  
\sum\limits_{m=m_1}^{m_2} P^{ex}_m .
\label{eq:a2}
\end{equation}  
This ensures the same normalization of the phenomenological and experimental distributions over the fitted 
range of bins $<m_1,m_2>$.  
The value of $\chi^2$ is computed according to (\ref{eq:a1}) from the normalized distributions in each iteration step.
The experimental errors $ \sigma_m $ are calculated from a quadratic sum of the statistical and systematic 
uncertainties. 
For the asymmetric systematic errors we used 
$\sigma_m=\sigma_m^{+}$ if $ P^{ph}_m>P^{ex}_m$ and $\sigma_m=\sigma_m^{-}$ if $ P^{ph}_m<P^{ex}_m$.
The results of the analysis with the CERN-MINUIT program are given in tables~1-7. 
The obtained values of $\chi^2/dof$ are quoted in the separate rows. The number of degrees of freedom, $dof$, 
is stated explicitly as number of the experimental points minus number of the free parameters and 
minus the first experimental point, if not included into the fit. 
 
Here we state the parameters of fits of the IPPI model to data discussed in the text.
We obtained the following values with the condition $\alpha_i=\alpha^i$:
$\chi^2/dof=$2133/(85-3) for  $m=15.1$, $k=0.926$ and $N=6$ (dashed line in figure \ref{Fig1}(a));
$\chi^2/dof=$117/(51-3) for  $m=2.7$, $k=3.438$ and $N=5$ (dashed line in figure \ref{Fig1}(b));
$\chi^2/dof=$49.6/(127-2-1) for $m=7.86$, $k=2.103$ and $N\equiv 6$ (dashed line in figure \ref{Fig2}(a));
$\chi^2/dof=$14.7/(68-3-1) for  $m=3.39$, $k=3.099$ and $N=4$ (dashed line in figure \ref{Fig2}(b)).

Allowing the probabilities $\alpha_i$ to be free parameters with the restriction $\alpha_{i+1}\le\alpha_i$, we obtained:
$\chi^2/dof=$91.7/(85-11) for $m=3.82$, $k=2.896$,
$ \alpha_1=0.4555$, $ \alpha_2=\alpha_3=0.1462$, $ \alpha_4=0.1069$, $ \alpha_5=\alpha_6=0.0477$,   
$ \alpha_7=0.0331$, $ \alpha_8=\alpha_9=0.0056$  and $ \alpha_{10}=0.0055$ 
(black dots in figure \ref{Fig1}(a));
$\chi^2/dof=$74.7/(51-5) for $m=4.01$, $k=2.868$, 
$ \alpha_1=0.677$, $ \alpha_2=0.158$, $ \alpha_3=0.157$ and $ \alpha_4=0.008$
(black dots in figure \ref{Fig1}(b));
$\chi^2/dof=$6.6/(127-13-1) for $m=4.184$, $k=3.258$,
$ \alpha_1=0.4566$, $ \alpha_2=\alpha_3=0.1750$, $ \alpha_4=0.0774$, $ \alpha_5=0.0601$,   
$ \alpha_6=0.0303$, $ \alpha_7=0.0147$, $ \alpha_8=0.0088$, $ \alpha_9=0.0006$
and $ \alpha_{10}=\alpha_{11}=\alpha_{12}=0.0005$  
(dotted line in figure \ref{Fig2}(a));
$\chi^2/dof=$11.5/(68-4-1) for $m=4.29$, $k=2.833$, 
$ \alpha_1=0.621$, $ \alpha_2=0.228$ and $ \alpha_3=0.151$
(dotted line in figure \ref{Fig2}(b)).

\end{document}